\documentclass[aps,prd,twocolumn,showpacs,groupedaddress,floatfix,
               amsmath,amssymb]{revtex4-1}

\usepackage{graphicx}%
\usepackage{color}%
\usepackage{dcolumn}
\usepackage{bm}

\def\urltilda{\kern -.15em\lower .7ex\hbox{\~{}}\kern .04em}

\begin{document}

\title{Fluctuating Surface Currents: A New Algorithm for 
       Efficient Prediction of Casimir Interactions among 
       Arbitrary Materials in Arbitrary Geometries}

%!!!!!!!!!!!!!!!!!!!!!!!!!!!!!!!!!!!!!!!!!!!!!!!!!!
%! authors section, take 2
%!!!!!!!!!!!!!!!!!!!!!!!!!!!!!!!!!!!!!!!!!!!!!!!!!!
\author{M. T. Homer Reid$^{1}$%
        \footnote{Electronic address: \texttt{homereid@mit.edu}}%
        \footnote{URL: \texttt{http://www.mit.edu/\urltilda homereid}},
       Jacob White$^{1,2}$,
       and Steven G. Johnson$^{1,3}$}
\affiliation{ $^{1}$Research Laboratory of Electronics,
              Massachusetts Institute of Technology, 
              Cambridge, MA 02139, USA \\
              $^{2}$Department of Electrical Engineering and Computer Science,
              Massachusetts Institute of Technology, 
              Cambridge, MA 02139, USA \\
              $^{3}$Department Of Mathematics,
              Massachusetts Institute of Technology, 
              Cambridge, MA 02139, USA}

\date{\today}

%%%%%%%%%%%%%%%%%%%%%%%%%%%%%%%%%%%%%%%%%%%%%%%%%%
%%%%%%%%%%%%%%%%%%%%%%%%%%%%%%%%%%%%%%%%%%%%%%%%%%
% homemade command definitions %%%%%%%%%%%%%%%%%%%
%%%%%%%%%%%%%%%%%%%%%%%%%%%%%%%%%%%%%%%%%%%%%%%%%%
%%%%%%%%%%%%%%%%%%%%%%%%%%%%%%%%%%%%%%%%%%%%%%%%%%
%\input{SpecialCommands}

\newcommand{\vb}[1]{\mathbf{#1}} 
\newcommand{\sups}[1]{^{\hbox{\scriptsize{#1}}}}
\newcommand{\supt}[1]{^{\hbox{\tiny{#1}}}}
\newcommand{\subs}[1]{_{\hbox{\scriptsize{#1}}}}
\newcommand{\subt}[1]{_{\hbox{\tiny{#1}}}}
\newcommand{\vbhat}[1]{\vb{\hat #1}}
\newcommand{\mc}[1]{\mathcal{#1}}
\newcommand{\bmc}[1]{\boldsymbol{\mathcal{#1}}}
\newcommand{\DD}{\boldsymbol{\mathfrak{D}}}
\newcommand{\eescat}{^{\hbox{\tiny{EE}\scriptsize{,scat}}}}
\newcommand{\emscat}{^{\hbox{\tiny{EM}\scriptsize{,scat}}}}
\newcommand{\mescat}{^{\hbox{\tiny{ME}\scriptsize{,scat}}}}
\newcommand{\mmscat}{^{\hbox{\tiny{MM}\scriptsize{,scat}}}}
\newcommand\Tr{\hbox{Tr }}
\newcommand{\pard}[2]{\frac{\partial #1}{\partial #2}}

\newcommand{\BG}{\boldsymbol{\Gamma}}
\newcommand{\bg}{\boldsymbol{\gamma}}

\newcommand{\MInt}{\hat M}
\newcommand{\NInt}{\hat N}
\newcommand{\MExt}{\check M}
\newcommand{\NExt}{\check N}

\newcommand{\xInt}{\vb{\hat x}}
\newcommand{\xExt}{\vb{\check x}}

\newcommand{\vbMInt}{\vb{\hat M}}
\newcommand{\vbNInt}{\vb{\hat N}}
\newcommand{\vbMExt}{\vb{\check M}}
\newcommand{\vbNExt}{\vb{\check N}}

\newcommand{\vbRho}{\boldsymbol{\rho}}

\newcommand{\EEe}{^{\hbox{\tiny{EE}\scriptsize{,$e$}}}}
\newcommand{\MEe}{^{\hbox{\tiny{ME}\scriptsize{,$e$}}}}
\newcommand{\EMe}{^{\hbox{\tiny{EM}\scriptsize{,$e$}}}}
\newcommand{\MMe}{^{\hbox{\tiny{MM}\scriptsize{,$e$}}}}

\newcommand{\EEn}{^{\hbox{\tiny{EE}\scriptsize{,$n$}}}}
\newcommand{\MEn}{^{\hbox{\tiny{ME}\scriptsize{,$n$}}}}
\newcommand{\EMn}{^{\hbox{\tiny{EM}\scriptsize{,$n$}}}}
\newcommand{\MMn}{^{\hbox{\tiny{MM}\scriptsize{,$n$}}}}
\newcommand{\PQn}{^{\hbox{\tiny{PQ}\scriptsize{,$n$}}}}

\newcommand{\EEr}{^{\hbox{\tiny{EE}\scriptsize{,$r$}}}}
\newcommand{\MEr}{^{\hbox{\tiny{ME}\scriptsize{,$r$}}}}
\newcommand{\EMr}{^{\hbox{\tiny{EM}\scriptsize{,$r$}}}}
\newcommand{\MMr}{^{\hbox{\tiny{MM}\scriptsize{,$r$}}}}
\newcommand{\PQr}{^{\hbox{\tiny{PQ}\scriptsize{,$r$}}}}

\newcommand{\EE}{^{\hbox{\tiny{EE}}}}
\newcommand{\MM}{^{\hbox{\tiny{MM}}}}
\newcommand{\ME}{^{\hbox{\tiny{ME}}}}
\newcommand{\EM}{^{\hbox{\tiny{EM}}}}

\newcommand\ler{L^{\text{\tiny{E}},r}}
\newcommand\lmr{L^{\text{\tiny{M}},r}}
\newcommand\vble{\mathbf{L}^{\text{\tiny{E}}}}
\newcommand\vblm{\mathbf{L}^{\text{\tiny{M}}}}
\newcommand\vbler{\mathbf{L}^{\text{\tiny{E}},r}}
\newcommand\vblmr{\mathbf{L}^{\text{\tiny{M}},r}}
\newcommand\vblee{\mathbf{L}^{\text{\tiny{E}},e}}
\newcommand\vblme{\mathbf{L}^{\text{\tiny{M}},e}}

\newcommand\eeo{^{ \text{{\tiny EE,1}}}}
\newcommand\emo{^{ \text{{\tiny EM,1}}}}
\newcommand\meo{^{ \text{{\tiny ME,1}}}}
\newcommand\mmo{^{ \text{{\tiny MM,1}}}}
\newcommand\eet{^{ \text{{\tiny EE,2}}}}
\newcommand\emt{^{ \text{{\tiny EM,2}}}}
\newcommand\met{^{ \text{{\tiny ME,2}}}}
\newcommand\mmt{^{ \text{{\tiny MM,2}}}}
\newcommand\eee{^{ \text{{\tiny EE}},e}}
\newcommand\eme{^{ \text{{\tiny EM}},e}}
\newcommand\mee{^{ \text{{\tiny ME}},e}}
\newcommand\mme{^{ \text{{\tiny MM}},e}}
\newcommand{\numeq}[2]{\begin{equation} #2 \label{#1} \end{equation}}

\newcommand{\GZO}{G_0^{\vb 1}}
\newcommand{\GZT}{G_0^{\vb 2}}
\newcommand{\GO}{G^{\vb 1}}
\newcommand{\GT}{G^{\vb 2}}
\newcommand{\CO}{C^{\vb 1}}
\newcommand{\CT}{C^{\vb 2}}
\newcommand{\Vbar}{\overline{V}}
\newcommand{\ibar}{\overline{\mathcal{I}}}
\newcommand{\PP}[2]{\partial_{#1}\partial_{#2}}
\newcommand{\PPP}[3]{\partial_{#1}\partial_{#2}\partial_{#3}}
\newcommand{\nn}{\nonumber \\}

%%%%%%%%%%%%%%%%%%%%%%%%%%%%%%%%%%%%%%%%%%%%%%%%%%
%%%%%%%%%%%%%%%%%%%%%%%%%%%%%%%%%%%%%%%%%%%%%%%%%%
% abstract
%%%%%%%%%%%%%%%%%%%%%%%%%%%%%%%%%%%%%%%%%%%%%%%%%%
%%%%%%%%%%%%%%%%%%%%%%%%%%%%%%%%%%%%%%%%%%%%%%%%%%
\begin{abstract}
This paper presents a new method 
for the efficient numerical computation of Casimir 
interactions between objects of arbitrary geometries, 
composed of materials with arbitrary frequency-dependent 
electrical properties. Our method formulates the Casimir 
effect as an interaction between effective electric and 
magnetic current distributions on the surfaces of material 
bodies, and obtains Casimir energies, forces, and torques 
from the spectral properties of a matrix that 
quantifies the interactions of these surface currents. 
The method can be formulated and understood in two 
distinct ways: \textbf{(1)} as a consequence of 
the familiar \textit{stress-tensor} approach to Casimir 
physics, or, alternatively, \textbf{(2)} as a particular 
case of the \textit{path-integral} approach to Casimir 
physics, and we present
both formulations in full detail.
In addition to providing an algorithm for computing
Casimir interactions in geometries that could not be 
efficiently handled by any other method, the framework 
proposed here thus achieves an explicit unification of 
two seemingly disparate approaches to computational Casimir 
physics. 
\end{abstract}

\maketitle

%%%%%%%%%%%%%%%%%%%%%%%%%%%%%%%%%%%%%%%%%%%%%%%%%%
%%%%%%%%%%%%%%%%%%%%%%%%%%%%%%%%%%%%%%%%%%%%%%%%%%
% begin main text %%%%%%%%%%%%%%%%%%%%%%%%%%%%%%%%
%%%%%%%%%%%%%%%%%%%%%%%%%%%%%%%%%%%%%%%%%%%%%%%%%%
%%%%%%%%%%%%%%%%%%%%%%%%%%%%%%%%%%%%%%%%%%%%%%%%%%
%\input{IntroductionSection}

\section{Introduction}
\label{IntroductionSection}

This paper presents a new method 
for the efficient numerical computation of Casimir 
interactions between objects of arbitrary geometries, 
composed of materials with arbitrary frequency-dependent 
electrical properties. Our method formulates the Casimir 
effect as an interaction between effective electric and 
magnetic currents on the surfaces of material 
bodies, and obtains Casimir energies, forces, and torques 
from the spectral properties of a matrix that 
quantifies the interactions of these surface currents.
Our final formulas for Casimir quantities---equations 
(\ref{MasterFSCFormulas}) below---may be derived in two
distinct ways: \textbf{(a)} by integrating the 
Maxwell stress tensor over a closed bounding surface,
as is commonly done in purely numerical approaches
to Casimir computation~\cite{Johnson2011}, but with the 
distinction that here we evaluate the surface integral
\textit{analytically}; or, alternatively,
\textbf{(b)} by evaluating a path-integral expression
for the Casimir energy, as is commonly done in 
quasi-analytical approaches to Casimir physics~\cite{Rahi10},
but with the distinction that here we are not restricted
to the use of geometry-specific special functions.
In this paper, we present these two distinct derivations
of our master formulas (\ref{MasterFSCFormulas}) and
compare our new approach to existing computational Casimir 
methods. A free, open-source software package implementing
our method is available~\cite{scuff-em}; the technical details 
of this and other numerical implementations of our method will 
be discussed elsewhere.

Results obtained using our new technique have appeared in print
before~\cite{Reid2009, Reid2011A, Levin2010, Pan2011, McCauley2011},
and Refs.~\cite{Reid2009, Reid2011A} briefly sketched the 
path-integral derivation of our method, but omitted many details.
The purposes of the present paper are to furnish a complete 
presentation of the path-integral derivation and to present 
the alternative stress-tensor derivation, which has not 
appeared in print before. By arriving at identical 
formulas---our master formulas, equation (\ref{MasterFSCFormulas})---
from the two seemingly disparate starting points of 
path integrals and stress tensors, we explicitly demonstrate
the equivalence of these two formulations of Casimir
physics.

In particular, our demonstration of this equivalence furnishes
an alternative demonstration that the Maxwell stress tensor 
in dispersive media---questionable under ordinary circumstances---is
in fact valid in the thermodynamic context, as has been
argued on other grounds by Pitaevskii~\cite{Pitaevskii2006},
and by Philbin~\cite{Philbin2010,Philbin2011} in the context
of the canonical quantization of macroscopic electromagnetism.
An algebraic equivalence similar to ours, but relating a 
path-integral expression to the energy density instead 
of the Maxwell stress tensor, was demonstrated in 
Ref.~\cite{Milton2008A}, which used this equivalence to
explain why the dispersive energy density (which is valid in 
ordinary electrodynamics only for negligible 
dissipation~\cite{Jackson1998})
is the appropriate quantity to consider in the context of 
thermal and quantum fluctuations. Our work does for 
the stress tensor what Ref.~\cite{Milton2008A} does for
the energy density. 
(An alternative approach to relating
the stress-tensor picture to the energy viewpoint
was suggested in Ref.~\cite{Bimonte09}, but details were omitted; 
also, the method was restricted to geometries that 
admit a separating plane between objects, whereas the
method of this paper has no such restriction and
is applicable even to geometries containing objects
with interpenetrating features.)

Although Casimir physics has been with us for some seven
decades~\cite{Casimir1948}, the past fifteen years have 
witnessed a renaissance of interest in the field, 
driven by laboratory observations of Casimir phenomena 
in an increasingly complex variety of geometric and material 
configurations~\cite{Lamoreaux97, Capasso07, Munday09, 
Mohideen10, Rodriguez2011A}.
Whereas the theoretical methods used in the original Casimir 
prediction~\cite{Casimir1948} were restricted to the case
of simple geometries and idealized materials, recent experimental 
progress has spurred the development of theoretical techniques 
for predicting Casimir forces among bodies of \textit{arbitrary} 
shapes and material properties. Such general-purpose
Casimir methods have typically pursued one of
two general strategies.

A first approach~\cite{Rahi10, Genet2003, MaiaNeto2005, Lambrecht2006, 
Kenneth2008, Dalvit2010, Messina2011}
seeks to exploit geometrical symmetries by approximating Casimir 
quantities as expansions in \textit{special functions} (solutions 
of the scalar or vector Helmholtz equation in various coordinate 
systems)
that encode global geometric information in a concise way. 
(Techniques of this sort are often known as 
\textit{spectral methods}~\cite{Graham2009}.)
These methods have the virtue of yielding compact expressions 
relating Casimir energies, forces, and torques to linear-algebra 
operations (matrix inverse, determinant, and trace) on matrices 
describing the interactions of the global basis functions. As 
is commonly true for spectral methods, the expressions are 
rapidly convergent (in the sense of obtaining accurate 
numerical results with low-dimensional truncations 
of the matrices) for highly symmmetric geometries, 
but less well-suited to \textit{asymmetric} configurations,
where the very geometric specificity encoded in the 
closed-form Helmholtz solutions becomes more curse
than blessing and requires the special-function 
expansions to be carried out to high orders for even
moderate numerical precision.

An alternative approach is a numerical implementation of 
the \textit{stress-tensor} formulation of Casimir physics 
pioneered by Lifshitz et al.~\cite{DLP1961, LifshitzPitaevskii}.
Here the Casimir force on a body is obtained by 
integrating the Maxwell stress tensor---suitably averaged 
over thermal and quantum-mechanical fluctations---over a 
fictitious bounding surface surrounding the body; in
modern numerical 
approaches~\cite{Rodriguez2007,PasqualiMaggs2008,XiongChew2009, Johnson2011} 
the integral is evaluated by numerical cubature 
(that is, as a weighted sum of integrand samples)
with values of the stress tensor at each cubature point 
computed by solving numerical electromagnetic scattering problems. 
As compared to the special-function approach, this technique 
has the virtue of great generality, as it allows one to take 
advantage of the 
wide range of existing numerical techniques for solving 
scattering problems involving arbitrarily complex geometries 
and materials. The drawback is that the spatial integral over 
the bounding surface adds a layer of conceptual and 
computational complexity that is absent from the special-function 
approach. 

In this paper we show how the best features of these two 
approaches may be combined to yield a new technique for Casimir 
computations. Our \textit{fluctuating-surface-current}
(FSC) approach expresses Casimir 
energies, forces, and torques among bodies of arbitrary
geometries and material properties in terms of 
interactions among effective electric and magnetic 
currents flowing on the object surfaces.
The method borrows techniques from surface-integral-equation
formulations of electromagnetic scattering~\cite{Harrington93}
to represent objects entirely in terms of their surfaces---thus 
retaining the full flexibility of the numerical stress-tensor 
method in handling arbitrarily complex asymmetric geometries---but 
bypasses the unwieldy numerical cubatures of the usual stress-tensor 
approach to obtain Casimir energies, forces, and 
torques \textit{directly} from linear-algebra
operations (matrix inverse, determinant and trace) 
on matrices describing the interactions of the surface 
currents, thus retaining the conceptual simplicity 
and computational ease of the usual
special-function approach.

The FSC formulas for the zero-temperature 
Casimir energy, force, and torque are
%====================================================================%
\begin{subequations}
\begin{align}
\mathcal{E}
&=\frac{\hbar}{2\pi}\int_0^\infty \, d\xi \,
  \log \frac{\det \vb M(\xi)}{\det \vb M_\infty(\xi)}
\\[5pt]
\mathcal{F}_i
&=-\frac{\hbar}{2\pi}\int_0^\infty \, d\xi \,
   \Tr \Big\{ \vb M^{-1}(\xi) \cdot \pard{\vb M(\xi)}{\vb r_i} \Big\}
\\[5pt]
\mathcal{T}
&=-\frac{\hbar}{2\pi}\int_0^\infty \, d\xi \,
   \Tr \Big\{ \vb M^{-1}(\xi) \cdot \pard{\vb M(\xi)}{\theta} \Big\}
\end{align}
\label{MasterFSCFormulas}
\end{subequations}
where the precise form of the matrix $\vb M(\xi)$ is given in 
Section \ref{BEMReviewSection}.
%Here the $\xi$ 
%integrals extend over the positive imaginary frequency 
%axis, and the partial derivative in the
%force (torque) equation is taken with respect to a rigid 
%displacement (rotation) of the object in question.
Readers familiar with scattering-matrix methods for Casimir
computations~\cite{Rahi10, Lambrecht2006, Kenneth2008, Dalvit2010,
  Messina2011} will note the striking similarity of our equation
(\ref{MasterFSCFormulas}a) to the Casimir energy formulas reported in
those works (such as equation 5.13 of Ref.~\cite{Rahi10}); in both
cases, the Casimir energy is obtained by integrating over the
imaginary frequency axis, with an integrand expressed as a ratio of
matrix determinants. The difference lies in the meaning of the
matrices in the two cases; whereas the matrix in typical 
scattering-matrix Casimir methods descibes the interactions
of \textit{incoming and outgoing wave solutions of Maxwell's equations},
the matrix in our equations (\ref{MasterFSCFormulas}) describes
the interactions of \textit{surface currents} flowing on the 
boundaries of the interacting objects in a Casimir geometry.
This distinction has important ramifications for the 
convenience and generality of our method.

In traditional scattering-matrix Casimir methods, the matrix that
enters equations like (\ref{MasterFSCFormulas}) describes 
interactions among the elements of a basis of known solutions of
Maxwell's equations propagating to and from the interacting bodies.
Such treatments afford a highly efficient description of 
scattering in the handful of geometries for which analytical 
solutions are available---such as incoming and outgoing 
spherical waves for spherical scatterers, left- and 
right-traveling plane waves for planar geometries, 
cylindrical wave for cylinders, etc.---but may be particularly
\textit{inefficient} for describing more general objects, 
as, for example, if one attempts to describe scattering
from a cube using a basis of spherical waves.
Moreover, practical implementations of these methods require
significant retooling to accommodate new shapes of objects; 
if, for example, having formulated the method for spheres, 
one wishes instead to treat spheroids, one must recompute 
Maxwell solutions in a new coordinate system and reformulate
the matrices in equations like (\ref{MasterFSCFormulas}) 
to describe the interactions of these new solutions.

In contrast, the matrix in our equations (\ref{MasterFSCFormulas}) 
describes the interactions of surface currents flowing on the 
surfaces of the interacting objects in a Casimir geometry, as 
discussed in detail in Section \ref{BEMReviewSection}. A crucial
advantage of this description is that the basis we use to
represent surface currents is arbitrary; the basis functions
are \textit{not} required to solve the wave equation or 
any other equation, and the choice of basis is thus liberated
from the underlying physics of the problem---we are free to 
choose a basis that efficiently represents any given geometry.
One convenient choice---though by no means the only 
possibility---is a basis of \textit{localized} functions
conforming to a nonuniform surface-mesh discretization
(Figure~\ref{BEMCartoon}), where the mesh may be automatically
generated for arbitrarily complex 
geometries~\cite{Frey2000}.
A particular advantage of this type of basis is that, once
we have implemented our method using basis functions of this
type, we can apply it to arbitrary geometries with
almost no additional effort; in particular, having applied 
the method to spheres, it is essentially effortless
to apply it to cubes (Section~\ref{ApplicationsSection}).

The objective of this paper is to provide two separate derivations of
the master formulas (\ref{MasterFSCFormulas}), one based on the
stress-tensor formalism and making no reference to path integration,
and a second based on path integrals and making no reference to stress
tensors. These derivations contain a number of theoretical innovations
beyond the practical usefulness of the method itself; in particular,
in the stress-tensor derivation we state and prove a new integral
identity involving the homogeneous dyadic Green's functions of
Maxwell's equations (Appendix \ref{IntegralIdentityAppendix}), 
while in the path-integral derivation we introduce a new 
\textit{surface-current} representation of the Lagrange 
multipliers that constrain functional integrations over the 
electromagnetic field, which we expect to be a tool of 
general use in quantum field theory.

The Casimir method described in this paper is closely
related to the surface-integral-equation (SIE) formulation
of classical electromagnetic theory. Although well-known 
in the engineering literature~\cite{Harrington93},
this technique has not been extensively discussed in 
the Casimir literature, and for this reason we begin
in Section \ref{BEMReviewSection} with a brief review
of SIE theory.
Although the majority of this section
is a review of 
standard material, the explicit expressions for 
dyadic Green's functions that we derive in
\ref{BEMDGFSubsection}
have not, to our knowledge, appeared in print before.
In Sections 
\ref{StressTensorFSCSection} 
and
\ref{PathIntegralFSCSection},
which constitute the centerpiece 
of the paper, we present two separate derivations of the 
master FSC formulas (\ref{MasterFSCFormulas}); one 
derivation starts from the stress-tensor approach to 
Casimir physics (Section \ref{StressTensorFSCSection}), 
while an independent
derivation starts from a path-integral expression for the 
Casimir energy (Section \ref{PathIntegralFSCSection}). 
In Section \ref{PartialTraceEqualitySection}, we note
an important practical simplification that follows
from the structure of the matrices in equations
(\ref{MasterFSCFormulas}). 
In Section \ref{ApplicationsSection} we validate
our method by using it to reproduce known results, then
illustrate its generality by applying it 
to geometries that would be difficult to address using 
existing Casimir methods. 
(Further examples of the utility of 
our method may be found in 
Refs.~\cite{Reid2009, Reid2011A, Levin2010, Pan2011, McCauley2011}.)
Our conclusions are presented in Section \ref{ConclusionsSection}, 
and a number of technical details are relegated to the Appendices.
Technical details of practical numerical implementations,
as well as additional computational applications, will be 
discussed elsewhere.

\section{A Review of the Surface-Integral-Equation 
         Formulation of Classical Electromagnetism}
\label{BEMReviewSection}

%####################################################################%
%####################################################################%
%####################################################################%
\begin{figure*}
\begin{center}
\resizebox{\textwidth}{!}{\includegraphics{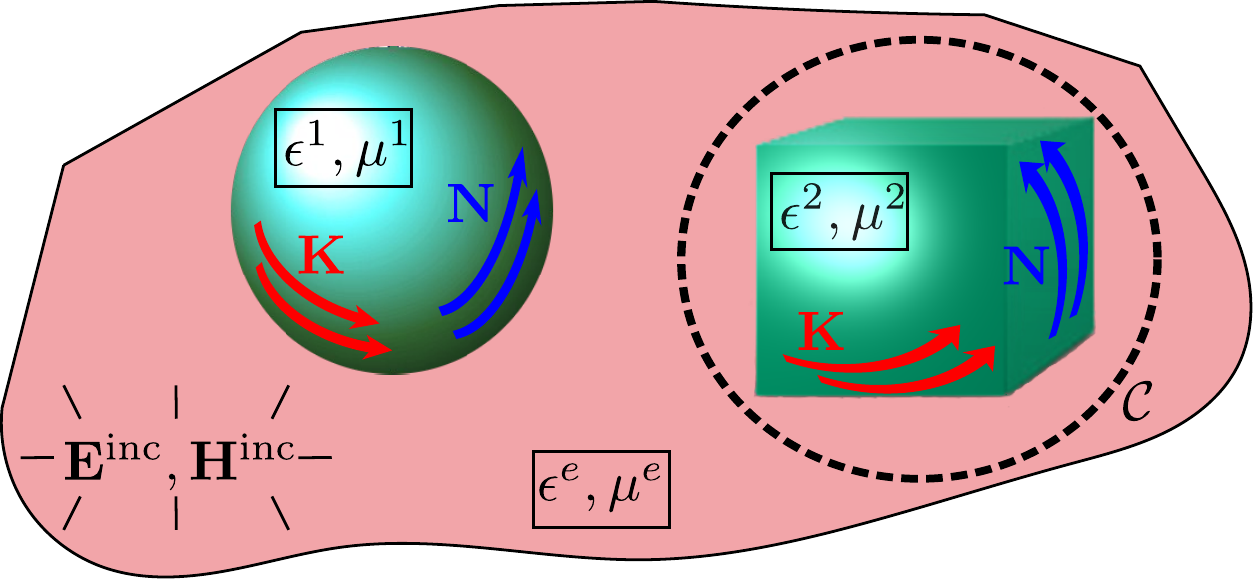}}
\caption{Schematic depiction of a scattering geometry in
the surface-integral-equation picture. A 
collection of arbitrarily-shaped homogeneous bodies,
with frequency-dependent relative electrical properties 
$\{\epsilon^r, \mu^r\}$, is embedded in a homogeneous 
medium with electrical properties $\{\epsilon^e,\mu^e\}$.
Incident radiation, characterized by 
electric and magnetic fields 
$\mathbf E\sups{inc}, \mathbf{H}\sups{inc}$, 
impinges on the objects to induce \textit{surface currents};
for perfectly conducting objects we have only
electric surface currents ($\vb K$), while for
general objects we have equivalent
electric and magnetic ($\vb N$) surface currents.
The goal of surface-integral-equation methods is to
solve for the surface-current distributions in terms 
of the incident fields, after which we can compute the 
scattered fields anywhere in space from the surface
currents. [The dotted line indicates a fictitious 
bounding contour $\mathcal{C}$ surrounding one of 
the objects over which we integrate the Maxwell 
stress tensor to compute the Casimir force on that 
object (Section \ref{StressTensorFSCSection}.)]}
\label{ScatteringCartoon}
\end{center}
\end{figure*}
%####################################################################%
%####################################################################%
%####################################################################%

Computational Casimir physics is intimately related to
the theory of classical electromagnetic scattering, and 
many practical methods for predicting Casimir interactions 
are based on well-known techniques for solving scattering 
problems. Among the classical scattering methods that have
been appropriated for Casimir purposes are the T-matrix
method~\cite{Waterman1965, Kenneth2008, Rahi2009}, 
the method of reflection coefficients~\cite{Lambrecht2006},
and the numerical finite-difference 
method~\cite{Rodriguez2007, PasqualiMaggs2008}.
The method discussed in this paper derives from 
yet another well-known approach to scattering problems,
namely, the method of \textit{surface integral equations}
(SIEs). (SIE techniques were first used for
Casimir studies in Ref.~\cite{Reid2009}, while
Refs.~\cite{XiongChew2009,XiongChew2010}
presented an SIE-based implementation of the 
numerical stress-tensor method.)
As background for the remainder of this paper, in 
this section we review the well-known SIE procedure.

Surface-integral techniques have a long history in
electromagnetic theory~\cite{Rengarajan2000},
dating back to the equivalence principles of 
Love~\cite{Love1901} and 
Schelkunoff~\cite{Schelkunoff1936}
and the Stratton-Chu equations~\cite{StrattonChu}
from the first half of the 20th century.
Numerical implementation of SIEs---known 
as the ``boundary-element method'' (BEM) or 
the ``method of moments''---emerged in the
1970s as an 
alternative to other computational procedures such as 
the finite-difference method (FDM) and the finite-element 
method (FEM)~\cite{Harrington1961, Medgyesi1994}.
Whereas the FDM and the FEM proceed by numerically solving a 
spatially \textit{local} form of Maxwell's equations, and 
thus allow treatment of materials with essentially arbitrary
spatial variation of the dielectric permittivity and magnetic 
permeability, the SIE approach takes advantage of the 
analytically known solutions of Maxwell's equations in 
homogeneous media, and thus in practice is most readily 
applicable to piecewise-homogeneous material configurations.
For this reason, while the FDM and FEM have the advantage of 
being able to treat a wider class of materials, the SIE method
exhibits significant practical advantages for the 
piecewise-homogeneous geometries typically encountered in 
Casimir studies.  

To fix ideas and notation for the remainder of this paper,
we here review the SIE formulation of electromagnetic 
scattering problems, beginning in 
Section \ref{PECBEMSubsection} with the simplest case 
of perfectly electrically conducting (PEC) scatterers,
and then generalizing in Section \ref{PMCHWBEMSubsection}
to the case of arbitrary materials.

The material of these two subsections
is well-known and
entirely standard within the computational electromagnetism
literature, and is reviewed here only for completeness.
However, in Section \ref{BEMDGFSubsection} we extend
the SIE formalism one step beyond what is usually done
to write explicit expressions [equations (\ref{BEMDGFPEC})
and (\ref{BEMDGFPMCHW})] for scattering dyadic Green's
functions in terms of the SIE matrices and the homogeneous 
dyadic Green's functions (DGFs). Although these expressions are
straightforward consequences of the standard BEM procedure
outlined in Sections IIA-B, to our knowledge they are 
appearing here for the first time.

Throughout this section we will refer to the 
scattering situation depicted schematically in
Figure \ref{ScatteringCartoon}, in which a collection
of homogeneous scatterers (with frequency-dependent
relative electrical properties $\{\epsilon^r, \mu^r\}$) 
is embedded
in a homogeneous medium (electrical properties 
$\{\epsilon^e, \mu^e\}$) and irradiated by incident radiation
characterized by an incident electric field $\mathbf E\sups{inc}$.

%%%%%%%%%%%%%%%%%%%%%%%%%%%%%%%%%%%%%%%%%%%%%%%%%%%%%%%%%%%%%%%%%%%%%%
%%%%%%%%%%%%%%%%%%%%%%%%%%%%%%%%%%%%%%%%%%%%%%%%%%%%%%%%%%%%%%%%%%%%%%
%%%%%%%%%%%%%%%%%%%%%%%%%%%%%%%%%%%%%%%%%%%%%%%%%%%%%%%%%%%%%%%%%%%%%%
\subsection{The SIE Method For PEC Bodies}
\label{PECBEMSubsection}

\begin{table*}
\begin{center}
\fbox{
\begin{tabular}{lp{0.4\textwidth}p{0.05\textwidth}lp{0.4\textwidth}}
%--------------------------------------------------------------------%
%- row 1 
%--------------------------------------------------------------------%
  %
  % left column 
  %
  $\mathcal{O}_r$    & $r$th homogeneous object
                       (exterior medium is $\mathcal{O}_e$)
  &\hspace{0.1in}&
%
  %
  % right column
  %
  $\xi$    & Imaginary frequency $(\omega=i\xi)$
\\[6pt]
%--------------------------------------------------------------------%
%- row 2 
%--------------------------------------------------------------------%
  %
  % left column 
  %
  $\partial \mathcal{O}_r$    & Surface of $\mathcal{O}_r$
  &\hspace{0.1in}&
%
  %
  % right column
  %
  $\kappa^r$ & Imaginary wavenumber in $\mathcal{O}_r$
               $\big(=\xi\sqrt{\epsilon_0 \epsilon^r \mu_0 \mu^r})$
\\[6pt]
%--------------------------------------------------------------------%
%- row 2.5 
%--------------------------------------------------------------------%
  %
  % left column 
  %
  $\epsilon_0, \mu_0$ & Permittivity, permeability of vacuum
  &\hspace{0.1in}&
%
  %
  % right column
  %
  $Z_0, Z^r$   & $Z_0=\sqrt\frac{\mu_0}{\epsilon_0}$, 
                 \quad
                 $Z^r=\sqrt\frac{\mu^r}{\epsilon^r}$ 
\\[6pt]
%--------------------------------------------------------------------%
%- row 3 
%--------------------------------------------------------------------%
  %
  % left column 
  %
  $\epsilon^r, \mu^r$ & Relative permittivity, permeability of
                        $\mathcal{O}_r$
  &&
%
  %
  % right column 
  %
  $\vb K(\vb r), \vb N(\vb r)$ & Electric, magnetic surface currents
\\[6pt]
%--------------------------------------------------------------------%
%- row 4 
%--------------------------------------------------------------------%
  %
  % left column 
  %
  $\BG^{\text{\tiny PQ},r}$ & \textit{Homogeneous} dyadic Green's
                               function for the medium interior to
                               $\mathcal{O}_r$; gives the P-field
                               due to a Q-current, where 
                               P,Q $\in$ \{E,M\} for electric and
                               magnetic fields and currents

  &&
%
  %
  % right column 
  %
  $\vb f_\alpha(\vb r)$  & $\alpha$th element in a set of
                           tangential-vector--valued basis functions
                           defined on object surfaces
\\[6pt]
%--------------------------------------------------------------------%
%- row 5 
%--------------------------------------------------------------------%
  %
  % left column 
  %
  $\bmc{G}^\text{{\tiny PQ}}$ & Scattering part of \textit{inhomogeneous}
                                dyadic Green's function; gives the 
                                scattered P-field due to a Q-current 
                                in the presence of material
                                inhomogeneties.
  &&
%
  %
  % right column 
  %
  $k_\alpha, n_\alpha$  & Expansion coefficients for electric
                          and magnetic surface currents 
                          in the $\{\vb f_\alpha\}$ basis
\\[15pt]
%--------------------------------------------------------------------%
%- row 6 
%--------------------------------------------------------------------%
  %
  % left column 
  %
  $\vb G(\kappa; \vb r)$ & solution of 
                           $\Big[\nabla \times \nabla \times 
                                 \,-\, \kappa^2\Big]\vb G =  
                            \delta(\vb r)\vb{1}$. 

                           \smallskip

                           Related to $\BG$ via
                           $\BG\EE=-Z\kappa \vb G, 
                            \BG\MM=-\frac{\kappa}{Z} \vb G.$

  &&
%
  %
  % right column 
  %
  $\vb M$  & surface-current interaction matrix,
             eqs. (\ref{BEMSystemPEC}, \ref{BEMSystemPMCHW})
\\[8pt]
%--------------------------------------------------------------------%
%- row 6 
%--------------------------------------------------------------------%
  %
  % left column 
  %
  $\vb C(\kappa; \vb r)$ & $\vb C = \frac{1}{\kappa}\nabla\times \vb G$

                           \smallskip

                           Related to $\BG$ via 
                           $\BG\ME=-\BG\EM=\kappa \vb C.$
  &&
%
  %
  % right column 
  %
%
  $\vb W$  & $\vb M^{-1}$
\end{tabular}}
\end{center}
\caption{A glossary of symbols used in this paper.}
\label{Glossary}
\end{table*}
We first consider the case in which the scattering objects in 
Figure \ref{ScatteringCartoon} are perfect conductors.
An incident field impinging on PEC bodies induces a tangential 
electric current distribution $\vb K(\vb x)$ on the body surfaces, 
which gives rise to a scattered field according to
\begin{equation}
  \vb E\sups{scat}(\vb x)
  = \int \BG\EEe(\vb x, \vb x^\prime) \cdot \vb K(\vb x^\prime)
         \, d\vb x^\prime;
\label{EScatBEMPEC}
\end{equation}
here the integral extends over the surfaces of the bodies
and $\BG\EEe$ is the homogeneous DGF for the exterior
medium. (Our notation for DGFs is summarized in Appendix
\ref{DGFAppendix}; throughout this section we work at a 
single frequency and suppress frequency arguments to 
$\vb E, \BG,$ and $\vb K$.) For a given incident 
field $\vb E\sups{inc}$ we can solve for $\vb K$ by 
requiring that the total (incident $+$ scattered) field 
satisfy the appropriate boundary condition, 
which for PEC bodies is simply that the total tangential 
$\vb E$-field vanish for all points $\vb x$ on the body 
surfaces:
$$ \Big[ \vb E\sups{scat}(\vb x) + \vb E\sups{inc}(\vb x) 
   \Big] \times \vbhat n(\vb x) = 0.
$$
[Here taking the cross product with $\vbhat n(\vb x)$,
the outward-pointing surface normal at $\vb x$, is simply
a convenient way of extracting the tangential components 
of a vector.]
Inserting (\ref{EScatBEMPEC}) yields an integral equation for 
$\vb K(\vb x):$
\begin{equation}
 \left[ \int 
        \BG\EEe(\vb x, \vb x^\prime) \cdot \vb K(\vb x^\prime)\,
        d\vb x^\prime 
 \right] \times \vbhat n(\vb x) 
 = -\vb E\sups{inc}(\vb x) \times \vbhat n(\vb x).
\label{BCBEMPEC}
\end{equation}
Equation (\ref{BCBEMPEC}) is known as the ``electric-field
integral equation (EFIE)''~\cite{Harrington93}.

\begin{figure*}
\begin{center}
\resizebox{\textwidth}{!}{\includegraphics{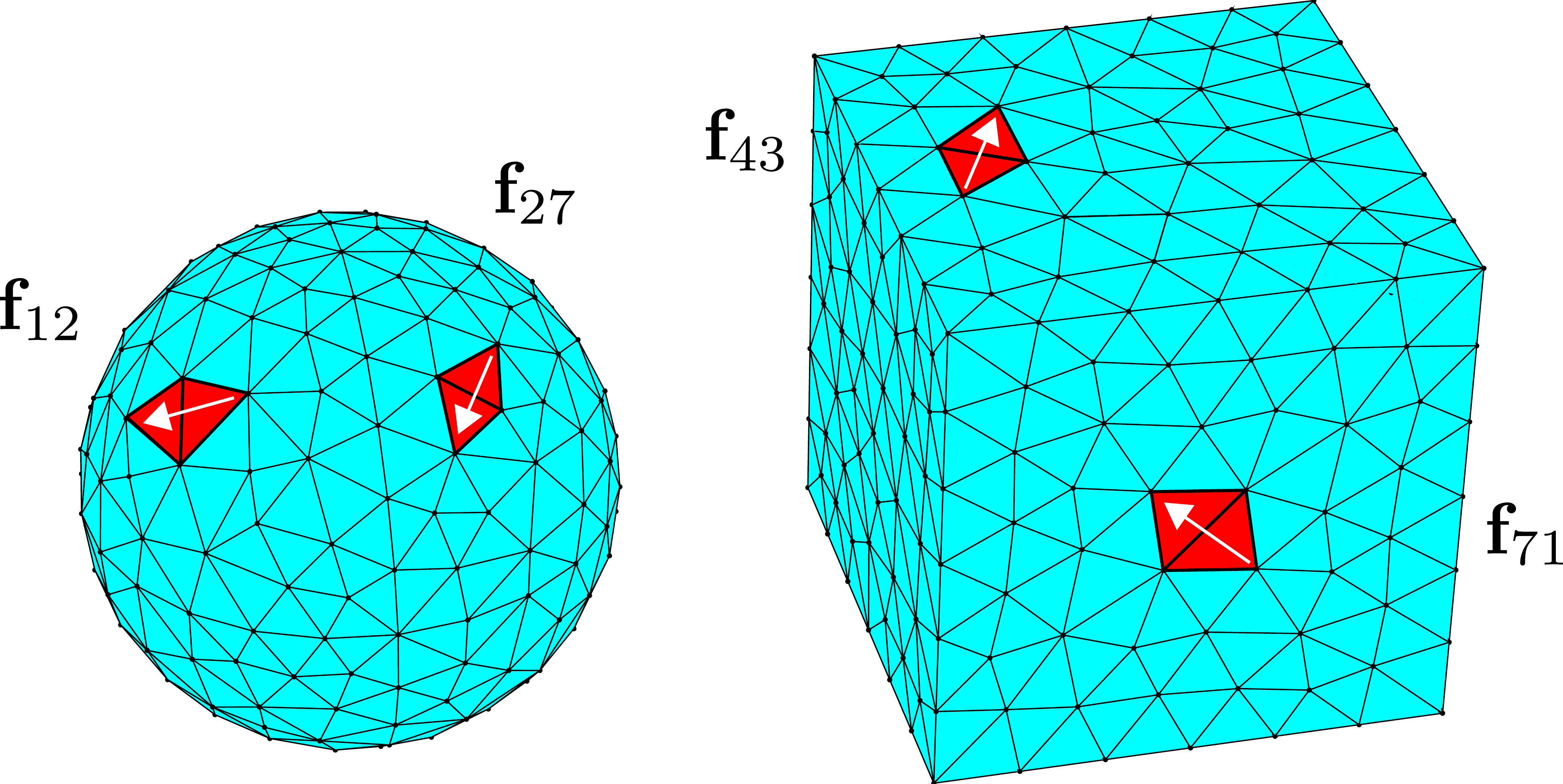}}
\label{BEMCartoon}
\caption{One possible choice of expansion functions for tangential
  currents on object surfaces is obtained by discretizing object
  boundaries into unions of small flat panels and introducing
  \textit{localized} basis functions describing elemental currents
  sourced and sunk at panel vertices: these are ``RWG'' basis
  functions, associated with each \emph{edge} (pair of triangles) in
  the mesh~\cite{RWG}.}
\end{center}
\end{figure*}

Thus far all we have done is to restate the problem in 
an integral-equation form. The next step is 
to \textit{discretize} this integral equation
by introducing a finite set of tangential vector-valued 
basis functions $\{\vb f_\alpha(\vb x)\}$, defined on
the surfaces of the bodies, which serve a dual purpose as 
expansion functions for surface currents and test functions
for boundary conditions.
As noted in Section \ref{IntroductionSection},
an advantage of SIE methods is that they place
\textit{no restriction} on these basis functions;
in particular, the $\{\vb f_\alpha\}$ need not
solve the wave equation or any other equation
and need not encapsulate any global information
about the scattering geometry.
Of course, if symmetries are present then we 
\textit{may} wish to choose the $\{\vb f_\alpha\}$
in a way that reflects them---we might choose 
vector spherical harmonics for a spherical 
scatterer, say, or a Fourier basis for a 
planar scatterer---but nothing in the SIE 
formulation \textit{requires} such a choice, 
and we are equally free to choose the 
$\{\vb f_\alpha\}$ to be arbitrary polynomials, 
piecewise-linear functions, or any other 
functions we like.
For scatterers of complex geometries, a
particularly convenient strategy is
to discretize object surfaces into small
flat panels and take the $\{\vb f_\alpha\}$ 
to describe elemental currents sourced and sunk 
at panel vertices~\cite{RWG}, as depicted in 
Figure \ref{BEMCartoon}.
The localized basis functions that result from 
such a procedure are known as ``boundary elements,'' 
and SIE implementations based on them are commonly
known as ``boundary-element methods'' (BEM) or the 
``method of moments.''

%####################################################################%
%####################################################################%
%####################################################################%
%####################################################################%
%####################################################################%
%####################################################################%

Having chosen a set of basis functions, the surface electric 
current distribution is approximated as a finite expansion
in the $\{\vb f_\alpha\}$:
%====================================================================%
\begin{equation}
 \vb K(\vb x) = \sum k_\alpha \vb f_\alpha(\vb x).
 \label{KExpansion}
\end{equation}
%====================================================================%
This expansion is then inserted into (\ref{BCBEMPEC}),
and the inner product of that equation is taken with each 
member in the set $\{\vb f_\alpha\}$, yielding one equation
for each of the unknown coefficients $k_\alpha.$ 
Collecting these equations yields a linear system 
of the form
\begin{equation}
\vb M \vb k = \vb v,
\label{BEMSystemPEC}
\end{equation}
where $\vb k$ is the vector of $k_\alpha$ coefficients,
the elements of the RHS vector $\vb v$ 
describe the interactions of the basis functions with the 
incident field,
%====================================================================%
\begin{subequations}
\begin{align}
v_{\alpha} 
  &=-\int_{\sup \vb f_\alpha} 
     \vb f_\alpha(\vb x) \cdot \vb E\sups{inc}(\vb x) \, d\vb x
\\
  &\equiv -\Big\langle\vb f_\alpha \Big| \vb E\sups{inc}\Big\rangle,
\end{align}
\label{VElementsPEC}
\end{subequations}
%====================================================================%
(where $\sup \vb f_\alpha$ is the support of basis function 
$\vb f_\alpha$), 
and the elements of the $\vb M$ matrix describe the interactions
of the basis functions with each other through the exterior medium:
%====================================================================%
\begin{subequations}
\begin{align}
\vb M_{\alpha\beta} 
  &=\int_{\sup \vb f_\alpha} d\vb x \,
    \int_{\sup \vb f_\beta} d\vb x^\prime \,
    \vb f_\alpha(\vb x) \cdot \BG\EEe(\vb x, \vb x^\prime) 
    \cdot \vb f_\beta(\vb x^\prime) 
\\
  &\equiv\Big\langle \vb f_\alpha \Big| \BG\EEe \Big| \vb f_\beta 
         \Big\rangle.
\end{align}
\label{MElementsPEC}
\end{subequations}
%====================================================================%
The linear system (\ref{BEMSystemPEC}) may be solved
for the surface-current expansion coefficients $\{k_\alpha\}$,
after which we can compute the components of the 
scattered field at an arbitrary point in the exterior medium
using the discretized form of equation (\ref{EScatBEMPEC}):
\begin{subequations}
\begin{align}
E_i\sups{scat}(\vb x) 
  &= \sum_\alpha k_\alpha \int_{\sup \vb f_\alpha} 
     \Gamma_{ij}\EEe(\vb x, \vb x^\prime) f_{\alpha j}(\vb x^\prime)
     \,d\vb x^\prime
\\
  &\equiv \sum_{\alpha} k_{\alpha} 
      \Big\langle \BG_{i}\EEe(\vb x) \Big| \vb f_{\alpha} \Big\rangle.
\end{align}
\label{EScatBEMPEC2}
\end{subequations}
[The second lines of equations (\ref{VElementsPEC}),
(\ref{MElementsPEC}), and (\ref{EScatBEMPEC2}) 
define some useful shorthand for commonly-encountered integrals;
in (\ref{EScatBEMPEC2}b) note that the contracted index and 
the integrated argument to $\BG$ are suppressed, while the 
uncontracted index and non-integrated argument are written out.]

%%%%%%%%%%%%%%%%%%%%%%%%%%%%%%%%%%%%%%%%%%%%%%%%%%%%%%%%%%%%%%%%%%%%%%
%%%%%%%%%%%%%%%%%%%%%%%%%%%%%%%%%%%%%%%%%%%%%%%%%%%%%%%%%%%%%%%%%%%%%%
%%%%%%%%%%%%%%%%%%%%%%%%%%%%%%%%%%%%%%%%%%%%%%%%%%%%%%%%%%%%%%%%%%%%%%
\subsection{The SIE Method For General Bodies}
\label{PMCHWBEMSubsection}

For PEC bodies, the mathematics of the SIE procedure neatly
mirrors the physics of the actual situation. Indeed, for good 
conductors at moderate frequencies it really \textit{is} true
that the physical induced currents are confined near the
object surfaces; the surface current distribution $\vb K(\vb x)$ 
for which we solve in an SIE method thus has a direct physical 
interpretation as an induced surface current.
 
The situation is more complicated for general (non-PEC) objects,
for here the physical induced currents are no longer confined to 
the surfaces, but instead extend throughout the bulk of
the object. The obvious extension of the procedure outlined
above would be to introduce a \textit{volume} discretization
and solve a system analogous to (\ref{BEMSystemPEC}) for 
the coefficients in an expansion of a volume current density 
$\vb J(\vb x)$. Such a procedure, while retaining the intuitive
interpretation of the quantity computed as a physical current 
density, would suffer from poor complexity scaling, as the 
number of unknowns [and thus the dimension of the linear system 
corresponding to (\ref{BEMSystemPEC})] would scale like the
volume, not the surface area, of the scattering objects.

An alternative approach is to abandon the strategy of solving
for the physical volume sources and to solve instead for equivalent 
\textit{surface} sources that give rise to the same scattered fields.
The mathematical machinery underlying this approach is a vector
generalization of Green's theorem known as the \textit{Stratton-Chu}
equations~\cite{StrattonChu}, which relate the $\vb E$ and $\vb H$ 
fields in the interior of a region to the tangential components
of the fields on the boundary of that region. More precisely, 
let $\partial \mc O_r$ be the surface of the $r$th object 
in our geometry, and for points $\vb x$ on $\partial \mc O_r$ 
define two tangential vector fields according to 
\begin{equation}
  \vb K\sups{eff}(\vb x) 
   \equiv 
   \vbhat n(\vb x) \times \vb H(\vb x),
   \qquad
   \vb N \sups{eff}(\vb x) 
   \equiv
   \vb E(\vb x) \times \vbhat n(\vb x)
\label{KNEff}
\end{equation}
where $\vbhat{n}$ is the outward-pointing normal 
to $\partial \mc O_r$ at $\vb x$ and where $\vb E$ and $\vb H$ 
are the total fields at that that point. The Stratton-Chu 
equations are then the following surface-integral expressions 
for the fields inside and outside $\partial \mathcal{O}_r$:
%====================================================================%
\begin{subequations}
\begin{align}
 \vb E\sups{in}
%--------------------------------------------------------------------%
&=-\int_{\partial \mc O_r} 
  \Big\{ \BG\EEn
          \cdot \vb K\sups{eff}
               +
          \BG\EMn
          \cdot \vb N\sups{eff}
  \Big\} \, d\vb x^\prime
\\
%--------------------------------------------------------------------%
 \vb H\sups{in}
&=-\int_{\partial \mc O_r }
  \Big\{ \BG\MEn
          \cdot \vb K\sups{eff}
          +
          \BG\MMn
          \cdot \vb N\sups{eff}
  \Big\} \, d\vb x^\prime
\end{align}
\label{StrattonChuInside}
\end{subequations}
%====================================================================%
\begin{subequations}
\begin{align}
 \vb E\sups{out}
%--------------------------------------------------------------------%
&=\vb E\sups{inc}
 +\int_{ \cup \partial \mc O_r }
  \Big\{ \BG\EEe
          \cdot \vb K\sups{eff}
               +
          \BG\EMe
          \cdot \vb N\sups{eff}
   \Big\} \, d\vb x^\prime
\\
%--------------------------------------------------------------------%
 \vb H\sups{out}
&=\vb H\sups{inc}
 +\int_{ \cup \partial \mc O_r}
  \Big\{ \BG\MEe
          \cdot \vb K\sups{eff}
          +
          \BG\MMe
          \cdot \vb N\sups{eff}
  \Big\} \, d\vb x^\prime
\end{align}
\label{StrattonChuOutside}
\end{subequations}
[In equations (\ref{StrattonChuInside}--\ref{StrattonChuOutside}),
the $r$ and $e$ superscripts on $\BG$ label the homogeneous DGFs
for the medium interior to $\partial \mc O_r$ and the 
exterior medium, respectively;
the spatial arguments to $\vb E, \BG, \vb K,$ and $\vb N$ are 
as in equation (\ref{EScatBEMPEC}), but are suppressed here to
save space.]

Note the following differences between expressions
(\ref{StrattonChuInside}) and (\ref{StrattonChuOutside})
for the interior and exterior fields: 
\textbf{(a)} the surface integrals in the two cases differ
in sign, arising from the reversal of direction of 
the surface normal in (\ref{KNEff}); \textbf{(b)} 
the $\BG$ dyadics in (\ref{StrattonChuInside}) are those
for the homogeneous medium interior to $\mc O_r$, 
while in (\ref{StrattonChuOutside}) we instead have those
for the the exterior medium; 
\textbf{(c)} in (\ref{StrattonChuInside}) we 
integrate over the single surface $\partial \mc O_r$, while in
(\ref{StrattonChuOutside}) the integral is over the 
union of all object surfaces,
$\cup \, \partial \mc O_r$ 
(which we may think of as the boundary of the exterior medium, 
$\cup\,\partial \mc O_r = \partial \mc O_e$);
\textbf{(d)} the incident fields contribute to expressions 
(\ref{StrattonChuOutside}) for the exterior fields, but are 
absent from expressions (\ref{StrattonChuInside}) for the 
interior fields.

Although the tangential vector fields defined by (\ref{KNEff})
are simply the components of the $\vb E$ and $\vb H$ fields and 
do not correspond to physical source densities, nonetheless
the form of equations (\ref{StrattonChuInside}--\ref{StrattonChuOutside})
suggests interpreting $\vb K\sups{eff}$ and $\vb N\sups{eff}$
as effective electric and magnetic surface current densities, 
which, if known, would allow computation of the fields anywhere 
in space, just as knowledge of the physical surface current 
$\vb K$ suffices in the PEC case to determine uniquely the 
full scattered field. To emphasize this analogy,
we will henceforth drop the ``eff'' designation from $\vb K$ and 
$\vb N$.

As in the PEC case, the $\vb K$ and $\vb N$ distributions are 
determined by requiring that the total fields satisfy appropriate
boundary conditions. For non-PEC bodies these are simply that 
the tangential components of the total fields be continuous
across material boundaries; for a point $\vb x$ on the surface of 
a body we have 
%====================================================================%
\begin{subequations}
\begin{align}
 \Big[\vb E\sups{out} (\vb x) - \vb E\sups{in}(\vb x) 
 \Big] \times \vbhat n(\vb x)
&=0
\\
%--------------------------------------------------------------------%
 \Big[ \vb H\sups{out} (\vb x) - \vb H\sups{in}(\vb x) 
 \Big] \times \vbhat n(\vb x)
&=0.
\end{align}
\label{BCBEMPMCHW}
\end{subequations}
%====================================================================%
Inserting (\ref{StrattonChuInside}-\ref{StrattonChuOutside}) into
(\ref{BCBEMPMCHW}) leads to integral equations for $\vb K$ and $\vb N$
that generalize equation (\ref{BCBEMPEC}) for non-PEC bodies.
As in the PEC case, the next step is to discretize these integral 
equations by approximating the electric and magnetic surface 
currents as expansions in a finite set of tangential vector-valued 
basis functions defined on the object surfaces,
\begin{equation}
  \vb K(\vb x) = \sum k_\alpha \vb f_\alpha(\vb x),
   \quad
   \vb N(\vb x) = -\sum n_\alpha \vb f_\alpha(\vb x),
\label{KNExpand}
\end{equation}
and testing the integral equations obtained from 
(\ref{BCBEMPMCHW}) with each basis function.
(The minus sign in the magnetic surface-current 
expansion is a useful convention that leads to a 
symmetric linear system~\cite{Medgyesi1994}.)
The result of this procedure is a linear system of the 
same general form as (\ref{BEMSystemPEC}),
but now enlarged to exhibit a 2x2 block structure:
%====================================================================%
\begin{equation}
 \left(\begin{array}{cc}
   \vb M\supt{EE} & \vb M\supt{EM} \\[5pt]
   \vb M\supt{ME} & \vb M\supt{MM}
 \end{array}\right)
 \left(\begin{array}{c}
   \vb k \\[5pt] \vb n
 \end{array}\right)
= 
 \left(\begin{array}{c}
   \vb v\supt{E} \\[5pt] \vb v\supt{M}
 \end{array}\right).
\label{BEMSystemPMCHW}
\end{equation}
%====================================================================%
In equation (\ref{BEMSystemPMCHW}), the elements of the RHS vector 
describe the interactions of the basis functions with the incident
electric and magnetic fields [compare equation (\ref{VElementsPEC})],
\begin{equation}
 \left(\begin{array}{c}
   v\supt{E}_\alpha \\[5pt] v\supt{M}_\alpha
 \end{array}\right)
=-
 \left(\begin{array}{c}
         \big\langle \vb f_\alpha \big| \vb E\sups{inc} \big\rangle
        \\[5pt]
         \big\langle \vb f_\alpha \big| \vb H\sups{inc} \big\rangle
 \end{array}\right),
\label{VElementsPMCHW}
\end{equation}
while the elements of the $\vb M$ matrix describe the basis functions
interacting with each other both through the exterior medium and
through the medium interior to one of the scattering bodies.
For example, the elements of the $\vb M\supt{EE}$ block are
\begin{equation}
  M\supt{EE}_{\alpha\beta} = 
   \Big\langle \vb f_\alpha 
   \Big| \BG\EEe + \BG\EEr \Big|
   \vb f_\beta \Big\rangle
\label{MEEAlpha}
\end{equation}
and similarly for the other blocks. (The $\BG\EEr$ term here
is present only if basis functions
$\vb f_\alpha$ and $\vb f_\beta$ are defined on the surface of 
the same object $\mathcal{O}_r$, while the $\BG\EEe$ term is
present even for basis functions defined on the surfaces of 
different objects.)

After solving (\ref{BEMSystemPMCHW}), the scattered fields at
an arbitrary point $\vb x$ are obtained, in analogy to equation
(\ref{EScatBEMPEC2}), from the expansions
\begin{subequations}
\begin{align}
E_i\sups{scat}(\vb x) 
  &=\sum_{\alpha} 
    \left\{ k_{\alpha} 
            \Big\langle \BG_{i}\EEe(\vb x) \Big| \vb f_{\alpha} \Big\rangle
           -n_{\alpha}
            \Big\langle \BG_{i}\EMe(\vb x) \Big| \vb f_{\alpha} \Big\rangle
    \right\}
\\
H_i\sups{scat}(\vb x) 
  &=\sum_{\alpha} 
    \left\{ k_{\alpha} 
            \Big\langle \BG_{i}\MEe(\vb x) \Big| \vb f_{\alpha} \Big\rangle
           -n_{\alpha}
            \Big\langle \BG_{i}\MMe(\vb x) \Big| \vb f_{\alpha} \Big\rangle
    \right\}.
\end{align}
\label{EHScatBEMPMCHW}
\end{subequations}
(These are the scattered fields in the exterior region; the expressions
for fields in the interior of object $r$ are similar, but involve
the homogeneous DGFs $\BG\PQr$ for the medium interior to object $r$.)

%%%%%%%%%%%%%%%%%%%%%%%%%%%%%%%%%%%%%%%%%%%%%%%%%%%%%%%%%%%%%%%%%%%%%%
%%%%%%%%%%%%%%%%%%%%%%%%%%%%%%%%%%%%%%%%%%%%%%%%%%%%%%%%%%%%%%%%%%%%%%
%%%%%%%%%%%%%%%%%%%%%%%%%%%%%%%%%%%%%%%%%%%%%%%%%%%%%%%%%%%%%%%%%%%%%%
\subsection{Explicit SIE Expressions for Dyadic Green's Functions}
\label{BEMDGFSubsection}

The discretized SIE method reviewed in the previous two subsections
is typically employed as a numerical technique, with the linear
systems (\ref{BEMSystemPEC}) and (\ref{BEMSystemPMCHW}) 
solved using methods of computational linear algebra and 
the scattered fields (\ref{EScatBEMPEC2}) and (\ref{EHScatBEMPMCHW}) 
evaluated numerically. In this paper, in contrast, we will use 
the SIE formalism in a somewhat unusual way, by carrying the 
analytical development one step further than is commonly done.
By exploiting the formal solution of equations
(\ref{BEMSystemPEC}) and (\ref{BEMSystemPMCHW}), we will obtain
useful expressions for scattering dyadic Green's functions
in terms of the formal inverse of the SIE matrix $\vb M.$
These expressions will then be used in
Section \ref{StressTensorFSCSection}
to derive compact FSC expressions relating Casimir quantities
to linear-algebraic manipulations of the $\vb M$ 
matrix. Although these final expressions will ultimately be 
evaluated numerically, the analytical expressions derived
in this subsection are an important ingredient in their 
derivation by stress-tensor 
methods. (The expressions derived in this subsection
are not needed for the path-integral derivation of the FSC
Casimir formulas.)

\subsubsection*{The PEC Case}

The scattering dyadic Green's function
$\mc G\supt{EE}_{ij}(\vb x, \vb x^\prime)$ 
is the scattered electric field at $\vb x$
due to a point electric source at $\vb x^\prime$
(Appendix \ref{DGFAppendix}); here we will
need the case in which both $\vb x$ and $\vb x^\prime$
lie in the exterior medium.
To compute this quantity using the SIE technique
of Section \ref{PECBEMSubsection},
we take the incident field to be the field of a 
unit-strength $j$-directed point electric current 
source at a point $\vb x^\prime$ in the exterior
medium, which is simply 
$$ E\sups{inc}_i(\vb x) = \Gamma\EEe_{ij}(\vb x, \vb x^\prime).$$
Then the elements of the RHS vector in (\ref{BEMSystemPEC}) are 
\begin{equation}
 v_{\alpha} = -\Big\langle \vb f_\alpha \Big| \BG\EEe_{j}(\vb x^\prime)
               \Big\rangle,
\label{VAlphaFormal}
\end{equation}
while the coefficients in the expansion of the scattered field 
may be obtained as the formal solution of
(\ref{BEMSystemPEC}),
\begin{equation}
  k_\alpha = \sum_\beta W_{\alpha\beta} V_\beta
\label{KAlphaFormal}
\end{equation}
(where $\vb W=\vb M^{-1}$ is the inverse SIE matrix).
Inserting (\ref{KAlphaFormal}) and (\ref{VAlphaFormal}) into 
(\ref{EScatBEMPEC2}), the scattered field 
at $\vb x$---which is just the scattering DGF we are seeking
to compute---is 
%====================================================================%
\begin{subequations}
\begin{equation}
 \mc G\supt{EE}_{ij}(\vb x, \vb x^\prime)
  = -\sum_{\alpha\beta} 
          \Big \langle \BG_i\EEe(\vb x) 
          \Big| \vb f_\alpha \Big \rangle
          W_{\alpha\beta}
          \Big  \langle \vb f_\beta \Big|
          \BG_j\EEe(\vb x^\prime) \Big\rangle.
\end{equation}
\label{BEMDGFPEC}
We will also need the magnetic-magnetic DGF $\mc G\supt{MM}$, which
is the scattered \textit{magnetic} field due to a point
\textit{magnetic} source. This is obtained in easy analogy 
to the above by \textbf{(a)} taking the incident field to
be the field of a magnetic point source instead of an electric
point source, which has the effect of substituting $\BG\supt{EM}$ for
$\BG\supt{EE}$ in (\ref{VAlphaFormal}); 
and \textbf{(b)} computing the scattered magnetic field instead
of the scattered electric field, which has the effect of 
substituting $\BG\supt{ME}$ for $\BG\supt{EE}$ in (\ref{EScatBEMPEC}).
The result is 
\begin{equation}
 \mc G\supt{MM}_{ij}(\vb x, \vb x^\prime)
  = -\sum_{\alpha\beta} 
          \Big \langle \BG_i\MEe(\vb x) 
          \Big| \vb f_\alpha \Big \rangle
          W_{\alpha\beta}
          \Big  \langle \vb f_\beta \Big|
          \BG_j\EMe(\vb x^\prime) \Big\rangle.
\end{equation}
\label{BEMDGFPEC}
\end{subequations}
%====================================================================%

\subsubsection*{The General Case}

To obtain explicit expressions for scattering DGFs in 
general geometries, we mimic the procedure followed above,
but now using the general SIE formalism outlined in 
Section \ref{PMCHWBEMSubsection} instead of the PEC formalism
of Section\ref{PECBEMSubsection}.
To compute $\mc G\supt{EE}$, 
we again take the incident field to be the field of a
unit-strength $j$-directed point electric source at 
a point $\vb x^\prime$ in the exterior medium, in which 
case the elements of the RHS of equation 
(\ref{BEMSystemPMCHW}) are
\begin{equation}
 \left(\begin{array}{c}
 v\supt{E}_{\alpha} \\[5pt] v\supt{M}_\alpha
 \end{array}\right)
 =
 \left(\begin{array}{c}
         \big\langle \vb f_\alpha \big| \BG\EEe_{j}(\vb x^\prime)
         \big\rangle
        \\[5pt]
         \big\langle \vb f_\alpha \big| \BG\MEe_{j}(\vb x^\prime)
         \big\rangle
        \end{array}
 \right)
\label{VAlphaFormalPMCHW}
\end{equation}
The expansion coefficients that enter into equation
(\ref{EHScatBEMPMCHW}) are given, in analogy to 
equation (\ref{KAlphaFormal}), by the formal 
solution of (\ref{BEMSystemPMCHW}):
\begin{equation}
 \left(\begin{array}{c} 
 k_\alpha \\[5pt] n_\alpha
 \end{array}\right)
= \sum_{\beta} 
 \left(\begin{array}{cc}
   W\supt{EE}_{\alpha\beta} & W\supt{EM}_{\alpha\beta} \\[5pt]
   W\supt{ME}_{\alpha\beta} & W\supt{MM}_{\alpha\beta}
 \end{array}\right)
 \left(\begin{array}{c} 
 V\supt{E}_\beta \\[5pt] V\supt{M}_\beta
 \end{array}\right)
\label{KNAlphaFormal}
\end{equation}
Inserting (\ref{VAlphaFormalPMCHW}) and (\ref{KNAlphaFormal})
into (\ref{EHScatBEMPMCHW}), and proceeding similarly 
for the magnetic-magnetic case, then yields the generalization of 
equation (\ref{BEMDGFPEC}) to non-PEC geometries:
%====================================================================%
\begin{subequations}
\begin{align}
&\mc G\supt{EE}_{ij}(\vb x, \vb x^\prime)=
\\
&-\sum_{\alpha\beta}
%--------------------------------------------------------------------%
 \left(\begin{array}{c}
         \big\langle \BG\EEe_{i}(\vb x) \big| \vb f_\alpha
         \big\rangle
        \\[8pt]
         -\big\langle \BG\EMe_{i}(\vb x) \big| \vb f_\alpha
         \big\rangle
 \end{array}\right)
%--------------------------------------------------------------------%
 \cdot
  \left( 
    \vphantom{ \begin{array}{c}\big\langle\\[5pt]\big\langle\end{array}}
    W_{\alpha\beta}
  \right)
 \cdot
%--------------------------------------------------------------------%
 \left(\begin{array}{c}
         \big\langle \vb f_\beta \big| \BG\EEe_{j}(\vb x^\prime)
         \big\rangle
        \\[8pt]
         \big\langle \vb f_\beta \big| \BG\MEe_{j}(\vb x^\prime)
         \big\rangle
 \end{array}\right)
\nonumber \\[10pt]
&\mc G\supt{MM}_{ij}(\vb x, \vb x^\prime)=
\\
&-\sum_{\alpha\beta}
%--------------------------------------------------------------------%
 \left(\begin{array}{c}
         \big\langle \BG\MEe_{i}(\vb x) \big| \vb f_\alpha
         \big\rangle
        \\[8pt]
         -\big\langle \BG\MMe_{i}(\vb x) \big| \vb f_\alpha
         \big\rangle
 \end{array}\right)
%--------------------------------------------------------------------%
 \cdot
  \left( 
    \vphantom{ \begin{array}{c}\big\langle\\[5pt]\big\langle\end{array}}
    W_{\alpha\beta}
  \right)
 \cdot
%--------------------------------------------------------------------%
 \left(\begin{array}{c}
         \big\langle \vb f_\beta \big| \BG\EMe_{j}(\vb x^\prime)
         \big\rangle
        \\[8pt]
         \big\langle \vb f_\beta \big| \BG\MMe_{j}(\vb x^\prime)
         \big\rangle
 \end{array}\right).
\nonumber
\end{align}
\label{BEMDGFPMCHW}
\end{subequations}

Equations (\ref{BEMDGFPEC}) and (\ref{BEMDGFPMCHW}) are the most
important results of this section of the paper.
The crucial property of these expressions is that they 
present the inhomogeneous Green's function in a fully-factorized
in which factors depending on $\vb x$ are separated from those 
depending on $\vb x^\prime$.
In thise sense, equations (\ref{BEMDGFPEC}) are similar to
Green's-function expansions for special geometries commonly encountered 
in the literature, such as spherical-harmonic expansions for spherical
geometries or Bessel-function expansions for cylindrical 
geometries~\cite{Jackson1998, Rahi2009}; the difference, of course, is that 
(\ref{BEMDGFPEC}) is applicable to \textit{arbitrary} geometries, 
with the geometric information encoded in the $\vb W$ matrix and the 
basis functions $\{\vb f_\alpha\}$.

\section{Stress-Tensor Derivation of the FSC Casimir Formulas}
\label{StressTensorFSCSection}

The stress-tensor approach to Casimir physics relates
Casimir forces to classical dyadic Green's
functions (DGFs). This technique was pioneered by 
Dzyaloshinskii, Lifshitz, and Pitaevskii (DLP) in the 
1950s~\cite{DLP1961, LifshitzPitaevskii} and has remained an important 
computational strategy ever since~\cite{Boyer1974, Schwinger1978};
in particular, modern numerical algorithms for computing Casimir 
forces between bodies of complex geometries have tended to use
the stress-tensor approach, with values for the relevant DGFs 
computed numerically~\cite{Rodriguez2007, PasqualiMaggs2008,
XiongChew2009, Johnson2011}. 
Here, after briefly reviewing the formalism relating 
Casimir forces to DGFs (Section \ref{StressTensorReviewSubsection}),
we will show that the concise SIE expressions for 
the DGFs that we derived in Section \ref{BEMDGFSubsection}
afford a significant simplification of the usual 
computational procedure.
In particular, we show that the surface integral
of the stress tensor, which in previous work has typically 
been evaluated by numerical cubature, may in fact be evaluated 
\textit{analytically} for an arbitrary closed surface
of integration, leaving behind a simple expression
relating the Casimir force to the trace of a certain matrix.

\subsection{A Review of Stress-Tensor Casimir Physics}
\label{StressTensorReviewSubsection}

In the stress-tensor approach, the $i$-directed Casimir force on a body 
is obtained by integrating the expectation value of the Maxwell stress 
tensor over a closed bounding surface surrounding the body:
%====================================================================%
\begin{equation}
\mathcal{F}_i=\int_0^\infty \frac{d\xi}{\pi} F_i(\xi),
\label{FFiXi}
\end{equation}
%====================================================================%
\begin{equation}
 F_i(\xi)=\oint_{\mathcal{C}}\Big\langle T_{ij}(\xi; \vb x)\Big\rangle
 \, \vbhat{n}_j(\vb x) \, d\vb x.
\label{FiXi}
\end{equation}
%====================================================================%
Here the integration surface $\mathcal{C}$ may be the surface of the body
in question or any fictitious closed surface in space bounding the body
(as in Figure \ref{ScatteringCartoon}),
and the expectation value is taken with respect to quantum and 
thermal fluctuations. The expectation value of $T_{ij}$ is next
written in terms of the components of the electric and magnetic fields,
%====================================================================%
\begin{align}
 \Big\langle T_{ij} \Big\rangle
&=  \epsilon \Big\langle E_i E_j \Big\rangle
   +\mu \Big\langle H_i H_j \Big\rangle
\nonumber \\
&\qquad
   -\frac{\delta_{ij}}{2}
    \bigg[ \epsilon \Big\langle E_k E_k \Big \rangle
          +\mu \Big\langle H_k H_k \Big \rangle
    \bigg]. 
\label{Tij}
\end{align}
%====================================================================%
[Here it is understood that $\epsilon=\epsilon_0\epsilon^e$ 
and $\mu=\mu_0\mu^e$ are the (spatially constant) permittivity 
and permeability of the exterior medium at the frequency in 
question; $\epsilon^e,\mu^e$ are the dimensionless 
\textit{relative} quantities.]
Finally, the fluctuation-dissipation theorem is invoked
to relate the expectation values of products of field
components to scattering DGFs~\cite{DLP1961, LifshitzPitaevskii};
at temperature $T=0$, the relations read
%====================================================================%
\begin{subequations}
\begin{align}
 \Big<E_i(\xi, \vb x) E_j(\xi, \vb x^\prime)\Big>
&=-\hbar \xi \mc G_{ij}\EE(\xi; \vb x, \vb x^\prime)
\\
%--------------------------------------------------------------------%
 \Big<H_i(\xi, \vb x) H_j(\xi, \vb x^\prime)\Big>
&=-\hbar \xi \mc G_{ij}\MM(\xi; \vb x, \vb x^\prime)
\end{align}
\label{FDT}
\end{subequations}
%====================================================================%
where, as discussed in Appendix \ref{DGFAppendix}, 
$\mc G\EE(\xi, \vb x, \vb x^\prime)$ is the 
scattered portion of the electric field at $\vb x$ due
to an electric current source at $\vb x^\prime$, all
quantities having time dependence $\propto e^{+\xi t};$
similarly, $\mc G\MM$ gives the scattered \textit{magnetic} 
field due to a \textit{magnetic} current source.
(In the original work, DLP wrote 
$\nabla \times \nabla \times \bmc G\supt{EE}$
in place of $\bmc G\supt{MM}$; the equivalence
of the two quantities has been discussed e.g. in 
Ref.~\cite{Johnson2011}.)

Inserting (\ref{Tij}) and (\ref{FDT}) into (\ref{FiXi}) yields
an expression for the Casimir force-per-unit-frequency
in terms of scattering DGFs:
\begin{equation}
  F_i(\xi) =
   -\frac{\hbar \xi}{\pi}
   \oint_{\mathcal{C}} \bigg\{
    \epsilon \mc G_{ij}\supt{EE}
   +\mu \mc G_{ij}\supt{MM}
   -\frac{\delta_{ij}}{2}
    \Big[ \epsilon \mc G_{kk}\supt{EE}
         +\mu \mc G_{kk}\supt{MM}
    \Big]\bigg\} \vbhat n_j d\vb x.
\label{CasimirStressTensor}
\end{equation}
%where e.g. $\mc G\supt{EE}_{ij}$ is short for 
%$\mc G\EE_{ij}(\xi; \vb x, \vb x).$
%[As discussed in Appendix \ref{DGFAppendix}, 
%$\mc G\EE_{ij}(\xi, \vb x, \vb x^\prime)$ 
%is the \textit{scattering part}
%of the dyadic Green's function and is finite and well-behaved
%in the limit $\vb x^\prime \to \vb x.$]

Equation (\ref{CasimirStressTensor}) is the starting point of
many numerical Casimir studies, as it reduces the computation
of Casimir forces to the computational of classical DGFs.
In principle, the DGFs in question may be computed using
any of the myriad available numerical techniques for 
classical scattering problems; to date, numerical Casimir 
investigations using both the
finite-difference method~\cite{Rodriguez2007,PasqualiMaggs2008}
and the discretized SIE method reviewed in 
Section \ref{BEMReviewSection}~\cite{XiongChew2009, XiongChew2010}
have appeared.
In these studies, the surface integral in (\ref{CasimirStressTensor})
is evaluated by numerical cubature, with the values of the 
integrand at each cubature point computed by solving
numerical scattering problems.

Here we proceed in a different direction. Instead of 
taking equation (\ref{CasimirStressTensor}) as the 
jumping-off point for a numerical investigation, we
will continue the analytical development one step 
further by inserting our explicit SIE expressions 
(\ref{BEMDGFPEC}) and (\ref{BEMDGFPMCHW}) into 
equation (\ref{CasimirStressTensor}) and analyzing 
the result. As we will see, this step will allow us 
to evaluate the surface integral in (\ref{CasimirStressTensor}) 
\textit{analytically}, eliminating the need for 
numerical cubature and resulting in a compact 
matrix-trace formula for the Casimir force. 

Because the thrust of the argument is easiest to present
in the simplest case of perfectly electrically 
conducting (PEC) bodies, we begin with that case
in Section \ref{PECStressTensorDerivationSection}, 
leaving the treatment of general materials to 
Section \ref{DielectricStressTensorDerivationSection}.

%
%For simplicity, we begin in Section \ref{PECStressTensorDerivationSection}), generalization
%first the case of 
%PEC objects embedded in vacuum 
%to general material configurations 
%(Section \ref{PECStressTensorDerivationSection}); the generalization
%
\subsection{Stress-Tensor Derivation of FSC Formulas for PEC Objects}
\label{PECStressTensorDerivationSection}

In Section \ref{BEMDGFSubsection} we derived explicit SIE
expressions for the DGFs that enter into the integrand 
of (\ref{CasimirStressTensor}); for the case of PEC 
scatterers, the relevant expressions are equations
(\ref{BEMDGFPEC}). Our strategy here will be to insert 
these expressions into (\ref{CasimirStressTensor}) and 
analyze the result; to facilitate this procedure, it is 
convenient first to write equations (\ref{BEMDGFPEC}) in a slightly
different form by \textbf{(a)} expressing the four $\BG$ dyadics
in terms of the two $\vb G$ and $\vb C$ dyadics 
(Appendix \ref{DGFAppendix}), and \textbf{(b)} writing out
inner products like $\langle \vb f| \BG \rangle$ explicitly
as integrals over the supports of the basis function $\vb f$
[compare equations (\ref{VElementsPEC}), (\ref{MElementsPEC}),
and (\ref{EScatBEMPEC2})]. 
Then the quantities that enter into the integrand of 
(\ref{CasimirStressTensor}) are 
%====================================================================%
\begin{widetext}
\begin{subequations}
\begin{align} 
 \epsilon \mc G\supt{EE}_{ij}(\vb x, \vb x)
&=-\epsilon\sum_{\alpha\beta} 
           \Big \langle \BG_i\EEe(\vb x) 
           \Big| \vb f_\alpha 
           \Big \rangle
           W_{\alpha\beta}
           \Big  \langle \vb f_\beta \Big|
           \BG_j\EEe(\vb x^\prime) 
           \Big\rangle
\nonumber\\
&=-\mu_0 \mu^e (\kappa^e)^2
  \sum_{\alpha\beta} W_{\alpha\beta}
   \left\{     
   \int_{\sup \vb f_\alpha} \,
              G_{ik}(\vb r_\alpha, \vb x) \,
              f_{\alpha k}(\vb r_\alpha) \,
              d\vb r_\alpha  
   \right\} 
   \left\{
   \int_{\sup \vb f_\beta} \,
          f_{\beta \ell}(\vb r_\beta) \,
          G_{\ell j}(\vb x, \vb r_\beta)\,
          d\vb r_\beta
   \right\}
\\[10pt]
 \mu \mc G\supt{MM}_{ij}(\vb x, \vb x)
&=-\mu\sum_{\alpha\beta} 
           \Big \langle \BG_i\MEe(\vb x) 
           \Big| \vb f_\alpha 
           \Big \rangle
           W_{\alpha\beta}
           \Big  \langle \vb f_\beta \Big|
           \BG_j\EMe(\vb x^\prime) 
           \Big\rangle
\nonumber\\
&=+\mu_0 \mu^e (\kappa^e)^2
  \sum_{\alpha\beta} W_{\alpha\beta}
   \left\{     
   \int_{\sup \vb f_\alpha} \,
              C_{i k}(\vb r_\alpha, \vb x) \,
              f_{\alpha k}(\vb r_\alpha) \,
              d\vb r_\alpha  
   \right\} 
   \left\{
   \int_{\sup \vb f_\beta} \,
          f_{\beta \ell }(\vb r_\beta) \,
          C_{\ell j}(\vb x, \vb r_\beta)\,
          d\vb r_\beta
   \right\}.
\end{align}
\label{eGuG}
\end{subequations}
%====================================================================%
(Here $\kappa^e=\sqrt{\epsilon^{e} \mu^{e}}\cdot \xi$ is the 
imaginary wavenumber of the exterior medium, and we have suppressed
the dependence of the $\vb G$ and $\vb C$ tensors on $\kappa^e$.)
Note that both of these expressions have the same form: a sum over 
basis functions $\vb f_\alpha$ and $\vb f_\beta$, with a summand 
involving integrations over the supports of the basis functions. 
Indeed, equations (\ref{eGuG}a) and (\ref{eGuG}b)
are identical up to the different kernel functions ($\vb G$ or $\vb C$)
that enter into the integrals over basis functions. Note also that the 
variable $\vb x$, which is the integration variable in the surface 
integral in (\ref{CasimirStressTensor}), appears in (\ref{eGuG}) 
\textit{only} through these kernel functions.
This implies that, after inserting (\ref{eGuG})
into (\ref{CasimirStressTensor}), we will again have
a sum of terms of this same form---a sum over basis
functions, with the summand involving integrals over the basis 
functions---and, moreover, that many of the factors
in this summand will be independent of the integration 
variable $\vb x$ in (\ref{CasimirStressTensor}) and may thus be
pulled outside the surface integral, which will now contain
only factors of $\vb G$ and $\vb C$. 
The result is
$(Z_0=\sqrt{\mu_0/\epsilon_0}, Z^e=\sqrt{\mu^e/\epsilon^e})$
%====================================================================%
\begin{equation}
  F_i(\xi) =  \frac{\hbar}{\pi} \sum_{\alpha\beta}
   W_{\alpha\beta}
   \cdot Z_0 Z^e \kappa^e 
   \int_{\sup \vb f_\alpha} \, d\vb r_\alpha \,
   \int_{\sup \vb f_\beta} \, d\vb r_\beta \,
   \Big\{ f_{\alpha k}(\vb r_\alpha) 
           \mathcal{I}_{ik\ell}(\vb r_\alpha, \vb r_\beta)
           f_{\beta \ell}(\vb r_\beta)
   \Big\}
\label{FiXi1}
\end{equation}
%====================================================================%
where, as anticipated, the surface integral is now contained inside
the definition of the $\mathcal{I}$ kernel:
%====================================================================%
\begin{align*}
&\mathcal{I}_{ik\ell}(\vb r_\alpha, \vb r_\beta) 
   \equiv
      (\kappa^e)^2
      \oint_{\mathcal C} 
      \left\{  G_{ik}(\vb r_\alpha, \vb x) G_{\ell j}(\vb x,\vb r_\beta) 
              -C_{ik}(\vb r_\alpha, \vb x) C_{\ell j}(\vb x,\vb r_\beta) 
      \right.
\\&\hspace{2in}
      \left. 
             -\frac{\delta_{ij}}{2}
              \Big[
               G_{mk}(\vb r_\alpha, \vb x) G_{\ell m}(\vb x,\vb r_\beta) 
             - C_{mk}(\vb r_\alpha, \vb x) C_{\ell m}(\vb x,\vb r_\beta) 
              \Big]
      \right\}\vbhat{n}_j d\vb x.
\end{align*}
%====================================================================%
%[In equation (\ref{FiXi1}) we have used the vacuum and relative
%wave impedances 
%(Appendix \ref{DGFAppendix})].
The fact that $\vb W$ is a symmetric matrix 
($W_{\alpha\beta}=W_{\beta\alpha}$) allows us to rewrite 
equation (\ref{FiXi1}) to read
%====================================================================%
\begin{equation}
  F_i(\xi) =  \frac{\hbar}{2\pi} \sum_{\alpha\beta}
   W_{\alpha\beta}
   \cdot Z_0Z^e \kappa^e
   \int_{\sup \vb f_\alpha} \, d\vb r_\alpha \,
   \int_{\sup \vb f_\beta} \, d\vb r_\beta \,
   \Big\{ f_{\alpha k}(\vb r_\alpha) 
           \overline{\mathcal{I}}_{ik\ell}(\vb r_\alpha, \vb r_\beta)
           f_{\beta \ell}(\vb r_\beta)
   \Big\}
\label{FiXi2}
\end{equation}
%====================================================================%
where we have defined a symmetrized version of the $\mathcal{I}$ 
kernel:
$$ \overline{\mathcal{I}}_{ik\ell}(\vb r, \vb r^\prime)\equiv 
   \mathcal{I}_{ik\ell}(\vb r, \vb r^\prime)
  +\mathcal{I}_{i\ell k}(\vb r^\prime, \vb r).
$$
The point of this step is that, as demonstrated in 
Appendix \ref{IntegralIdentityAppendix}, \textit{the surface integral
in the definition of the $\overline{\mathcal{I}}$ kernel
may be evaluated in closed form}, for \textit{any} topological
two-sphere $\mathcal{C}$, with the result
%====================================================================%
\begin{equation}
 \overline{\mathcal{I}}_{ik\ell}(\vb r, \vb r^\prime)
 =\begin{cases} 
     0, \qquad &\text{if $\vb r, \vb r^\prime$ 
                      lie both inside or both outside $\mathcal{C}$}
     \\[8pt]
     \displaystyle{
     \frac{\partial}{\partial \vb r_i\supt{I}}
     G_{k\ell}(\vb r\supt{I}, \vb r\supt{E})}
               &\text{if $\vb r, \vb r^\prime$ 
                      lie on opposite sides of $\mathcal{C}$}
  \end{cases} 
\label{IBarClosedForm}
\end{equation}
%====================================================================%
where, in the second case, $\vb r\supt{I}$ ($\vb r\supt{E}$) is 
whichever of $\vb r, \vb r^\prime$ lies in the interior (exterior)
of $\mathcal{C}$.

Armed with the dichotomy (\ref{IBarClosedForm}), we can now analyze
the quantity in curly brackets in (\ref{FiXi2}). Recall that
the bounding contour $\mathcal{C}$ encloses one of the objects
in our Casimir geometry; call this object $\mathcal{O}_1$
and the remaining objects $\mathcal{O}_{2,3,\cdots}$. Equation
(\ref{IBarClosedForm}) then tells us that the curly-bracketed term
in (\ref{FiXi2}) vanishes except when precisely one of the 
basis functions $\{\vb f_\alpha, \vb f_\beta\}$ lies on the
surface of object $\mathcal{O}_1$. When this condition is 
satisfied, the integral over basis functions in (\ref{FiXi2})
reads 
%====================================================================%
\begin{align*}
& -Z_0 Z^e \kappa^e 
   \int_{\sup \vb f_\alpha} \, d\vb r_\alpha \,
   \int_{\sup \vb f_\beta} \, d\vb r_\beta \,
   \left\{ f_{\alpha k}(\vb r_\alpha)
          \Big[\frac{\partial}{\partial \vb r_{\alpha i}}
                G_{k\ell}(\vb r_\alpha, \vb r_\beta)
          \Big]
          f_{\beta \ell}(\vb r_\beta)
   \right\}
\\
&= \Big \langle \vb f_{\alpha} 
   \Big|
   \frac{\partial}{\partial \vb r_{\alpha i}} \BG\EEe 
   \Big|
    \vb f_\beta \big \rangle
\intertext{But this is nothing but the derivative of the $\alpha,\beta$ 
           element of the SIE matrix (\ref{MElementsPEC}) with respect 
           to a rigid infinitesimal displacement of object $\mathcal{O}_1$ 
           in the $i$ direction,}
&= \frac{\partial}{\partial \vb r_i} M_{\alpha\beta}.
\end{align*}
\end{widetext}
%====================================================================%
Inserting this into (\ref{FiXi2}), we find that the 
imaginary-frequency-$\xi$ contribution to the Casimir force is given simply by 
\begin{align}
F_i(\xi) 
&= \frac{\hbar}{2\pi} \sum_{\alpha\beta} W_{\alpha\beta} 
   \cdot \Big[\frac{\partial}{\partial \vb r_i} M_{\alpha\beta}\Big]
\nonumber \\
&= \frac{\hbar}{2\pi} \Tr \Big[ \vb M^{-1} \cdot 
              \frac{\partial}{\partial \vb r_i} \vb M\Big]
\label{PECTraceForce}
\end{align}
(where we have recalled the definition $\vb W = \vb M^{-1}$), and
inserting this into (\ref{FFiXi}) we obtain 
the FSC formula for the Casimir force, 
equation (\ref{MasterFSCFormulas}b).
To obtain the FSC formula for the Casimir energy,
we note that 
the Casimir force on an object is minus the 
derivative of the energy with respect to a rigid
displacement of that object; using the standard 
identity 
$$ \frac{\partial}{\partial \vb r_{i}} \log \det \vb M 
 = \text{Tr}\Big[ \vb M^{-1} \cdot
   \frac{\partial}{\partial \vb r_{i}} \vb M\Big],
$$
and choosing the zero of energy to correspond
to the energy of the configuration in which all 
objects are removed to infinite separations
(for which configuration we denote the SIE matrix
by $\vb M_\infty$), we
recover equation (\ref{MasterFSCFormulas}a).
Finally, equation (\ref{MasterFSCFormulas}c) follows from
taking derivatives with respect to a rigid \textit{rotation}
instead of a rigid displacement.

This completes the stress-tensor derivation of the FSC 
formulas for the case of PEC objects.

\subsection{Stress-Tensor Derivation of FSC Formulas for General Objects} 
\label{DielectricStressTensorDerivationSection}

The derivation of the FSC formulas for general objects is now
a straightforward generalization of the procedure for PEC objects. 
Again we start with equation (\ref{CasimirStressTensor}), and again 
we insert in this equation the factorized expressions
for scattering DGFs that we derived in Section 
\ref{BEMDGFSubsection}; the difference is that
for non-PEC objects we must now use the more
complicated expressions (\ref{BEMDGFPMCHW}).
Mimicing the discussion following equations
(\ref{eGuG}) above now leads to a modified 
version of equation 
(\ref{FiXi2}) in which the $\mathcal{I}$ kernel
is promoted to a $2\times 2$ matrix:
%====================================================================%
\begin{widetext}
\begin{equation}
  F_i(\xi) = +\frac{\hbar}{2\pi} \Tr \sum_{\alpha\beta}
   \left(\begin{array}{cc}
    W\supt{EE}_{\alpha\beta} & W\supt{EM}_{\alpha\beta} 
    \\[8pt]
    W\supt{ME}_{\alpha\beta} & W\supt{MM}_{\alpha\beta}
   \end{array}\right)
   \int_{\sup \vb f_\alpha}  d\vb r_\alpha
   \int_{\sup \vb f_\beta}  d\vb r_\beta 
   \Big\{ f_{\alpha k}(\vb r_\alpha) 
          \left(\begin{array}{cc}
           Z_0 Z^e \kappa^e 
           \overline{\mathcal{I}}_{ik\ell}(\vb r_\alpha, \vb r_\beta) &
           \kappa^e 
           \overline{\mathcal{J}}_{ip\ell}(\vb r_\alpha, \vb r_\beta) 
           \\[8pt]
           -\kappa^e 
           \overline{\mathcal{J}}_{ip\ell}(\vb r_\alpha, \vb r_\beta) &
           \frac{\kappa^e}{Z_0 Z^e} 
           \overline{\mathcal{I}}_{ip\ell}(\vb r_\alpha, \vb r_\beta)
          \end{array}\right)
           f_{\beta \ell}(\vb r_\beta)
   \Big\}
\label{FiXi3}
\end{equation}
with Tr denoting a $2\times 2$ matrix trace and the 
$\overline{\mathcal{J}}$ kernel defined in analogy to 
$\overline{\mathcal{I}}$:
%====================================================================%
$$ \overline{\mathcal{J}}_{ik\ell}(\vb r, \vb r^\prime)\equiv 
   \mathcal{J}_{ik\ell}(\vb r, \vb r^\prime)
  +\mathcal{J}_{i\ell k}(\vb r^\prime, \vb r)
$$
\begin{align*}
&\mathcal{J}_{ik\ell}(\vb r_\alpha, \vb r_\beta) 
   \equiv
      (\kappa^e)^2
      \oint_{\mathcal C} 
      \left\{  G_{ik}(\vb r_\alpha, \vb x) C_{\ell j}(\vb x,\vb r_\beta) 
              +C_{ik}(\vb r_\alpha, \vb x) G_{\ell j}(\vb x,\vb r_\beta) 
      \right.
\\&\hspace{2in}
      \left. 
             -\frac{\delta_{ij}}{2}
              \Big[
               G_{mk}(\vb r_\alpha, \vb x) C_{\ell m}(\vb x,\vb r_\beta) 
             + C_{mk}(\vb r_\alpha, \vb x) G_{\ell m}(\vb x,\vb r_\beta) 
              \Big]
      \right\}\vbhat{n}_j d\vb x.
\end{align*}
Again in analogy to $\overline{\mc I}$, the surface integrals 
in the definition of $\overline{\mc J}$ may be evaluated in closed
form to yield 
%====================================================================%
\begin{equation}
 \overline{\mathcal{J}}_{ik\ell}(\vb r, \vb r^\prime)
 =\begin{cases} 
     0, \qquad &\text{if $\vb r, \vb r^\prime$ 
                      lie both inside or both outside $\mathcal{C}$}
     \\[8pt]
     \displaystyle{
     \frac{\partial}{\partial \vb r_i\supt{I}}
     C_{k\ell}(\vb r\supt{I}, \vb r\supt{E})}
               &\text{if $\vb r, \vb r^\prime$ 
                      lie on opposite sides of $\mathcal{C}$}
  \end{cases} 
\label{JBarClosedForm}
\end{equation}
%====================================================================%
and, armed with (\ref{IBarClosedForm}) and (\ref{JBarClosedForm}), 
it is now easy to identify the integral over basis functions
in (\ref{FiXi3}) as nothing but the derivative of the SIE matrix:

\begin{equation}
   \int_{\sup \vb f_\alpha}  d\vb r_\alpha
   \int_{\sup \vb f_\beta}  d\vb r_\beta 
   \Big\{ f_{\alpha k}(\vb r_\alpha) 
          \left(\begin{array}{cc}
           Z_0 Z^e \kappa^e 
           \overline{\mathcal{I}}_{ik\ell}(\vb r_\alpha, \vb r_\beta) &
           \kappa^e 
           \overline{\mathcal{J}}_{ip\ell}(\vb r_\alpha, \vb r_\beta) 
           \\[8pt]
           -\kappa^e 
           \overline{\mathcal{J}}_{ip\ell}(\vb r_\alpha, \vb r_\beta) &
           \frac{\kappa^e}{Z_0 Z^e} 
           \overline{\mathcal{I}}_{ip\ell}(\vb r_\alpha, \vb r_\beta)
          \end{array}\right)
           f_{\beta \ell}(\vb r_\beta)
   \Big\}
=
\frac{\partial}{\partial{\vb r_i}} 
   \left(\begin{array}{cc}
    M\supt{EE}_{\alpha\beta} & M\supt{EM}_{\alpha\beta} 
    \\[8pt]
    M\supt{ME}_{\alpha\beta} & M\supt{MM}_{\alpha\beta}
   \end{array}\right).
\label{DerivativeMatrix2}
\end{equation}
Inserting (\ref{DerivativeMatrix2}) into (\ref{FiXi3}) now
simply reproduces equation (\ref{PECTraceForce}) 
with the $\vb M$ matrix understood to refer to the 
general-material SIE matrix in equation 
(\ref{BEMSystemPMCHW}).

\end{widetext}

\section{Path-Integral Derivation of the FSC Casimir Formulas}
\label{PathIntegralFSCSection}

It is remarkable that the path-integral approach to Casimir physics, 
which bears little superficial resemblance to the stress-tensor 
formalism of the previous section, may nonetheless be used to furnish a 
separate and entirely independent derivation of the same
FSC formulas that we derived above using stress-tensor ideas.
In this section, after first reviewing the well-known formalism
for obtaining Casimir energies from constrained path integrals
(Section \ref{PathIntegralReviewSubsection}), we present this
alternate derivation
(Section \ref{PathIntegralFSCDerivationSubsection}).

The path-integral procedure presented here differs from 
typical path-integral treatments of Casimir phenomena in
at least two ways. First, whereas many authors 
write the action for the electromagnetic field in terms of 
the gauge-independent $\vb E$ and $\vb B$ 
fields~\cite{Rahi2009}, or in 
terms of the four-vector potential $A^\mu$ in a way that 
depends on a specific choice of gauge (often the ``temporal'' 
or ``Weyl'' gauge $A^0\equiv 0$~\cite{Amooghorban2011}), 
here we write the action 
in terms of $A^\mu$ with a Fadeev-Popov parameter that allows 
arbitrary gauge choices; we verify explicitly that the 
Fadeev-Popov parameter is absent from all final physical 
predictions. (This portion of our treatment is similar
to that of Ref.~\cite{Bordag1985}.)

Second, we introduce a new implementation of the 
constraint that the path integral
extend only over field configurations satisfying 
the boundary
conditions. Our representation emphasizes the continuity 
of the tangential $\vb E$ and $\vb H$ fields across the 
surfaces of the objects in a Casimir geometry,
and the Lagrange multipliers that we introduce to
enforce the constraints have an attractive physical 
interpretation as \textit{surface currents}, thus
establishing a connection to the SIE ideas reviewed
above. After integrating out the photon field, we are 
left with functional integrals over surface-current 
distributions, with an effective action describing the
interactions of these currents through the electrical
media interior and exterior to the objects; upon
discretization, this action turns out to involve 
precisely the same surface-current-interaction
matrix that appears in the SIE formulation of 
scattering reviewed in Section \ref{BEMReviewSection}.

%%%%%%%%%%%%%%%%%%%%%%%%%%%%%%%%%%%%%%%%%%%%%%%%%%%%%%%%%%%%%%%%%%%%%%
%%%%%%%%%%%%%%%%%%%%%%%%%%%%%%%%%%%%%%%%%%%%%%%%%%%%%%%%%%%%%%%%%%%%%%
%%%%%%%%%%%%%%%%%%%%%%%%%%%%%%%%%%%%%%%%%%%%%%%%%%%%%%%%%%%%%%%%%%%%%%
\subsection{A Review of Constrained Path-Integral Techniques
            for Casimir Energies}          

\label{PathIntegralReviewSubsection}

Path-integral formulations of field-fluctuation problems were 
pioneered by Bordag, Robaschik, and Wieczorek~\cite{Bordag1985}
and by Li and Kardar~\cite{LiKardarPRL1991,LiKardarPRA1991} and
have since been further developed by a number of authors 
(see ~\cite{Rahi2009, Dalvit2011} for extensive surveys of 
recent developments.) In this section we review the key steps 
in this approach. 

%The content of this section is not new, but 
%is reproduced here for completeness; in the following subsection
%(Section \ref{PathIntegralFSCDerivationSubsection}) we 

%.....................................................................
%.....................................................................
%.....................................................................
\subsubsection*{Casimir Energies from Constrained Path Integrals}

In the presence of material boundaries, the partition function 
for a quantum field $\phi$ (which may be scalar, vector, 
electromagnetic, or otherwise, but is here assumed bosonic) 
at inverse temperature $\beta$ takes the form 
%====================================================================%
\begin{equation}
Z(\beta)
   =\int \Big[\mathcal{D}\phi(\tau, \vb x)\Big]_C \, e^{-\frac{1}{\hbar}S_\beta[\phi]}
\label{Zbeta1}
\end{equation}
%====================================================================%
where the action $S_\beta$ is the 
spacetime integral of the Euclidean Lagrangian density 
for the $\phi$ field, 
%====================================================================%
\begin{equation}
S_\beta[\phi]
 =\int_0^{\hbar \beta} d\tau \int d\vb x \,\,
  \mathcal{L}_E\Big\{ \phi(\tau, \vb x)\Big\},
\label{Sbeta}
\end{equation}
%====================================================================%
and where the notation $\left[\cdots\right]_C$ in 
(\ref{Zbeta1}) indicates that this is a \textit{constrained}
path integral, in which the functional integration extends only
over field configurations $\phi$ satisfying the appropriate boundary 
conditions at all material boundaries.

If the boundary conditions are time independent \textit{and} 
the Lagrangian density contains no terms of higher than quadratic
order in $\phi$ and its derivatives, then it is convenient to
introduce a Fourier series in the Euclidean time variable,
%====================================================================%
$$ \phi(\tau, \vb x)=\sum_n \phi_n(\vb x) e^{-i\xi_n \tau}, \qquad
   \xi_n=\frac{2\pi n}{\hbar \beta},
$$
%====================================================================%
whereupon the path integral (\ref{Zbeta1}) factorizes into a
product of contributions from individual frequencies,
%====================================================================%
$$ Z(\beta) = \prod_n \mathcal{Z}(\beta; \xi_n), $$
%====================================================================%
%====================================================================%
\begin{equation}
   \mathcal{Z}(\beta; \xi_n) = 
   \int \Big[\mathcal{D}\phi_n(\vb x)\Big]_C \,\,
   e^{-S[\phi_n; \xi_n]},
 \label{mzbeta}
\end{equation}
%====================================================================%
with
%====================================================================%
$$ S\Big[\phi_n(\vb x); \xi_n\Big]
   =\beta \int d\vb x \, 
    \mathcal{L}_E\Big\{\phi_n(\vb x)e^{-i\xi_n \tau}\Big\}
$$
%====================================================================%
representing the contribution to the full action (\ref{Sbeta}) 
made only by those field configurations with Euclidean-time 
dependence $\sim e^{-i\xi_n \tau}.$ The free energy is 
then obtained as a sum over Matsubara frequencies,
%====================================================================%
\begin{equation}
   F =-\frac{1}{\beta} \ln \frac{Z(\beta)}{Z_\infty(\beta)}
     = -\frac{1}{\beta} \sum_{n=0}^\infty 
           \ln \frac{\mathcal{Z}(\beta, \xi_n)} 
                    {\mathcal{Z}_\infty(\beta, \xi_n)},
 \label{FreeEnergy}
\end{equation}
%====================================================================%
where $Z_\infty(\mathcal{Z}_\infty)$ is $Z(\mathcal{Z})$ 
evaluated with all material objects separated by infinite distances
[dividing out these contributions in (\ref{FreeEnergy}) is a 
useful convention that amounts to a choice of the zero of energy].
In the zero-temperature limit, the frequency sum becomes an 
integral, and the zero-temperature Casimir energy is 
%====================================================================%
\begin{equation}
\mathcal{E} = -\frac{\hbar}{2\pi} 
  \int_0^\infty d\xi \, \ln \frac{\mathcal{Z}(\xi)}{\mathcal{Z}_\infty(\xi)}.
\label{ZeroTemperatureEnergy}
\end{equation}
%====================================================================%
(Here and below we omit the $\beta$ argument to $\mathcal{Z}$).

\subsubsection*{Enforcing Constraints via Functional $\delta$-functions}

Equations (\ref{FreeEnergy}-\ref{ZeroTemperatureEnergy}) 
reduce the computation
of Casimir energies to the evaluation of constrained path integrals 
(\ref{mzbeta}). In most branches of physics, the path integrals
associated with physically interesting quantities are difficult
to evaluate because the action $S$ in the exponent contains 
interaction terms (terms of third or higher order in the 
fields and their derivatives). In Casimir physics, on the other 
hand, the action is not more than quadratic in $\phi$, and the 
difficulty in evaluating expressions like (\ref{mzbeta}) stems 
instead from the challenge of implementing the implicit 
constraint on the functional integration measure, 
arising from the boundary conditions and indicated 
by the $[\cdots]_C$ notation in (\ref{mzbeta}). 
 
The innovation of Bordag~\cite{Bordag1985} and of 
Li and Kardar~\cite{LiKardarPRL1991} 
was to represent these constraints \textit{explicitly} through the use 
of functional $\delta$ functions. If the boundary conditions on $\phi$
may be expressed as the vanishing of a set of quantities 
$\{L_{\alpha} \phi\}$, where $\left\{L_{\alpha}\right\}$ will generally 
be some family of linear integrodifferential operators indexed by a 
discrete or continuous label $\alpha$, then the constrained path 
integral may be written in the form 
%====================================================================%
\begin{align}
\mathcal{Z}(\xi_n) 
&=\int \Big[\mathcal{D}\phi_n(\vb x)\Big]_C e^{-S[\phi_n; \xi_n]} 
 \nonumber \\
&=\int \mathcal{D}\phi_n(\vb x)
  \prod_{\alpha} \delta\Big( L_{\alpha} \phi \Big)
 e^{-S[\phi_n; \xi_n]} 
  \label{unconstrained1}
\end{align}
%====================================================================%
where now the functional integration over 
$\phi_n$ is \textit{unconstrained.} A particularly convenient 
representation for the one-dimensional Dirac $\delta$ function 
is 
%====================================================================%
\begin{equation}
 \delta(u) = \int \frac{d\lambda}{2\pi} e^{i\lambda u},
\label{deltarep}
\end{equation}
%====================================================================%
where we may think of $\lambda$ as a Lagrange multiplier
enforcing the constraint that $u$ vanish. 
Inserting one copy of (\ref{deltarep}) for each $\delta$
function in the product in (\ref{unconstrained1}) yields
%====================================================================%
$$
\mathcal{Z}(\xi_n) 
=\int \mathcal{D}\phi_n(\vb x) 
  \int \prod_{\alpha} \frac{d\lambda_\alpha}{2\pi} \,
  e^{ -S[\phi_n; \xi_n]
      +i\sum_\alpha \lambda_\alpha L_\alpha \phi}.
$$
%====================================================================%
The final step is to evaluate the unconstrained integral over $\phi$;
since the exponent is quadratic in $\phi$, this can be done 
\textit{exactly} using standard techniques of Gaussian integration, 
yielding an expression of the form
\begin{equation}
\mathcal{Z}(\xi_n) =\Big\{ \# \Big\}
\int \prod_{\alpha} d\lambda_\alpha \,
 e^{-S\supt{eff}\{\lambda_\alpha\}}
\label{PathIntegralOverLambda}
\end{equation}
[where $\{\#\}$ is a constant that cancels in the ratios in 
(\ref{FreeEnergy}-\ref{ZeroTemperatureEnergy})].
The constrained functional integral over the field $\phi$ is thus 
replaced by a new integral over the set of Lagrange multipliers 
$\{\lambda_\alpha\}$, with an effective action $S\sups{eff}$ 
describing interactions mediated by the original fluctuating 
field $\phi$.

\subsubsection*{Representation of Boundary Conditions}

Equation (\ref{PathIntegralOverLambda}) makes clear that
the practical convenience of path-integral Casimir computations
is entirely determined by the choice of the Lagrange multipliers 
$\{\lambda_\alpha\}$ and the complexity of their effective 
action $S\sups{eff}$; these, in turn, depend on the details
of the boundary conditions imposed on the fluctuating field. 
For a given physical situation there 
may be multiple ways to express the boundary conditions, 
each of which will generally lead to a distinct expression 
for the integral in (\ref{PathIntegralOverLambda}).
Ultimately, of course, all choices must lead to equivalent 
results, but different choices may exhibit significant differences
in computational complexity and in the range of geometries that 
can be efficiently treated. Several different representations 
of boundary conditions and Lagrange multipliers have appeared 
in the literature to date.

The original work of Bordag et al.~\cite{Bordag1985} considered 
QED in the presence of superconducting boundaries, 
with the boundary conditions taken to be the vanishing
of the normal components of the dual field-strength tensor; 
in the notation of the previous section, 
$L_{\vb x}\phi=\hat{n}^\mu F_{\mu\nu}^*(\vb x),$ and the 
set of Lagrange multipliers $\{\lambda_{\vb x}\}$ 
constitutes a three-component auxiliary 
field defined on the bounding surfaces. 
The method is applicable to the computation of electromagnetic 
Casimir energies, but the treatment was restricted to the case 
of parallel planar boundaries.

Li and Kardar~\cite{LiKardarPRL1991,LiKardarPRA1991} 
considered a \textit{scalar} field satisfying Dirichlet or Neumann
boundary conditions on a prescribed boundary manifold. Here 
again the boundary conditions amount to the vanishing of 
a local operator applied to $\phi$, 
$L_{\vb x}\phi=\phi(\vb x)$ (Dirichlet) or 
$L_{\vb x}\phi=|\partial\phi/\partial n|_{\vb x}$ (Neumann),
and we have one Lagrange multiplier $\lambda(\vb x)$
for each point on the boundary manifold. In this case it is 
tempting to interpret $\lambda(\vb x)$ as a scalar source
density, confined to the boundary surfaces and with a
self-interaction induced by the fluctuations of the 
$\phi$ field. This formulation was capable, in principle,
of handling \textit{arbitrarily}-shaped boundary surfaces,
but was restricted to the case of scalar fields.

The technique of Refs.~\cite{LiKardarPRL1991,LiKardarPRA1991}
was subsequently reformulated~\cite{Emig2007,Emig2008,Rahi2009}
in a way that allowed extension to the case of the 
electromagnetic field. Whereas the original formulation
imposed a \textit{local} form of the boundary conditions---and
took the Lagrange multipliers $\lambda(\vb x)$ to be local
surface quantities---the revised formulation abandons the 
surface-source picture in favor of an alternative viewpoint
emphasizing incoming and outgoing electromagnetic waves.
This approach associates one Lagrange multiplier 
$\lambda_\alpha$ to each \textit{multipole} term in a 
multipole expansion of the EM field,
with the choice of multipole basis (spherical, cylindrical, 
etc.) governed by the symmetries of the problem; the 
effective action $S\sups{eff}$ then describes the 
interactions among multipoles.

The virtue of multipole expansions is that, for certain 
geometries, a small number of multipole coefficients may suffice 
to solve many problems of interest to high accuracy. This has long 
been understood in domains such as electrostatics and scattering 
theory, and in recent years has been impressively demonstrated in 
the Casimir context as well~\cite{Emig2007,Emig2008,Rahi2009}, 
where multipole expansions have been used to obtain rapidly convergent 
and even \textit{analytically} tractable series for Casimir energies in
certain special geometries. The trick, of course, is that the 
very definition of the multipoles already encodes a significant 
amount of information about the geometry, thus requiring relatively 
little additional work to pin down what more remains to be said in 
any particular situation.

But this blessing becomes a curse when we seek a unified
formalism capable of treating all geometries on an equal
footing. The very geometric specificity of the multipole description,
which so streamlines the treatment of compatible or 
nearly-compatible geometries, has the opposite effect of 
\textit{complicating} the treatment of incompatible geometries; 
thus, whereas a basis of spherical multipoles might allow highly 
efficient treatment of interacting spheres or nearly-spherical
bodies, it would be a particularly \textit{unwieldy} 
choice for the description of cylinders, tetrahedra, or 
parallelepipeds. Of course, for each new geometric configuration
we could simply redefine our multipole expansion and
correspondingly re-implement the full arsenal of computational 
machinery (a strategy pursued for a dizzying array of geometries 
in Ref.~\cite{Rahi2009}), but such a procedure contradicts the 
spirit of a single, general-purpose scheme into which we simply 
plug an arbitrary experimental geometry and turn a crank.

Instead, the goal of designing a more general-purpose 
implementation of the path-integral Casimir paradigm
leads us to seek a representation of the boundary 
conditions that, while inevitably less efficient than 
spherical multipoles for spheres (or cylindrical 
multipoles for cylinders, or ...) has the flexibility
to handle all manner of surfaces within a single
computational framework.
This is one motivation for the fluctuating-surface-current 
(FSC)
approach to Casimir computations, whose path-integral
derivation we now discuss.

%%%%%%%%%%%%%%%%%%%%%%%%%%%%%%%%%%%%%%%%%%%%%%%%%%%%%%%%%%%%%%%%%%%%%%
%%%%%%%%%%%%%%%%%%%%%%%%%%%%%%%%%%%%%%%%%%%%%%%%%%%%%%%%%%%%%%%%%%%%%%
%%%%%%%%%%%%%%%%%%%%%%%%%%%%%%%%%%%%%%%%%%%%%%%%%%%%%%%%%%%%%%%%%%%%%%
\subsection{Path-Integral Derivation of the FSC Casimir Formulas}
\label{PathIntegralFSCDerivationSubsection}

As noted above, key features of the path-integral treatment
presented here include an unusual choice of action for the 
electromagnetic field and the introduction of surface currents 
as Lagrange multipliers constraining the photon field.
After discussing these points in 
Sections \ref{EuclideanLagrangianSection} and 
\ref{BoundaryConditionsSection}, respectively,
we show in Section \ref{ConstrainedPathIntegralSection}
how together they allow us to evaluate the constrained 
path integral for the Casimir energy to obtain equation
(\ref{MasterFSCFormulas}a).

\subsubsection{Euclidean Lagrangian for the electromagnetic field}
\label{EuclideanLagrangianSection}

The usual (Minkowski-space) Lagrangian for the electromagnetic field is 
%====================================================================%
$$ 
   S=\int \frac{d\omega}{2\pi} \, \int d\vb x \, 
     \mathcal{L}(\omega, \vb x),
$$
$$ \mathcal{L}(\omega, \vb x)
   =\frac{1}{2}
    \Big(\epsilon(\omega, \vb x) |\vb E(\omega,\vb x)|^2 
                    -\mu(\omega, \vb x) |\vb H(\omega,\vb x)|^2 
    \Big).
$$
%====================================================================%
Rewriting $\vb E$ and $\vb H$ in terms of the four-vector potential
$A^\mu$, integrating by parts, and rotating to Euclidean space 
via the prescription 
$\{ \omega, A^0, A^{0*}\} \to \{ i\xi, iA^0, iA^{0*}\}$
yields a Euclidean action density of the form 
%====================================================================%
\begin{align*}
\mathcal{L}\subs{E}(\xi, \vb x)
=&\frac{\epsilon(i\xi, \vb x)}{2}
  \Big( -\xi^2 A^{i*}A^i
        -i\xi A^{0*} \partial_i A^i
\nonumber \\&\qquad \qquad \qquad
        -i\xi A^{i*} \partial_i A^0
        +A^{0*} \partial_i\partial_i A^{0}
  \Big)
\nonumber \\
 &+\frac{1}{2\mu(i\xi, \vb x)}
  \Big( A^{i*} \partial_j \partial_j A^{i}
       -A^{i*} \partial_i \partial_j A^{j}
  \Big)
\nonumber
\end{align*}
%====================================================================%
or, introducing a convenient matrix-vector notation,
%====================================================================%
\begin{widetext}
\begin{equation}
\mathcal{L}^\prime\subs{E}(\xi, \vb x)
= \frac{1}{2}
   \left( \begin{array}{c} \mc A^0 \\ \mc A^1 \\ \mc A^2 \\ \mc A^3 \end{array} \right)^\dagger
   \left[ \begin{array}{c}
          \vspace{0.1in} \\
          \large{   \DD_1(\xi) - \DD_2(\xi) } \\
          \vspace{0.1in} 
          \end{array}
  \right]
  \left( \begin{array}{c} \mc A^0 \\ \mc A^1 \\ \mc A^2 \\ \mc A^3 \end{array} \right)
\label{LXix1} 
\end{equation}
%====================================================================%
where we have defined
%====================================================================%
\begin{equation}
  \left(\begin{array}{c}
   \mathcal{A}^0 \\
   \mathcal{A}^1 \\
   \mathcal{A}^2 \\
   \mathcal{A}^3
   \end{array}\right)
  =
   \left(\begin{array}{c}
   \sqrt{\epsilon\mu}\cdot A^0 \\
   A^1 \\ 
   A^2 \\ 
   A^3 \\ 
   \end{array}\right)
\label{NewA}
\end{equation}
and
%====================================================================%
$$
\DD_1
 = \left(\begin{array}{cccc}
   -\epsilon\xi^2 + \frac{1}{\mu}\nabla^2 & 0 & 0 & 0 \\
   0 & -\epsilon\xi^2 + \frac{1}{\mu}\nabla^2 & 0 & 0 \\
   0 & 0 & -\epsilon\xi^2 + \frac{1}{\mu}\nabla^2 & 0 \\
   0 & 0 & 0 & -\epsilon\xi^2 + \frac{1}{\mu}\nabla^2 
   \end{array}\right),
\qquad
%--------------------------------------------------------------------%
\DD_2
=\left(\begin{array}{cccc}
%--------------------------------------------------------------------%
   -\epsilon\xi^2 
   & i\sqrt\frac{\epsilon}{\mu} \xi \partial_x
   & i\sqrt\frac{\epsilon}{\mu} \xi \partial_y
   & i\sqrt\frac{\epsilon}{\mu} \xi \partial_z \\
%--------------------------------------------------------------------%
   i\sqrt\frac{\epsilon}{\mu} \xi \partial_x
   &\frac{1}{\mu}\partial_x^2  
   &\frac{1}{\mu}\partial_x \partial_y 
   &\frac{1}{\mu}\partial_x \partial_z  \\
%--------------------------------------------------------------------%
   i\sqrt\frac{\epsilon}{\mu} \xi \partial_y
   &\frac{1}{\mu}\partial_y \partial_x
   &\frac{1}{\mu}\partial_y^2
   &\frac{1}{\mu}\partial_y \partial_z  \\
%-------------------------------01------------------------------------%
   i\sqrt\frac{\epsilon}{\mu} \xi \partial_z
   &\frac{1}{\mu}\partial_z \partial_x
   &\frac{1}{\mu}\partial_z \partial_y 
   &\frac{1}{\mu}\partial_z^2
   \end{array}\right).
$$
\end{widetext}
%====================================================================%
The new four-vector field $\mathcal{A}^\mu$ defined by (\ref{NewA}) 
will be the field over which we path-integrate, and equation (\ref{LXix1}) 
is almost, but not quite, the quantity that enters into the exponent
of the constrained path-integral expression (\ref{mzbeta}). To complete
the story, we must add a Fadeev-Popov gauge fixing term, which we do 
in analogy to the usual QED procedure~\cite{PeskinSchroeder}  
by simply displacing the coefficient of $\DD_2$ term in (\ref{LXix1})
away from unity to ensure that the matrix in square brackets has no
zero eigenvalues. Our final Euclidean action is
\begin{align}
\mathcal{L}\subs{E}(\xi, \vb x)
 &=\mathcal{A}^\mu \Big[ \DD_1(\xi) - 
                          \left(1 - \frac{1}{\alpha\subt{FP}}\right)\DD_2(\xi)
                    \Big]_{\mu\nu}
    \mathcal{A}^\nu
\nonumber\\
 &\equiv \bmc{A} \cdot \DD(\xi) \cdot \bmc{A} 
\label{EuclideanAction}
\end{align}
where the Faddeev-Popov gauge-choice parameter $\alpha\subt{FP}$
may be chosen to have any finite value and is absent from all 
final physical predictions, as will be explicitly verified below 
[see equations (\ref{LULSandwich1}-\ref{LULSandwich2})]. 
Following the general procedure reviewed in 
Section \ref{PathIntegralReviewSubsection}, we can now write the
Casimir energy at inverse temperature $\beta$ in the form
%====================================================================%
\begin{equation}
   \mathcal{E}
   =-\frac{1}{\beta}\ln \frac{Z(\beta)}{Z_\infty(\beta)}
   =-\frac{1}{\beta}\sum_{n=0}^\infty 
     \ln \frac{\mathcal{Z}(\beta,\xi_n)}{\mathcal{Z}_\infty(\beta,\xi_n)}
%   \qquad\Big(\xi_n \equiv \frac{2\pi n}{\hbar \beta}\Big)
  \label{EBetaPathIntegral}
\end{equation}
%====================================================================%
\begin{equation}
   \mathcal{Z}(\beta,\xi) =
   \int \Big[\mathcal{D} \mathcal{A}^\mu \Big]\subt{C}
   e^{-\frac{\beta}{2} 
      \int \bmc{A} \cdot \DD(\xi) \cdot \bmc{A}\, d\vb x
     }
 \label{ZBetaPathIntegral}
\end{equation}
%====================================================================%
with the notation $[\cdots]\subt{C}$ indicating that the
functional integration ranges only over field configurations
that satisfy the boundary conditions in the presence of 
our interacting material objects.

%\begin{widetext}
%\end{widetext}

\subsubsection{Boundary conditions enforced by surface-current Lagrange multipliers}
\label{BoundaryConditionsSection}

%####################################################################%
%####################################################################%
%####################################################################%
\begin{figure*}
\begin{center}
\resizebox{\textwidth}{!}{\includegraphics{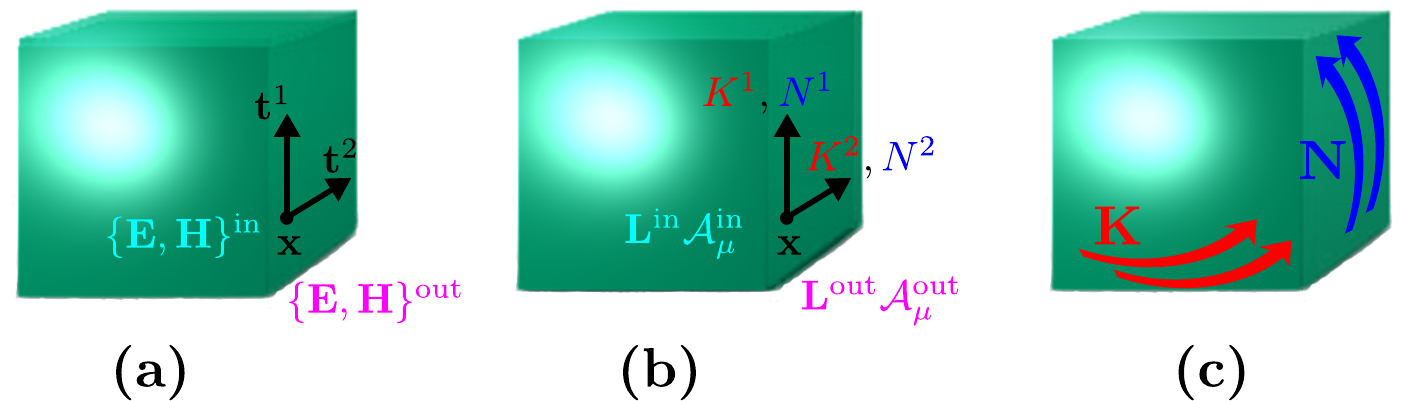}}
\caption{Enforcing boundary conditions via surface-current Lagrange
multipliers. 
\textbf{(a)} Consider a single point $\vb x$ on the surface of an 
object in a Casimir geometry. The boundary conditions at $\vb x$, 
which must be satisfied in the constrained path integral 
(\ref{ZBetaPathIntegral}), are that the tangential components 
of the $\vb E$ and $\vb H$ fields be continuous as we pass from 
inside to outside the object [equation (\ref{EHBoundaryConditions})];
here $\{\vb t^1, \vb t^2\}$ are vectors tangent to the surface
at $\vb x$.
\textbf{(b)} We rewrite the boundary conditions in terms 
of differential operators $\vb{L}\supt{E,M}$ operating
on the $\mathcal{A}$ field, and we introduce Lagrange multipliers
$\{K^1,K^2,N^1,N^2\}$
to enforce the boundary conditions at $\vb x$; specifically,
$K^1,K^2$ enforce the tangential $\vb E$-field continuity
at $\vb x$,  while 
$N^1,N^2$ enforce the tangential $\vb H$-field continuity
[equation (\ref{LagrangeMultipliersAtX})].
\textbf{(c)} Repeating this procedure for all points on 
the object surface, we obtain Lagrange multiplier 
\textit{fields} $\vb K(\vb x), \vb N(\vb x)$, which 
have an obvious interpretation as the electric and magnetic
\textit{surface currents} of Figure \ref{ScatteringCartoon}.
Integrating the photon field out of the path integral 
then yields an effective action describing the interactions
of these surface currents [equations (\ref{SEffDef}), 
(\ref{EffectiveActionSR}), and (\ref{EffectiveActionSE})], 
leading ultimately to our fluctuating-surface-current
formulas for the Casimir energy.
}
\label{SurfaceCurrentLagrangeMultiplierFigure}
\end{center}
\end{figure*}
%####################################################################%
%####################################################################%
%####################################################################%

In a scattering geometry consisting of one or more homogeneous bodies
bodies embedded in a homogeneous medium, the boundary conditions on the
electromagnetic field are simply that the tangential $\vb E$ and
$\vb H$ fields be continuous across all material boundaries:
if $\vb x$ is a point on the surface of an object, then we require
%====================================================================%
\begin{subequations}
%--------------------------------------------------------------------%
\begin{align} 
 \vb t^a \cdot 
 \Big[\vb E\sups{in}(\vb x) - \vb E\sups{out}(\vb x)\Big] 
 &=0 \\[4pt]
 \vb t^a \cdot 
  \Big[\vb H\sups{in}(\vb x) - \vb H\sups{out}(\vb x)\Big] 
 &=0
\end{align}
%--------------------------------------------------------------------%
\label{EHBoundaryConditions}
\end{subequations}
%====================================================================%
\noindent where $\{\vb E, \vb H\}\sups{in,out}$ are the fields evaluated
just inside and just outside the object surface at $\vb x$, 
and where $\vb t^a$ ($a\in\{1,2\}$) are vectors tangent to 
the surface at $\vb x$ 
[Figure \ref{SurfaceCurrentLagrangeMultiplierFigure}\textbf(a)].
In terms of the modified four-vector potential 
$\mc A^\mu$, these conditions may be 
written in the form
\begin{subequations}
\begin{align}
 t^a_i 
 \Big\{ L\supt{E,in}_{i\mu}(\vb x) - L\supt{E,out}_{i\mu}(\vb x)\Big\}
 \mc A^\mu(\vb x) &= 0 
\\[4pt]
 t^a_i 
 \Big\{ L\supt{M,in}_{i\mu}(\vb x) - L\supt{M,out}_{i\mu}(\vb x)\Big\}
 \mc A^\mu(\vb x) &= 0 
\end{align}
\label{LxA}
\end{subequations}
where $\vbler$ and $\vblmr$ are differential operators
that operate on $\mc A^\mu$ to yield the components
of the $\vb E$ and $\vb H$ fields in region $r$.
(We are here using a shorthand in which the $A^\mu$
fields in the different regions, $\mc A^{\mu,\text{in}}$ and 
$\mc A^{\mu,\text{out}}$, are abbreviated simply as $\mc A^{\mu}$  
and pulled outside the braces.)
In a homogeneous region with spatially constant relative
permittivity and permeability
$\epsilon(\xi, \vb x)=\epsilon^r(\xi), \mu(\xi, \vb x)=\mu^r(\xi),$
the $\vb L$ operators take the form
%====================================================================%
\renewcommand{\arraystretch}{1.5}
\begin{subequations}
\begin{align}
  \vbler
 &=\left(\begin{array}{cccc} 
      -\displaystyle{\frac{1}{\sqrt{\epsilon^r\mu^r}}\partial_x} 
      & i\xi  & 0 & 0     \\  
      -\displaystyle{\frac{1}{\sqrt{\epsilon^r\mu^r}}\partial_y} 
      & 0 & i\xi & 0     \\  
      -\displaystyle{\frac{1}{\sqrt{\epsilon^r\mu^r}}\partial_z} 
      & 0 & 0 & i\xi 
   \end{array}\right),
\\[10pt]
%\label{LEr} \\[5pt]
 \vblmr
 &=\frac{1}{\mu^r}
    \left(\begin{array}{cccc} 
     0 & 0          & -\partial_z & \partial_y  \\[6pt]
     0 & \partial_z & 0           & -\partial_x \\[6pt]
     0 &-\partial_y &  \partial_x & 0 \\
    \end{array}\right).
\end{align}
\label{LErLHr}
\end{subequations}
\renewcommand{\arraystretch}{1.0}
%====================================================================%
Equations (\ref{LxA}) are a set of four boundary 
conditions for each point $\vb x$ on the surfaces
of the material bodies in our geometry; in the 
language of Section \ref{PathIntegralReviewSubsection}, 
these are our
constraints $L_\alpha \phi$, and to each constraint
we now associate a Lagrange multiplier.
We use the symbols $K^a(\vb x)$ and $N^a(\vb x)$
($a=1,2$), respectively,
to denote the Lagrange multipliers associated
with constraints (\ref{LxA}a) and 
(\ref{LxA}b) at the single point $\vb x$
[Figure \ref{SurfaceCurrentLagrangeMultiplierFigure}\textbf(b)].
Then the $\delta$ functions that
enforce the boundary conditions (\ref{LxA}) at 
$\vb x$ are 
%====================================================================%
\begin{subequations}
\begin{align}
%--------------------------------------------------------------------%
 \delta\Big[  \vb E\supt{in}_\parallel(\vb x) 
             -\vb E\supt{out}_\parallel(\vb x)
       \Big]
&=\int \frac{d\vb K_{\vb x}}{(2\pi)^2} 
  e^{i\vb K_{\vb x} \cdot 
      \left[ \vb L\supt{E,in}_\mu 
             - \vb L\supt{E,out}_\mu 
      \right] \mathcal{A}^\mu(\vb x) 
    } 
\\[5pt]
%--------------------------------------------------------------------%
 \delta\Big[  \vb H\supt{in}_\parallel(\vb x)
             -\vb H\supt{out}_\parallel(\vb x)
       \Big]
&=\int \frac{d\vb N_{\vb x}}{(2\pi)^2} 
  e^{i\vb N_{\vb x} \cdot 
      \left[ \vb L\supt{M,in}_\mu 
            -\vb L\supt{M,out}_\mu 
      \right] \mathcal{A}^\mu(\vb x) 
    }
\end{align}
\label{LagrangeMultipliersAtX}
\end{subequations}
%====================================================================%
\noindent where we may think of
$\{\vb K_{\vb x},\vb N_{\vb x}\}=\sum_{a=1}^2 \{K^a,N^a\} \vb t^a$
as vectors in the tangent space to the boundary surface at 
$\vb x.$ 
Aggregating the corresponding $\delta$ functions
for all points on the surface of a single object,
we obtain \textit{functional} $\delta$-functions,
%====================================================================%
\begin{subequations}
\begin{align}
%--------------------------------------------------------------------%
\int \mathcal{D}\vb K(\vb x)
  e^{ i \int_{\partial \mc O} \vb K(\vb x) \cdot 
      \left[ \vb L\supt{E,in}_\mu 
             - \vb L\supt{E,out}_\mu 
      \right] \mathcal{A}^\mu(\vb x) d\vb x
    } 
\\[5pt]
%--------------------------------------------------------------------%
\int \mathcal{D}\vb N(\vb x)
  e^{ i \int_{\partial \mc O} \vb N(\vb x) \cdot 
      \left[ \vb L\supt{M,in}_\mu 
             - \vb L\supt{M,out}_\mu 
      \right] \mathcal{A}^\mu(\vb x) d\vb x
    } 
\end{align}
\label{FunctionalDelta}
\end{subequations}
where the integral in the exponent is over the surface 
$\partial \mc O$ of an object in our geometry, and where
the functional integrations $\int \mathcal{D} \vb K$, 
$\int \mathcal{D} \vb N$ extend over all possible tangential 
vector fields on $\partial \mc O$. 

Since $\vb K$ and $\vb N$ are tangential vector fields on 
$\partial \mc O$ that enforce the continuity of the tangential 
electric and magnetic fields, respectively, it is tempting 
to interpret these quantities as electric and magnetic surface
current densities, and with their introduction our
path-integral formalism begins to exhibit the first glimmers
of resemblance to the surface-integral-equation picture
reviewed in Section \ref{BEMReviewSection}.

\subsubsection{Evaluation of the Constrained Path Integral}
\label{ConstrainedPathIntegralSection}

In general we will have one copy of the functional 
$\delta$-functions (\ref{FunctionalDelta}) for the surface 
of each object in our geometry. Let $\{\vb K_r , \vb N_r\}$ 
denote the Lagrange-multiplier distributions
on the surface of the $r$th object; the constrained path
integral then reads
%====================================================================%
\begin{align}
%--------------------------------------------------------------------%
&\hspace{-0.0in} \mathcal{Z}(\beta, \xi)
=\int \Big[\mathcal{D} \mc A^\mu \Big]_C
  e^{ -\frac{\beta}{2}\int \bmc A \cdot\DD\cdot\bmc A\, d\vb x}
\nn
%--------------------------------------------------------------------%
&=\int \prod_r \mathcal{D}\vb K_r \mathcal{D}\vb N_r 
  \int \mathcal{D}\mc A^\mu\bigg\{ 
 e^{ -\frac{\beta}{2} \int \bmc A \cdot \DD \cdot \bmc A\, d\vb x } 
\label{UnconstrainedPathIntegralOverA}
\\
%--------------------------------------------------------------------%
&\hspace{0.3in} 
 \times
  e^{+i \sum_r \int_{\partial \mc O_r}
        \big\{ \vb K_r \cdot (\vbler - \vblee) 
              +\vb N_r \cdot (\vblmr - \vblme)
       \big\} \cdot \bmc A \, d\vb x
    }
  \bigg\}.
\nonumber
\end{align}
%====================================================================%
with $\int_{\partial \mc O_{r}}$ denoting integration over the 
surface of object $r$, and with the path-integration
over $\mathcal{A}$ in the second line now
\textit{unconstrained} 
[compare equation (\ref{unconstrained1})]. 
This is just a standard Gaussian functional integral, 
which we proceed to evaluate using standard 
techniques~\cite{PeskinSchroeder, Weinberg}.

To this end, it is convenient to think of breaking up the
functional integration over $\mc A^\mu$ into 
separate integrations over the fields in each object
and in the exterior region,
$$ \int \mathcal{D} \mc A^\mu 
 = \int \mathcal{D} \mc A_e^\mu
        \prod_r
        \mathcal{D} \mc A_r^\mu,
$$
where $\mc A^\mu_r$ is the field in the interior of region $r$ 
and $\mc A^\mu_e$ is the field in the exterior region.
The matrix $\DD(\vb x)$, which depends
on $\vb x$ through $\epsilon$ and $\mu$, is constant in each 
region due to the piecewise homogeneity of the geometry, while 
the operators $L^{\text{\tiny EM},r}$ only operate on the fields 
in region $r$.  The functional integral becomes
\begin{widetext}
\begin{align}
\mathcal{Z}(\beta, \xi)
&=\int \prod_r \mathcal{D}\vb K_r \, \mathcal{D}\vb N_r  \,
  \int
  \mathcal{D} \mc A^\mu_e \, \prod_r \mathcal{D} \mc A^\mu_r 
  \Big\{
  e^{ -\frac{\beta}{2} \int_{V_e} \bmc A_e\cdot\DD_e\cdot\bmc A_e\,
      -\frac{\beta}{2}\sum_r \int_{V_r} \bmc A_r\cdot\DD_r\cdot\bmc A_r\,
    }
\nonumber \\
&\hspace{2.00in}
  \times 
   e^{ +i\sum_r \int_{\partial \mc O_r}
         \big\{ [\vb K_r \cdot \vbler
                +\vb N_r \cdot \vblmr ]\cdot \bmc A_r 
               -[\vb K_r \cdot \vblee
                +\vb N_r \cdot \vblme ]\cdot \bmc A_e 
         \big\} d\vb x} \Big\}
\label{BigFunctionalIntegral}
\end{align}
%====================================================================%
%\begin{align}
%%--------------------------------------------------------------------%
%\mathcal{Z}(\beta, \xi)
%&=\int \prod_r \mathcal{D}\vb K_1 \, \mathcal{D}\vb N_r  \,
%  \mathcal{D} \mc A^\mu_e \, \prod_r \mathcal{D} \mc A^\mu_r 
%\bigg\{ 
%\nonumber\\
%%--------------------------------------------------------------------%
%&\hspace{-0.2in}
% e^{ -\frac{\beta}{2} \int_{V_e} \bmc A_e\cdot\DD_e\cdot\bmc A_e\,
%     -\frac{\beta}{2}\sum_r \int_{V_r} \bmc A_r\cdot\DD_r\cdot\bmc A_r\,
%   }
%\nonumber\\
%%--------------------------------------------------------------------%
%&\hspace{-0.35in}\times
%  e^{+i\sum_r \int_{\partial \mc O_r}
%        \big\{ [\vb K_r \cdot \vb L\supt{E,r}
%               +\vb N_r \cdot \vb L\supt{M,r}]\cdot \bmc A_r 
%              -[\vb K_r \cdot \vb L\supt{E,e}
%               +\vb N_r \cdot \vb L\supt{M,e}]\cdot \bmc A_e 
%        \big\} d\vb x
%    }
%  \bigg\}
%\label{BigFunctionalIntegral}
%\end{align}
\end{widetext}
%====================================================================%
with $\int_{V_r}$ denoting volume integration over the interior of region $r$.
Now performing the Gaussian functional integrations over the fields 
$\mc A_r^\mu$ immediately yields an expression of the form 
(\ref{PathIntegralOverLambda}):
%====================================================================%
\begin{equation}
\mathcal{Z}(\beta,\xi)
=\Big\{ \# \Big\}
 \int \prod_r \, \mathcal{D}\vb K_r \, \mathcal{D}\vb N_r 
  e^{-\frac{1}{\beta} S\sups{eff}}
\label{PathIntegralOverKN}
\end{equation}
where $\big\{ \# \big\}$ is an unimportant constant that cancels
upon taking the ratio in (\ref{EBetaPathIntegral}) (and which 
will be omitted from the equations below), and
where the effective action for the surface currents,
\begin{equation}
  S\sups{eff}= \sum_{r=1}^R  S_r\Big[ \vb K_r, \vb N_r \Big]
               +S_e\Big[ \big\{ \vb K_r, \vb N_r\big\}_{r=1}^R \Big],
\label{SEffDef}
\end{equation}
contains terms describing both 
the self-interactions and the mutual interactions of 
currents on the object surfaces, as we now discuss.

Consider first what happens when we integrate $\mc A_r^\mu$ 
out of (\ref{BigFunctionalIntegral}).
Because the exponent
of (\ref{BigFunctionalIntegral}) couples $\mc A_r^\mu$ only 
to $\vb K_r, \vb N_r$ (and not to currents on other objects 
$r^\prime \ne r$), integrating out $\mc A_r^\mu$ yields 
an effective action involving only $\vb K_r, \vb N_r$:
%====================================================================%
\begin{align*}
 \int \mathcal{D}\mc A_r^\mu \,
 & e^{ -\frac{\beta}{2}\int \bmc A_r \cdot \DD_r \cdot \bmc A_r 
       +i\int_{\partial \mc O_r} 
            [  \vb K_r \cdot \vbler
              +\vb N_r \cdot \vblmr
            ]\cdot \bmc A_r 
     }
\\
%--------------------------------------------------------------------%
&\hspace{-0.2in}=
 e^{ -\frac{1}{2\beta} \iint_{\partial \mc O_r}
       [  \vb K_r \cdot \vbler
         +\vb N_r \cdot \vblmr
       ]
       \cdot \vb \DD_1^{-1} \cdot
       [  \vb K_r \cdot \vbler
         +\vb N_r \cdot \vblmr
       ]
     }
\\
%--------------------------------------------------------------------%
&\hspace{-0.2in}\equiv
 e^{ -\frac{1}{\beta} S_r[\vb K_r, \vb N_r]}.
\end{align*}
%====================================================================%
The effective action $S_r$ describes the self-interactions of electric 
and magnetic 
currents on $\partial \mc O_r$ mediated by virtual photons propagating
through the interior of object $r$. More precisely, we have
%====================================================================%
\begin{align}
%--------------------------------------------------------------------%
&\hspace{-1.4in} S_r\Big[\vb K_r, \vb N_r\Big]
\label{EffectiveAction1} \\
=\frac{1}{2}\int_{\partial \mc O_r}d\vb x 
            \int_{\partial \mc O_r}d\vb x^\prime \bigg\{ 
\,& \vb K_r(\vb x) 
  \cdot 
  \bg\supt{EE,r}(\vb x, \vb x^\prime) 
  \cdot 
  \vb K_r(\vb x^\prime) 
\nonumber\\
%--------------------------------------------------------------------%
+& \vb K_r(\vb x) 
  \cdot 
  \bg\supt{EM,r}(\vb x, \vb x^\prime) 
  \cdot 
  \vb N_r(\vb x^\prime) \vphantom{\bigg\{ }
\nonumber\\
%--------------------------------------------------------------------%
+& \vb N_r(\vb x) 
  \cdot 
  \bg\supt{ME,r}(\vb x, \vb x^\prime) 
  \cdot 
  \vb K_r(\vb x^\prime) \vphantom{\bigg\{ }
\nonumber\\
%--------------------------------------------------------------------%
+& \vb N_r(\vb x) 
  \cdot 
  \bg\supt{MM,r}(\vb x, \vb x^\prime)
  \cdot 
  \vb N_r(\vb x^\prime) \bigg\}
\nonumber
%--------------------------------------------------------------------%
\end{align}
%====================================================================%
with the components of the tensor kernels given by
%====================================================================%
($\text{\small{P,Q}}\in\{\text{\small{E,H}}\}$)
\begin{equation}
\bg\supt{PQ,r}_{ij}
= L_{i\mu}^{\text{\tiny P},r} \mathfrak{D}^{-1}_{r\mu\nu} 
  L_{j\nu}^{\text{\tiny Q},r}.
\label{LDL}
\end{equation}
We will see presently that the $\bg\PQr$ matrices here 
turn out to be nothing but the usual dyadic Green's 
tensors $\BG\PQr$ for the homogeneous medium inside
object $\mathcal{O}_r.$

To see this, it is easiest to perform the 
matrix multiplications of equation (\ref{LDL})
in momentum space, where we have
\begin{widetext}
%====================================================================%
%\begin{align*}
$$
\vbler
%&=
=
 \left(\begin{array}{cccc}
 -i\frac{1}{\sqrt{\epsilon^r\mu^r}} k_x & i\xi & 0    & 0 \\ 
 -i\frac{1}{\sqrt{\epsilon^r\mu^r}} k_y & 0    & i\xi & 0 \\ 
 -i\frac{1}{\sqrt{\epsilon^r\mu^r}} k_z & 0    & 0    & i\xi
 \end{array}\right),
%\\\qquad
\qquad
\vblmr
=
 \frac{1}{\mu^r}
 \left(\begin{array}{cccc}
     0 & 0     & -ik_z & ik_y  \\[6pt]
     0 & ik_z  & 0     & -ik_x \\[6pt]
     0 & -ik_y & ik_x  & 0     \\
 \end{array}\right)
%\end{align*}
$$
and [cf. equation(\ref{EuclideanAction})]
%====================================================================%
\begin{align}
\DD_r^{-1}
&=\Big[ \DD_{r,1} - \left(1-\frac{1}{\alpha\supt{FP}}\right)\DD_{r,2} \Big]^{-1} 
\nonumber \\
&=-\left[ \frac{1}{\mu^r} \Big(\kappa^2 + |\vb k|^2\Big) \vb{1}
         -\Big( 1-\frac{1}{\alpha\subt{FP}} \Big)
          \vb \DD_{r,2}
   \right]^{-1}
\label{DDInverse0} \\
&=-\frac{\mu^r}{\kappa^2 + |\vb k|^2}
   \left[ \vb{1}
         +\mu^r\left(\frac{1-\alpha\subt{FP}}{\kappa^2 + |\vb k|^2}\right)
          \vb \DD_{r,2}
   \right].
\label{DDInverse}
\end{align}
%====================================================================%
\noindent where $\vb 1$ is the 4$\times$4 unit matrix, 
$\{\epsilon^r, \mu^r\}$ are the (spatially constant) permittivity 
and permeability of object r at imaginary frequency $\xi,$ 
$ \kappa=\sqrt{\epsilon^r \mu^r} \cdot \xi,$
and the momentum-space form of the $\DD_{r,2}$ matrix is
$$ 
   \DD_{r,2}= 
   -\frac{1}{\mu^r}
   \left(\begin{array}{cccc}
   \kappa^2   & \kappa k_x & \kappa k_y & \kappa k_z  \\
   \kappa k_x & k_x^2      & k_x k_y    & k_x k_z \\
   \kappa k_y & k_y k_x    & k_y^2      & k_y k_z \\
   \kappa k_z & k_z k_x    & k_z k_y    & k_z^2
  \end{array}\right).
$$
(The passage from the second to the third line of equation 
(\ref{DDInverse}) is a standard algebraic manipulation in quantum
field theory; see, e.g., equation (9.58) of 
Ref.~\cite{PeskinSchroeder}. The fact that we obtain such a 
concise form for the inverse of the matrix in square brackets
in (\ref{DDInverse0}) is due to the fact that $\DD_{r,2}$ is 
a rank-one matrix, and is known in numerical analysis as
the \textit{Sherman-Morrison} formula~\cite{Hager1989}.)

We now note the crucial fact that 
\textit{the second term in (\ref{DDInverse}) 
makes no contribution to the effective action for the surface 
currents.} Indeed, the contribution of this term to the 
$\bg$ kernels in (\ref{LDL}) involves triple matrix
products of the form 
$\vb L^{\text{\tiny P},r} 
 \cdot \DD_{r,2} \cdot 
 (\vb L^{\text{\tiny Q},r})\sups{T};$
but an explicit calculation reveals that $(c=1/\sqrt{\epsilon^r\mu^r})$
\begin{align}
   \vb L^{\text{\tiny E},r} \cdot \DD_{r,2} \cdot 
   (\vb L^{\text{\tiny E},r})\sups{T}
&=\frac{1}{\mu^r}
   \left(\begin{array}{cccc}
   -ic k_x & i\xi & 0    & 0 \\ 
   -ic k_y & 0    & i\xi & 0 \\ 
   -ic k_z & 0    & 0    & i\xi
   \end{array}\right)
   \cdot 
   \left(\begin{array}{cccc}
   \kappa^2   & \kappa k_x & \kappa k_y & \kappa k_z  \\
   \kappa k_x & k_x^2      & k_x k_y    & k_x k_z \\
   \kappa k_y & k_y k_x    & k_y^2      & k_y k_z \\
   \kappa k_z & k_z k_x    & k_z k_y    & k_z^2
  \end{array}\right)
    \cdot
   \left(\begin{array}{ccc}
     -ic k_x  & -ic k_y  & -ic k_z \\
         i\xi &         0  &         0 \\
            0 &       i\xi &         0 \\
            0 &          0 &      i\xi 
   \end{array}\right)
\\
&=\left(\begin{array}{ccc}
  0 & 0 & 0 \\ 
  0 & 0 & 0 \\ 
  0 & 0 & 0
 \end{array}\right)
\label{LULSandwich1}
\end{align}
and the other three possible $\vb L\cdot \DD \cdot \vb L$ products
also vanish identically:
\begin{equation}
   \vb L^{\text{\tiny E},r} \cdot \DD_{r,2} \cdot (\vb L^{\text{\tiny M},r})\sups{T}
   \quad=\quad
   \vb L^{\text{\tiny M},r} \cdot \DD_{r,2} \cdot (\vb L^{\text{\tiny E},r})\sups{T}
   \quad=\quad
   \vb L^{\text{\tiny M},r} \cdot \DD_{r,2} \cdot (\vb L^{\text{\tiny M},r})\sups{T}
   \quad=\quad 0.
\label{LULSandwich2}
\end{equation}
This furnishes the promised demonstration that the 
gauge-choice parameter $\alpha\subt{FP}$ makes
no appearance in the effective action (\ref{EffectiveAction1}), thus 
explicitly confirming the gauge invariance of our procedure.

Having verified that only the first term in (\ref{LDL}) contributes
to the $\bg$ kernels in (\ref{LDL}), these kernels are now easy to
evaluate. First,
%====================================================================%
\begin{align*}
\bg\EEr
 &=-\frac{\mu^r}{\kappa^2 + |\vb k|^2}
   \Big[\vbler \cdot \vb 1 \cdot (\vbler)\supt{T}\Big]
\\[5pt] 
 &=-\frac{\mu^r}{\kappa^2 + |\vb k|^2}
   \left(\begin{array}{cccc}
   -ic k_x & i\xi & 0    & 0 \\ 
   -ic k_y & 0    & i\xi & 0 \\ 
   -ic k_z & 0    & 0    & i\xi
   \end{array}\right)
   \cdot 
   \left(\begin{array}{ccc}
     -ic k_x  & -ic k_y  & -ic k_z \\
         i\xi &         0  &         0 \\
            0 &       i\xi &         0 \\
            0 &          0 &      i\xi 
   \end{array}\right)
\\[10pt]
&=\xi \cdot \frac{Z_0 Z^r}{\kappa(\kappa^2 + |\vb k|^2)}
  \left[
         \left(\begin{array}{ccc}
         \kappa^2 & 0 & 0 \\ 
         0 & \kappa^2 & 0 \\ 
         0 & 0 & \kappa^2 
         \end{array}\right)
         +
         \left(\begin{array}{ccc}
         k_x^2  & k_x k_y & k_x k_z \\
         k_yk_x & k_y^2   & k_y k_z \\
         k_zk_x & k_z k_y & k_z^2
         \end{array}\right)
   \right]
\end{align*}
But a quick comparison with the momentum-space forms of the dyadic 
Green's functions in Appendix \ref{DGFAppendix} reveals this to
be nothing but $\xi$ times the electric-electric dyadic Green's 
function for $\mathcal{O}_r,$ i.e.
$$ \bg\EEr(\xi; \vb k) = \xi \cdot \BG\EEr(\xi; \vb k)$$
or, transforming back to real space,
$$  \bg\EEr(\xi; \vb x,\vb x^\prime) 
  = \xi \cdot \BG\EEr(\xi; \vb x, \vb x^\prime)
$$
%====================================================================%
and the connection of our formalism to SIE methodology
begins to come into even sharper relief.

Next,
%====================================================================%
\begin{align}
\bg\EMr
 &=-\frac{\mu^r}{\kappa^2 + |\vb k|^2}
   \Big[\vbler \cdot \vb 1 \cdot (\vblmr)\supt{T}\Big]
\\
%--------------------------------------------------------------------%
 &=-\frac{1}{\kappa^2 + |\vb k|^2}
   \left(\begin{array}{cccc}
   -ic k_x & i\xi & 0    & 0 \\ 
   -ic k_y & 0    & i\xi & 0 \\ 
   -ic k_z & 0    & 0    & i\xi
   \end{array}\right)
   \cdot 
   \left(\begin{array}{ccc}
            0 & 0       & 0      \\
            0 &  ik_z   & -ik_y  \\
        -ik_z &     0   & ik_x   \\
         ik_y & -ik_x   & 0      \\
   \end{array}\right)
%--------------------------------------------------------------------%
\\
 &=-\frac{\xi}{\kappa^2 + |\vb k|^2}
   \left(\begin{array}{cccc}
   0   & -k_z &  k_y \\
   k_z &    0 & -k_x \\
  -k_y &  k_x &    0
   \end{array}\right)
\end{align}
%====================================================================%
and again comparing with Appendix \ref{DGFAppendix} reveals 
that we have simply 
$$ \bg\EMr = \xi \cdot \BG\EMr $$
Having established the obvious pattern, it is now a short step 
to confirm that the remaining two cases of the $\bg$ kernel
in (\ref{LDL}) are simply\,\footnote{A careful evaluation of 
$\bg\MMr$ reveals that the right-hand side of 
(\ref{MErMMr}) should be augmented by a $\delta$-function
term. This $\delta$ function is related to that which enters 
in the difference between the dyadic green's functions
$\nabla \times \BG\supt{EE} \times \nabla^\prime$ and
$\BG\supt{MM}$; as discussed in Ref.~\cite{Johnson2011}, 
such terms are geometry-independent and make no contribution 
to physical Casimir quantities, and for this reason we have
omitted this term in equation (\ref{MErMMr}).}
\begin{align}
 \bg\MEr &= \xi \cdot \BG\MEr
\\
 \bg\MMr &= \xi \cdot \BG\MMr
\label{MErMMr}
\end{align}
and equation (\ref{EffectiveAction1}), the portion of the 
effective action for surface currents that arises from 
integrating the photon field in the interior of object 
$\mathcal{O}_r$ out of (\ref{BigFunctionalIntegral}), reads
%====================================================================%
\begin{equation}
 S_r \Big[\vb K_r, \vb N_r \Big]
= \,\,\,\frac{\xi}{2}
  \int_{\partial \mathcal{O}_r} d\vb x \, 
  \int_{\partial \mathcal{O}{r}} d\vb x^\prime \,
   \left(\begin{array}{c} \vb K_r(\vb x) \\ \vb N_r(\vb x) \end{array}\right)
   \cdot
   \left(\begin{array}{cc}
    \BG\EEr & \BG\EMr \\
    \BG\MEr & \BG\MMr
   \end{array}\right)
   \cdot
   \left(\begin{array}{c} 
     \vb K_{r}(\vb x^\prime) \\ \vb N_{r}(\vb x^\prime) 
   \end{array}\right).
\label{EffectiveActionSR}
\end{equation}
%====================================================================%
Next, we consider integrating the photon field in the exterior 
region ($\mathcal{A}_e^\mu$) out of equation (\ref{BigFunctionalIntegral}). 
Although the computations proceed exactly as before, the
the resulting contribution to the effective action is slightly 
more complicated. Because the exponent of (\ref{BigFunctionalIntegral})
contains terms that couple $\mathcal{A}_e^\mu$ to the currents
on all object surfaces (unlike $\mathcal{A}_r^\mu$, which couples
only to surface currents on the single object $\mathcal{O}_r$),
the result of integrating out $\mathcal{A}_e^\mu$ will be 
an effective action describing the interactions of surface
currents on all object surfaces mediated    
by exchange of virtual photons propagating 
through the external medium:
%====================================================================%
\begin{equation}
 S_e \Big[\big\{\vb K_r, \vb N_r\big\}\Big]
= \,\,\,\frac{\xi}{2}\sum_{r, r^\prime}
  \int_{\partial \mathcal{O}_r} d\vb x \, 
  \int_{\partial \mathcal{O}_{r^\prime}} d\vb x^\prime \,
   \left(\begin{array}{c} \vb K_r(\vb x) \\ \vb N_r(\vb x) \end{array}\right)
   \cdot
   \left(\begin{array}{cc}
    \BG\EEe & \BG\EMe \\
    \BG\MEe & \BG\MMe
   \end{array}\right)
   \cdot
   \left(\begin{array}{c} 
     \vb K_{r^\prime}(\vb x^\prime) \\ \vb N_{r^\prime}(\vb x^\prime) 
   \end{array}\right).
\label{EffectiveActionSE}
\end{equation}
%====================================================================%
Combining (\ref{EffectiveActionSR}) and (\ref{EffectiveActionSE}),
the full path integral (\ref{PathIntegralOverKN}) now reads,
for the particular case of two objects,
%====================================================================%
\begin{equation}
\mathcal{Z}(\beta,\xi)
= \int\mathcal{D}\vb K_i \mathcal{D}\vb N_i
  \exp\left\{ -\frac{\xi}{2\beta} {\Huge\int}
%--------------------------------------------------------------------%
  \left(\begin{array}{c} 
   \vb K_1 \\ \vb N_1 \\ \vb K_2 \\ \vb N_2 
  \end{array}\right)\sups{T}
%--------------------------------------------------------------------%
  \cdot 
%--------------------------------------------------------------------%
  \left( 
   \begin{array}{ccc}
   &&\\[5pt]
   \parbox{0.1in}{\hspace{0.1in} }  & 
   \parbox{0.1in}{\huge $\BG$ }     & 
   \parbox{0.2in}{\hspace{0.2in}}   \\[5pt]
   &&
  \end{array}\right)
%--------------------------------------------------------------------%
  \cdot
%--------------------------------------------------------------------%
  \left(\begin{array}{c} 
   \vb K_1 \\ \vb N_1 \\ \vb K_2 \\ \vb N_2 
  \end{array}\right)
%--------------------------------------------------------------------%
  \right\}
\label{PathIntegralOverKN2}
\end{equation}
with
$$
  \left( \begin{array}{ccc}
   &&\\[5pt]
   \parbox{0.1in}{\hspace{0.1in} }  & 
   \parbox{0.1in}{\huge $\BG$ }     & 
   \parbox{0.2in}{\hspace{0.2in}}   \\[5pt]
   &&
  \end{array}\right)
=
  \left(\begin{array}{cccc}
   \BG\eeo + \BG\eee & \BG\emo + \BG\eme & \BG\eee & \BG\eme \\
   \BG\meo + \BG\mee & \BG\mmo + \BG\mme & \BG\mee & \BG\mme \\
   \BG\eee & \BG\eme & \BG\eet + \BG\eee & \BG\emt + \BG\eme \\
   \BG\mee & \BG\mme & \BG\met + \BG\mee & \BG\mmt + \BG\mme
  \end{array}\right).
$$
Note that the quantity $\BG \cdot \binom{\vb K}{\vb N}$ in 
(\ref{PathIntegralOverKN2}) is nothing but the left-hand
side of equation (\ref{BEMSystemPMCHW}).

Having elucidated the structure of the effective action for
surface currents, the remainder of our derivation is now 
straightforward. To evaluate the functional integral over sources in 
(\ref{PathIntegralOverKN2}), we approximate $\vb K$ and $\vb N$
as expansions in a finite set of tangential basis functions
representing surface currents flowing on the surface of the 
interacting Casimir objects, just as we did in equation 
(\ref{KNExpand}):
$$ 
   \vb K(\vb x)=\sum k_\alpha \vb f_\alpha(\vb x), 
   \qquad
   \vb N(\vb x)=- \sum n_\alpha \vb f_\alpha(\vb x).
$$
We now insert these expansions into (\ref{PathIntegralOverKN2})
and approximate the infinite-dimensional integrals over
$\vb K$ and $\vb N$ as \textit{finite-dimensional} integrals
over the $\vb k_\alpha$ and $\vb n_\alpha$ coefficients:
$$\int\mathcal{D}\vb K_i \mathcal{D}\vb N_i
  \quad\Longrightarrow \quad
  \mathcal{J} 
  \int \prod_\alpha d k_\alpha \, \prod_\alpha  dn_\alpha
$$
where $\mathcal{J}$, the Jacobian of the variable 
transformation, is an unimportant constant that cancels 
upon taking the ratio in (\ref{EBetaPathIntegral})
(and which will not be written out in the equations below).
Equation (\ref{PathIntegralOverKN2}) now becomes
simply 
\begin{align}
 \mathcal{Z}(\beta,\xi)
&= \int \prod_\alpha d k_\alpha \, \prod_\alpha  dn_\alpha 
   e^{-\frac{\xi}{2\beta}
       \binom{\vb k}{\vb n}\supt{T}
       \cdot 
       \vb M(\xi)
       \cdot 
       \binom{\vb k}{\vb n}
     },
\nonumber\\
\intertext{a finite-dimensional Gaussian integral which we evaluate
immediately to obtain}
&=\Big\{\#\Big\} \cdot \Big[ \det \vb M(\xi) \Big]^{-1/2} 
\label{ZBetaXiDetM}  
\end{align}
where $\vb M(\xi)$ is nothing but the SIE matrix
discussed in Section \ref{BEMReviewSection}
(and, once again, $\big\{\#\big\}$ is 
just an irrelevant constant into which the $\xi/2\beta$ 
prefactor in the exponent disappears).
Now finally inserting (\ref{ZBetaXiDetM}) into 
(\ref{ZBetaPathIntegral}) and (\ref{EBetaPathIntegral})
leads immediately back to our FSC formulae 
(\ref{MasterFSCFormulas}), and our derivation is complete.
\end{widetext}

\section{Equality of the Partial Traces}
\label{PartialTraceEqualitySection}

An important practical simplification of the FSC formulae
follows from the structure of the BEM matrices. Recall
from (\ref{MasterFSCFormulas}) that the quantity that
enters into the FSC expression for the Casimir force
is
\begin{align}
 \Tr \Big\{ \vb M^{-1} \cdot \frac{\partial \vb M}{\partial \vb r_i}
     \Big\}
&=
 \sum_{\alpha \beta}
  M^{-1}_{\alpha\beta} 
 \left[ \frac{\partial M_{\beta\alpha}}{\partial \vb r_i} \right]
\label{PartialTrace1}
\end{align}
(with a similar expression for the torque).
We will show that the trace in (\ref{PartialTrace1}) 
neatly splits into two equal subtraces, and thus that to
compute the full trace we need only sum a subset of 
the diagonal elements of the matrix in curly brackets
(and double the result). In practice this reduces 
the computational expense of the trace computation by 
a factor of two or greater.

To understand the physical intuition behind this 
simplification, note that the sum in (\ref{PartialTrace1})
runs over the elements of our basis of surface-current
expansion functions, which includes functions defined
on the surfaces of each of the objects in our 
Casimir geometry. In evaluating the portion of this sum 
contributed by basis functions defined on a single 
object surface, we are in effect computing a sort of 
surface integral over the surface of that object.
Intuitively we might expect that, to compute the
force on one object, it would suffice to 
evaluate this surface integral over the surface of 
that object alone, \textit{or} over the
surfaces of all other objects, but that we need not
do both. As we now show, this physical expectation is 
born out by the mathematics; to compute the full trace 
in (\ref{PartialTrace1}) we need only sum the 
contributions of basis functions on the surface of 
the object on which we are computing the force---and 
double the result.

As before, let the objects in our Casimir geometry be labeled
$\mathcal{O}_r,$ $r=1,2,\cdots$, with $\mathcal{O}_1$ the object
on which we are computing the Casimir force, and 
let $D_r$ be the dimension of the subblock of the $\vb M$ matrix 
corresponding to object $\mathcal{O}_r$ (that is, $D_r$ is the 
number of surface-current basis functions defined on the 
surface of $\mathcal{O}_r$).
The dimension of the full matrix is $D=\sum D_r$.

A rigid displacement of object $\mathcal{O}_1$ leaves
unchanged the interactions between all pairs of basis
functions save those pairs in which precisely one 
basis function lives on $\mathcal{O}_1$.
%
%on the interactions of basis functions on
%$\mathcal{O}_1$ with other basis functions on 
%$\mathcal{O}_1$, nor does it affect the interactions of
%basis functions on $\mathcal{O}_m$ with basis functions
%on $\mathcal{O}_n$ for $m,n\ge 2.$ Instead, the only
%interactions that are affected by such a displacement
%are those between basis functions on $\mathcal{O}_1$ 
%and basis functions on $\mathcal{O}_n$, $n\ge 2$. 
%This means that the elements of the derivative 
%matrix, $\partial M_{\alpha\beta} /\partial \vb r,$ 
%vanish unless we have $1,\alpha$ 
%We know that the only nonzero entries of the derivative 
%matrix $\partial \vb M/\partial r_i$ are those corresponding
%to pairs of basis functions of which precisely one is located
%on the surface of object $\mathcal{O}_1$. 
We can thus split the sum in (\ref{PartialTrace1}) into two pieces: 
\begin{align}
 \sum_{\alpha \beta}
  M^{-1}_{\alpha\beta} 
 \left[ \frac{\partial M_{\alpha\beta}}{\partial \vb r_i} \right]
=&
 \sum_{\alpha=1}^{D_1} \sum_{\beta=D_1+1}^{D}
  M^{-1}_{\alpha\beta} 
 \left[ \frac{\partial M_{\beta\alpha}}{\partial \vb r_i} \right]
\nn
\hspace{0.1in}
+&
 \sum_{\alpha=D_1+1}^{D} \sum_{\beta=1}^{D_1}
  M^{-1}_{\alpha\beta} 
 \left[ \frac{\partial M_{\beta\alpha}}{\partial \vb r_i} \right].
\end{align}
The first piece on the RHS here is the sum of the first $D_1$ 
diagonal elements of the matrix 
$\vb M^{-1}\cdot\frac{\partial{\vb M}}{d\vb r_i}$,
while the second piece is the sum of the remaining 
$D-D_1$ elements.
But from the fact that $\vb M$ is a symmetric matrix 
($M_{\alpha\beta}=M_{\beta\alpha}$) it now follows that 
the two pieces here are \textit{equal}, and thus to compute 
the full trace we need only sum the first $D_1$ diagonal 
elements of $\vb M^{-1}\cdot\frac{\partial{\vb M}}{d\vb r_i}$ 
(or the latter $D - D_1$ elements, if they are fewer) and double
the result, i.e.
$$ 
 \Tr \vb M^{-1} \cdot \frac{\partial \vb M}{\partial \vb r_i}
 =
 2\sum_{\alpha=1}^{D_1} 
  \sum_{\beta=D_1+1}^D
  M^{-1}_{\alpha\beta} 
 \left[ \frac{\partial M_{\alpha\beta}}{\partial \vb r_i} \right].
$$
In practice, a convenient way to evaluate this quantity is to
LU-factorize the matrix $\vb M$, solve the linear systems
$\vb M \cdot \vb X_m = \vb B_m$ where the vectors $\vb X_m$ are 
the first $D_1$ columns of 
$\frac{\partial \vb M}{\partial \vb r_i}$, 
then extract and sum the $m$th elements of the vectors $\vb B_m$
and double the result.
The equality of the partial traces then ensures that this
operation requires just $D_1$ linear solves, in 
contrast to the full $D$ solves that would be required
in the absence of the simplification.

%\input{ApplicationsSection}
%####################################################################%
%####################################################################%
%####################################################################%
\begin{figure*}
\begin{center}
\resizebox{\textwidth}{!}{\includegraphics{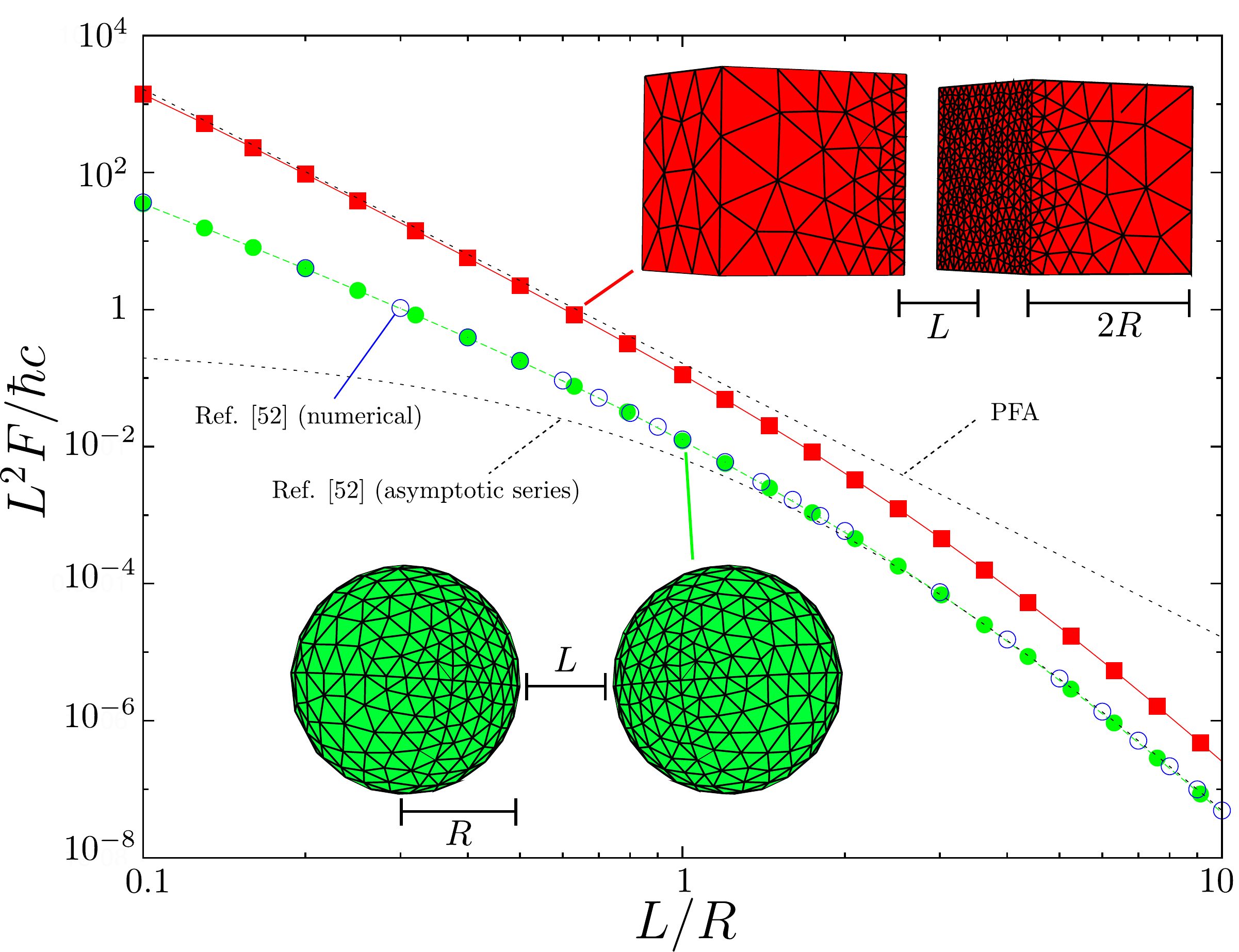}}
\caption{Casimir force between PEC spheres and between PEC cubes.
The filled green circles indicate sphere-sphere data computed
using the FSC method described in this paper, while
the hollow blue circles indicate sphere--sphere data computed
using a numerical implementation of the scattering-matrix 
method of Ref.~\cite{Emig2007}. 
(The lower dashed-line curve indicates the first four 
terms of the asymptotic series for the sphere-sphere 
Casimir force reported in Ref.~\cite{Emig2007}.)
The filled red squares indicate cube-cube data computed
using the FSC method described in this paper; for this
geometry, scattering-matrix methods and indeed almost
all existing Casimir methods would be unwieldy or impossible
to apply. The upper dashed-line curve indicates the 
proximity-force approximation (PFA) to the 
cube--cube force, 
$F\subs{PFA}=4R^2\cdot \frac{\pi^2 \hbar c}{240L^4}$ 
for cubes of face area $4R^2$.
}
\label{SpheresCubes}
\end{center}
\end{figure*}
%%%%%%%%%%%%%%%%%%%%%%%%%%%%%%%%%%%%%%%%%%%%%%%%%%
%%%%%%%%%%%%%%%%%%%%%%%%%%%%%%%%%%%%%%%%%%%%%%%%%%
%%%%%%%%%%%%%%%%%%%%%%%%%%%%%%%%%%%%%%%%%%%%%%%%%%

\section{Applications}
\label{ApplicationsSection}

\subsection{Casimir Forces between Metallic Spheres and Cubes}

To validate our new method and demonstrate its flexibility, 
we first calculate the Casimir force between pairs of metallic
particles of spherical and cubical shapes 
(Figure \ref{SpheresCubes}). 

For the sphere-sphere case, the Casimir force may be computed
using scattering-matrix methods based on a spherical-wave decomposition 
of the electromagnetic field~\cite{Emig2007,Canaguier2010},
with the Casimir energy expressed in terms of interactions among
waves labeled by the usual spherical indices $\{\ell,m\}.$
At large values of the sphere--sphere separation, 
only waves with small values of $\ell$ are relevant, 
and in this regime Ref.~\cite{Emig2007} obtained an 
asymptotic power series for the force,
whose first four terms we have plotted in 
Figure \ref{SpheresCubes} (lower dashed curve).
At smaller values of the sphere--sphere separation,
the sum over spherical waves may be evaluated numerically, 
as is done in Figure \ref{SpheresCubes} (hollow blue circles)
with waves up to $\ell=40$ retained for 
each of the two possible polarizations, 
corresponding to $2(\ell+1)^2=3362$ basis
functions for each sphere.

To perform the calculation using FSC techniques, we 
discretize the surfaces of the objects into small surface 
patches and expand surface currents using the localized
basis functions of Figure \ref{BEMCartoon}; in this 
case, the number of basis functions retained in the 
description of the surface currents is 2976 for each 
sphere. As illustrated in
Figure \ref{SpheresCubes}, the FSC calculation 
(solid red circles) reproduces the results of the 
spherical-wave calculation. Note that our choice
of surface-mesh basis functions allows us to 
concentrate more degrees of freedom in the regions
where we expect the surface source densities
to be most rapidly varying---namely, the regions
of each sphere that most closely approach the
other sphere---while simultaneously
using a relatively coarse-grained representation
of weakly interacting regions.
Thus already for the simple sphere-sphere geometry
our method begins to exhibit practical advantages over 
spherical-wave-basis methods, in which the resolution
can only be increased \textit{globally} rather 
than locally.

The FSC method really comes into its own in treating
objects that \textit{cannot} be efficiently described
by analytical Maxwell solutions, such as the case of
two cubes. To handle such a geometry using 
scattering-matrix methods, we would be forced either
to expand fields in and around the cubes in spherical
waves---an approach which would require retaining
basis functions up to inordinately large values of 
$\ell$ except in the long-distance 
limit~\cite{McCauleyChiral}---or to reformulate the 
method in a basis of Maxwell solutions for cubical 
scatterers, for which analytical expressions are 
not available. In contrast, the FSC approach handles 
the geometry with no more effort than is required 
for the sphere-sphere case (inset and solid blue 
circles in Figure~\ref{SpheresCubes}).

An additional limitation of the spherical-wave 
approach is that it is inherently restricted to 
separation distances large enough that the 
interacting objects may be enclosed in 
non-touching, nonoverlapping spheres.
For the cube--cube geometry pictured in
Figure (\ref{SpheresCubes}), this would
render the method inapplicable for
distances $L/R < 1.47$, excluding much
of the range plotted in the figure.

\subsection{Repulsion of an anisotropic nanoparticle 
from a square aperture in a thin metallic plate}

%####################################################################%
%####################################################################%
%####################################################################%
\begin{figure*}
\begin{center}
\resizebox{\textwidth}{!}{\includegraphics{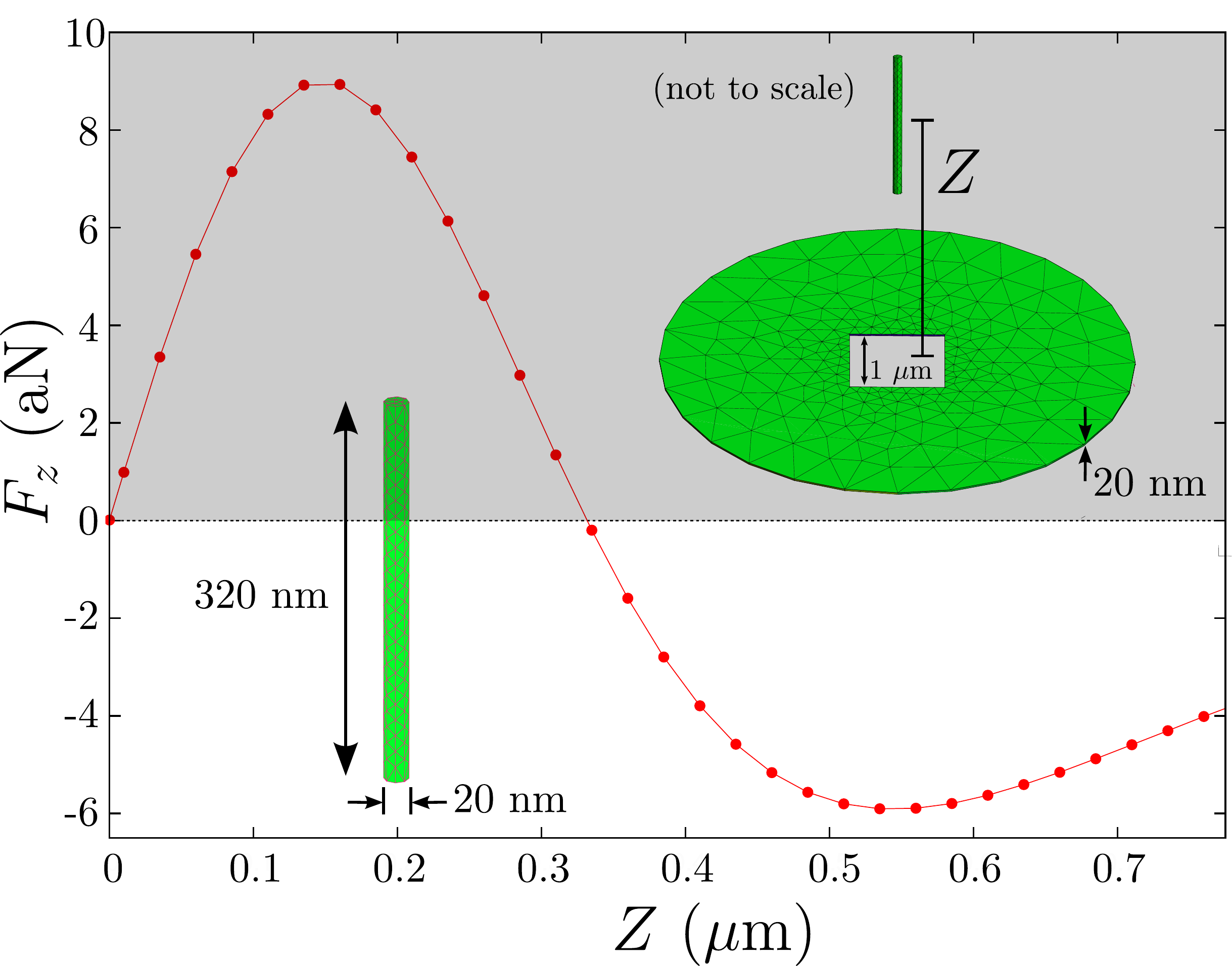}}
\caption{Casimir force on an elongated cylindrical nanoparticle 
above a square aperture in a thin plate. The upper inset (not to
scale) shows the particle--plate geometry, while the lower inset
is a close-up view of the surface mesh used to represent the
particle. The axis of the nanoparticle 
(the $z$-axis) is perpendicular to the plane of the plate; the center 
of the nanoparticle lies a distance $Z$ above the center of the plate. 
In the regime $0\le Z \le 330$ nm, the $z$-directed force on the 
nanoparticle is positive, i.e. the particle is repelled from 
the plate; the shaded portion of the graph indicates this
repulsive-force regime. For $Z>185$ the nanoparticle lies entirely
above the plate and the force is thus unambiguously repulsive. 
Both the nanoparticle and the plate are made of finite-conductivity 
gold (see text).
}
\label{PlateWithHoleFigure}

\end{center}
\end{figure*}

Reference~\cite{Levin2010} considered the Casimir
force on an elongated nanoparticle above an aperture 
in a thin metallic plate and predicted a regime in 
which the force on the nanoparticle is \textit{repulsive}. 
This work, as well as subsequent investigations of 
similar phenomena~\cite{Milton2012,Eberlein2011},
considered only \textit{circular} apertures; here we 
investigate the case of a \textit{square} aperture
(Figure \ref{PlateWithHoleFigure}). We consider 
the Casimir force on a cylindrical nanoparticle 
(lower inset), whose axis coincides with the 
axis of a thin (20 nm) plate with a 1 $\mu$m
square aperture, as a function 
of the distance $Z$ between the vertical center 
of the nanoparticle and the center of the thin 
plate.  Both cylinder and plate are made of 
real (lossy) gold, described by a 
relative dielectric function 
$\epsilon(\xi)=1+w_p^2/[\xi(\xi+\gamma)]$
with $\{w_p, \gamma\}=\{1.37\cdot 10^{16}, 
5.23\cdot 10^{13}\}$ rad/s.
When the center of the nanoparticle is vertically 
aligned with the center of the plate ($Z=0$), the 
$z$-directed Casimir force vanishes by symmetry; 
as the nanoparticle is displaced slightly in the 
positive $z$ direction it experiences first a 
\textit{repulsive} Casimir force (shaded region 
of plot) which peaks near the value of $Z$ at
which the nanoparticle first lies entirely above
the plate. As the nanoparticle is raised further
above the plate, the repulsive force decreases 
in magnitude and eventually crosses over
into an attractive force whose magnitude decays
at large $Z$ (unshared region of the plot).
 
The fact that the square-hole geometry reproduces
the repulsion phenomenon observed in the circular-hole
case is not particularly surprising, but we note
that this geometry exhibits several features which
would make it prohibitively expensive if not outright 
impossible to treat using any other Casimir method of which 
we are aware. In particular, the interpenetrating 
nature of the nanoparticle--plate configuration would
immediately render most scattering-matrix methods
inapplicable. Moreover, the drastic distance in length
scales between the tiny (20 nm diameter) cylinder
and the relatively large (5 $\mu$m radius) plate,
together with the absence of rotational symmetry,
would pose severe challenges to finite-difference methods.
In contrast, the FSC method easily accomodates
nonuniform surface meshes, allowing us to describe
the nanoparticle surface as a union of appropriately-sized
surface patches even as we use much larger patches
for the plate surface.

\section{Conclusions}
\label{ConclusionsSection}

The fluctuating-surface-current (FSC) approach to Casimir
physics, together with other recently-developed 
Casimir methods such as the finite-difference 
approaches~\cite{Rodriguez2007,PasqualiMaggs2008,Rodriguez2009,McCauley2010}, 
constitutes an advance in the development of Casimir 
algorithms that mirrors an earlier evolution in computational 
electromagnetism (EM). In the latter field, traditional 
special-function approaches such as Mie's method for
spherical scattering began to be complemented 
in the 1960s and 70s 
by a host of numerical techniques that 
expanded the range of geometries that could be 
accurately and efficiently treated. 
Today, numerical techniques for general geometries
coexist with geometry-specific special-function 
approaches to constitute a rich and varied arsenal 
of computational EM techniques appropriate for
almost any conceivable situation.

In the future, we expect a similar situation to 
prevail in the Casimir field. General-purpose methods 
such as the one presented here will not \textit{replace} 
scattering-matrix Casimir methods, any more than 
finite-difference Maxwell solvers have replaced 
the theory of Mie scattering; instead, methods such as 
our FSC technique will coexist with scattering-matrix 
methods, augmenting the toolbox of available methods
available for predicting Casimir interactions across 
the gamut of geometric and material configurations.

Among general-geometry numerical Casimir methods, 
the FSC approach is unique in obtaining compact 
determinant and trace formulas for Casimir 
quantities. All other general-purpose numerical
methods rely on numerical evaluation of a surface
integral for the Casimir force and torque (and on
even more unwieldy numerical \textit{volume} 
integrations for the Casimir energy). For this
reason, we expect the FSC approach to be the 
most efficient numerical Casimir method for the 
piecewise-homogeneous material configurations to 
which it applies.

In addition to the practical usefulness of the
FSC Casimir formulas, the two independent 
derivations that we have provided in this
paper contain a number of theoretical innovations
that we expect to find broader application. In 
particular, in the stress-tensor derivation we 
stated and proved a new integral identity 
involving the homogeneous dyadic Green's functions
of Maxwell's equations
(Appendix \ref{IntegralIdentityAppendix}), 
while in the path-integral derivation we introduced
a new type of Lagrange multiplier to constrain 
the functional integration over the photon field.
We hope the latter technique will prove to be a 
generally useful tool in quantum field theory;
one possible application beyond the Casimir realm 
is the boundary dependence of entanglement entropy
in the electromagnetic field, a subject recently 
addressed for scalar fields~\cite{Hertzberg2011}.

What challenges lie ahead for FSC Casimir computations?
If the evolution of computational Casimir physics
continues to mimic that of computational electromagnetism,
an obvious next step will be the development of
\textit{fast solvers}~\cite{Chew1997}---algorithms that exploit physical
insight to reduce the computational complexity of matrix 
manipulations in equations like (\ref{MasterFSCFormulas}) 
from $O(N^3)$ to a more tractable scaling such as 
$O(N\log N)$, where $N,$ the dimension of the matrix, is 
the number of surface-current expansion functions
$\{\vb f_\alpha\}$ retained in equations 
(\ref{KExpansion}) and (\ref{KNExpand}).
Although a number of algorithms are known for evaluating
matrix-vector products involving SIE matrices in 
$O(N \log N)$ time~\cite{Chew1997},
the question of how best to exploit such algorithms
to evaluate the determinant and trace in equations 
(\ref{MasterFSCFormulas}) is nontrivial. To date, all 
FSC Casimir calculations have considered
problems of moderate complexity ($N\lesssim 10^4$), 
for which dense-direct linear-algebra solvers are 
adequate; however, future problems may require
going beyond this regime, in which case fast 
solvers will be essential.

A separate challenge is to apply FSC techniques 
to the calculation of Casimir forces out of 
thermal equilibrium, as well as to the closely 
related problem of near-field radiative heat
transfer. These problems, which represent 
a logical next step in the study of 
fluctuation-induced forces beyond the equilibrium
Casimir case, have begun very recently to be 
investigated using 
scattering-matrix~\cite{Krueger2011A, Krueger2011B, Messina2011} and 
numerical~\cite{Rodriguez2011B} methods. 
A hybrid technique combining scattering-matrix
ideas with SIE-based numerical calculations
was proposed in~\cite{McCauley2011}, but
was restricted to a cylindrical-wave 
basis; to date there has been no 
basis-independent surface-current formulation 
of nonequilibrium fluctuation problems. Could 
the FSC technique presented in this paper 
be modified to apply to these problems? If so,
 what physical insight would the surface-current 
formulation lend, and what improvements 
could be achieved in the efficiency of 
practical calculations?

In this paper we have said almost nothing about the
practical challenges inherent in concrete numerical
implementations of the FSC formulas. Among these are
\textbf{(a)} how best to choose the surface-current 
basis functions $\{\vb f_\alpha\}$ for a given
geometry, 
\textbf{(b)} how to evaluate the multidimensional 
integrals that enter into the elements of the $\vb M$
matrix [such as equations (\ref{MElementsPEC})];
\textbf{(c)} how to compute the matrix determinant,
inverse, and trace in equations (\ref{MasterFSCFormulas}),
and 
\textbf{(d)} how to evaluate the imaginary-frequency
integrations in equations (\ref{MasterFSCFormulas}), 
as well as the Matsubara sums in their 
finite-temperature analogues.
All of these challenges will be addressed in 
subsequent publications.

\begin{acknowledgments}

We are grateful to Alejandro Rodriguez for providing 
the scattering-matrix sphere-sphere data of Figure \ref{SpheresCubes}.
The authors are grateful for support from the Singapore-MIT 
Alliance Computational Engineering flagship research program.
This work was supported in part by the Defense Advanced Research
Projects Agency (DARPA) under grant N66001-09-1-2070-DOD, by the Army
Research Office through the Institute for Soldier Nanotechnologies
(ISN) under grant W911NF-07-D-0004, and by the AFOSR Multidisciplinary
Research Program of the University Research Initiative (MURI) for
Complex and Robust On-chip Nanophotonics under grant FA9550-09-1-0704.

\end{acknowledgments}

%**********************************************************************
%**********************************************************************
%**********************************************************************
\appendix

\section{Homogeneous and Scattering Dyadic Green's Functions}
\label{DGFAppendix}

For reference, and to define our notation, we collect here
some well-known results concerning the dyadic Green's 
functions of classical electromagnetic 
theory~\cite{MorseFeshbach, Harrington1961}.

In the presence of known volume densities of electric and
magnetic current $\vb J(\vb x), \vb M(\vb x)$ at a fixed
imaginary frequency $\xi$, the components of the electric 
and magnetic fields are given by linear convolution relations 
of the form
%====================================================================%
\begin{align*}
E_i(\vb x) &= 
\int \Big\{  \Gamma\supt{EE}_{ij}(\vb \xi; \vb x, \vb x^\prime) J_j(\vb x^\prime)
             +\Gamma\supt{EM}_{ij}(\vb \xi; \vb x, \vb x^\prime) M_j(\vb x^\prime)
     \Big\} d\vb x^\prime \\
H_i(\vb x) &= 
\int \Big\{  \Gamma\supt{ME}_{ij}(\vb \xi; \vb x, \vb x^\prime) J_j(\vb x^\prime)
             +\Gamma\supt{MM}_{ij}(\vb \xi; \vb x, \vb x^\prime) M_j(\vb x^\prime)
     \Big\} d\vb x^\prime.
\end{align*}
%====================================================================%
These relations define the four dyadic Green's functions
$\boldsymbol{\Gamma}$.

In an infinite homogeneous medium with spatially constant
relative permeability and permittivity
$\epsilon(\xi; \vb x) = \epsilon^r(\xi)$,
$\mu(\xi; \vb x) = \mu^r(\xi)$, the four $\boldsymbol{\Gamma}$
functions may be expressed in terms of just two tensors:
%====================================================================%
\begin{subequations}
\begin{align}
%--------------------------------------------------------------------%
\BG\EEr(\xi, \vb r, \vb r^\prime)
&=
-Z_0 Z^r \kappa^r  \, \vb G(\kappa^r, \vb r- \vb r^\prime)
\\[5pt] 
%--------------------------------------------------------------------%
\BG\MEr(\xi, \vb r, \vb r^\prime) 
&=
\kappa^r \vb C(\kappa^r, \vb r- \vb r^\prime)
\\[5pt]
%--------------------------------------------------------------------%
\BG\EMr(\xi, \vb r, \vb r^\prime)
&=
-\kappa^r \, \vb C(\kappa^r, \vb r- \vb r^\prime)
\\[5pt]
%--------------------------------------------------------------------%
\BG\MMr(\xi, \vb r, \vb r^\prime)
&=
-\frac{\kappa^r}{Z_0 Z^r} \vb G(\kappa^r, \vb r- \vb r^\prime)
\end{align}
\label{BGfromGC}
\end{subequations}
%====================================================================%
$$\bigg(Z_0=\sqrt\frac{\mu_0}{\epsilon_0},
         \qquad
         Z^r=\sqrt\frac{\mu^r}{\epsilon^r}, 
         \qquad
         \kappa^r=\sqrt{\mu_0 \mu^r \epsilon_0 \epsilon^r}\cdot \xi
  \bigg)
$$
%====================================================================%
where $\vb G$, sometimes referred to as the ``photon Green's function,''
is the solution to the equation
\begin{equation}
 \Big[\nabla \times \nabla \times + \kappa^2 \Big]\vb G(\kappa; \vb r)
 =\delta(\vb r)\vb{1}
 \label{DyadicGFG}
\end{equation}
and $\vb C$ is defined by  
\numeq{DyadicGFC}
{\vb C = \frac{1}{\kappa} \nabla \times \vb G.}
(Note that $\vb G$ and $\vb C$ have dimensions of inverse
length, while the $\BG$ functions have dimensions of 
field/surface current density; for example, $\BG\supt{ME}$ has
dimensions of magnetic field / electric surface current density.)

Explicit expressions for the components of $\vb G$ and $\vb C$ are 
\begin{equation}
G_{ij} =
  \Big[\delta_{ij} -\frac{1}{\kappa^2} \partial_i \partial_j \Big]
  G_0, \qquad
  C_{ij} = -\frac{1}{\kappa} \varepsilon_{ijk} \partial_k G_0
\label{GCfromG0}
\end{equation}
where $G_0$ is the scalar Green's function for the Helmholtz equation,
%====================================================================%
\begin{equation}
G_0(\kappa; \vb r-\vb r^\prime) 
   =
  \frac{e^{-\kappa|\vb r-\vb r^\prime|}}{4\pi |\vb r - \vb r^\prime|}
\label{ScalarGF}
\end{equation}
%====================================================================%

which satisfies
$$ \Big[ \nabla^2 - \kappa^2 \Big] G_0(\kappa; \vb r-\vb r^\prime)
   =\delta(\vb r-\vb r^\prime).
$$
%%With these expressions, we can verify that equation (\ref{DyadicGFC}) 
%%is actually just the first half of a pair of reciprocal curl identities 
%%relating $\vb G$ and $\vb C:$
%%%====================================================================%
%%\begin{equation}
%%  \frac{1}{\kappa^r} \nabla \times \vb G=\vb C, 
%%\qquad
%%  \frac{1}{\kappa^r} \nabla \times \vb C=-\vb G.
%%\label{ReciprocalCurlIdentities}
%%\end{equation}
%%%====================================================================%
%%(As usual with tensors and dyadics, the vector notation here
%%is suggestive but vague; the precise meaning of (\ref{DyadicGFC}) is 
%%%====================================================================%
%%\begin{equation}
%%   \frac{1}{\kappa} \varepsilon_{iAB} \partial_A G_{Bj} = C_{ij},
%%   \qquad
%%   \frac{1}{\kappa} \varepsilon_{iAB} \partial_A C_{Bj} = -G_{ij}.)
%%\label{ReciprocalCurlIdentities2}
%%\end{equation}
%%====================================================================%
\subsection*{Momentum-Space Representations}
The momentum-space decomposition of the scalar Green's function
(\ref{ScalarGF}) is
$$
 G_0(\kappa; \vb r)
 =\int \frac{d\vb k}{(2\pi)^3}
  \tilde G_0(\kappa; \vb k) e^{i\vb k \cdot \vb r}
$$
with 
$$ \tilde G_0(\kappa; \vb k) = \frac{1}{\kappa^2 + |\vb k|^2}.$$
(In what follows we will generally omit the $\sim$ designation, 
relying on context to differentiate between real- and 
momentum-space functions.)

The momentum-space version of (\ref{GCfromG0}) reads
%====================================================================%
\begin{equation}
G_{ij} =
  \Big[\delta_{ij} + \frac{k_i k_j}{\kappa^2} \Big]
  G_0, \qquad
C_{ij} = -\frac{i}{\kappa} \varepsilon_{ijk} k_k G_0
\label{GCfromG0MomentumSpace}
\end{equation}
%====================================================================%
or, in matrix format,
%====================================================================%
\begin{subequations}
\begin{align}
\vb G(\vb k) 
&= \frac{1}{\kappa^2(\kappa^2 + |\vb k|^2)}
   \left[
         \left(\begin{array}{ccc}
         \kappa^2 & 0 & 0 \\ 
         0 & \kappa^2 & 0 \\ 
         0 & 0 & \kappa^2 
         \end{array}\right)
   \right. 
\\
&\hspace{1.0in}+
   \left.
         \left(\begin{array}{ccc}
         k_x^2  & k_x k_y & k_x k_z \\
         k_yk_x & k_y^2   & k_y k_z \\
         k_zk_x & k_z k_y & k_z^2
         \end{array}\right)
   \right]
\nonumber\\[10pt]
\vb C(\vb k) 
&= \frac{i}{\kappa(\kappa^2 + |\vb k|^2)}
    \left(\begin{array}{ccc}
    0    & -k_z & k_y \\ 
    k_z  &    0 &-k_x \\
   -k_y  &  k_x &   0 
    \end{array}\right).
\end{align}
\label{GCMomentumSpace}
\end{subequations}
%====================================================================%
\subsection*{Scattering Dyadic Green's Functions}
In a general inhomogeneous region, the dyadic Green's functions
may be expressed as the sum of two terms,
%====================================================================%
\begin{equation}
\BG\supt{EE}(\xi; \vb x, \vb x^\prime)
=
\BG\supt{EE,\textbf{x}}(\xi; \vb x - \vb x^\prime)
+ \bmc G\supt{EE}(\xi; \vb x, \vb x^\prime)
\label{DGFdecomposition}
\end{equation}
%====================================================================%
(and similarly for the other three $\BG$ functions);
here $\BG\supt{EE,{\textbf{x}}}$ is the 
\textit{homogeneous} DGF for an infinite medium with
constant $\epsilon^r,\mu^r$ set equal to their values 
at $\vb x$, and $\bmc G\supt{EE}$
is the \textit{scattering part} of the DGF, which 
describes the fields scattered from the inhomogeneities 
in the geometry. The first term in (\ref{DGFdecomposition})
is singular as $\vb x\to\vb x^\prime$, but the second
term is perfectly well-defined in that limit 
and is the quantity that enters into the 
fluctuation--dissipation expressions for the spectral
density of fluctuations in products of field components,
as discussed in Section \ref{StressTensorFSCSection}.

%Before proceeding we pause briefly to comment on our choice of 
%notation. The four dyadic Green's functions 
%$\BG\supt{PQ}$ 
%($\text{\small{P,Q}}\in\{\text{\small{E,M}}\}$)
%depend both on the
%imaginary frequency and on the homogeneous region in
%which they are defined, through the material-property functions
%$\{\epsilon^r(i\xi), \mu^r(i\xi)\}$ for that region.
%Thus these functions are properly thought of as functions of
%imaginary frequency, and they carry a superscript ($r$) 
%indexing the homogeneous region in which they are defined, i.e.
%$$\BG\supt{PQ}=\BG^{\text{\tiny{PQ},r}}(\xi,\vb r,\vb r^\prime).$$
%In contrast, the $\vb G$ and $\vb C$ dyadics are universal
%functions, defined mathematically by (\ref{GCfromG0}) and
%(\ref{ScalarGF}), that depend on $\epsilon$, $\mu,$ and $\xi$
%only through the single parameter $\kappa$, and hence these
%functions are 
%\begin{align*}
% \vb G &= \vb G(\kappa; \vb r, \vb r^\prime) \\
% \vb C &= \vb C(\kappa; \vb r, \vb r^\prime).
%\end{align*}

%\input{IntegralIdentityAppendix}

\section{Proof of Integral Identities}
\label{IntegralIdentityAppendix}
%####################################################################%
%####################################################################%
%####################################################################%
\begin{figure*}
\begin{center}
\begin{tabular}{|c|c|c|}\hline
  \resizebox{0.30\textwidth}{!}{\includegraphics{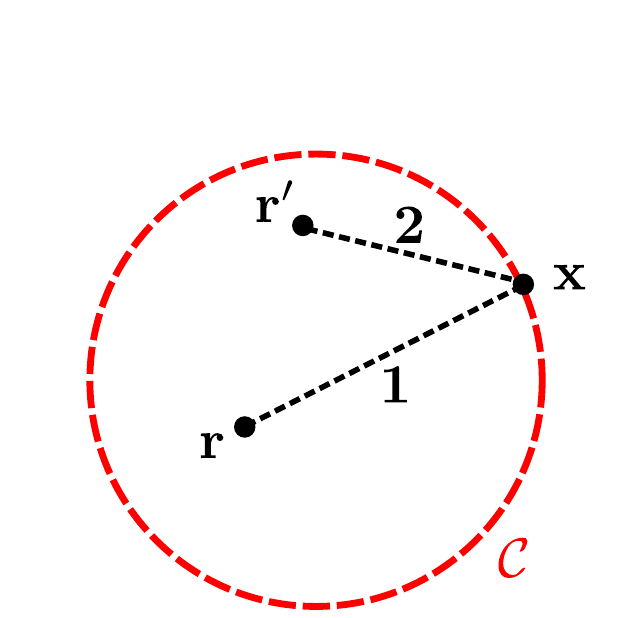}}
 &\resizebox{0.30\textwidth}{!}{\includegraphics{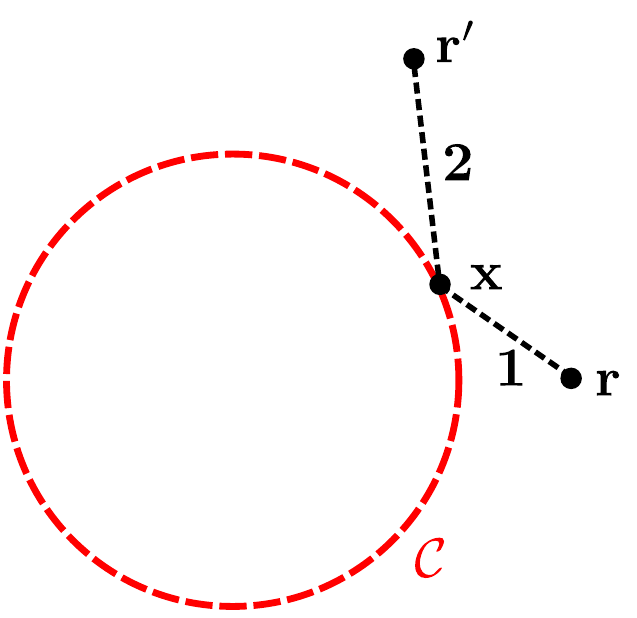}}
 &\resizebox{0.30\textwidth}{!}{\includegraphics{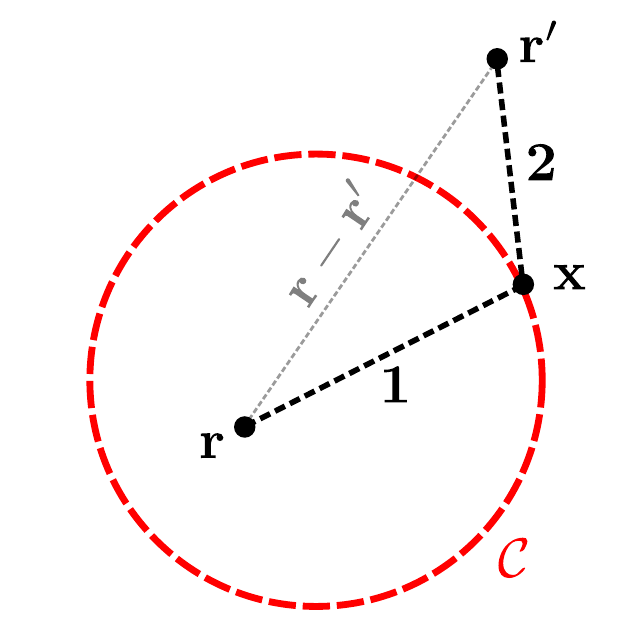}}
\\\hline
%--------------------------------------------------------------------%
   \parbox{0.30\textwidth}
     {\large $$ \oint_{\mathcal C} \Big\{ G(\vb 1) G(\vb 2) + \cdots \Big\} d\vb x
         \, = \, 
         \textcolor{red}{0}
      $$}
&
%--------------------------------------------------------------------%
   \parbox{0.30\textwidth}
     {\large $$ \oint_{\mathcal C} \Big\{ G(\vb 1) G(\vb 2) + \cdots \Big\} d\vb x
         \, = \, 
         \textcolor{red}{0}
      $$}
&
%--------------------------------------------------------------------%
   \parbox{0.36\textwidth}
     {\large 
       $$ \oint_{\mathcal C} \Big\{ G(\vb 1) G(\vb 2) + \cdots \Big\} d\vb x
         \, \propto\, 
%         \textcolor{red}{\nabla G(\vb r,\vb r^\prime)}
         \textcolor{red}{\frac{\partial G(\vb r,\vb r^\prime)}{\partial \vb r}}
       $$}
%--------------------------------------------------------------------%
\\\hline
\end{tabular}
\end{center}
\caption{Schematic summary of the integral identity 
(\ref{IBarResult}).
proved in
Appendix \ref{IntegralIdentityAppendix}.
We consider a closed bounding surface $\mathcal{C}$ and choose
two points $\vb r$ and $\vb r^\prime$, which may 
lie both inside (left panel), both outside (center panel),
or on opposite sides (right panel) of $\mathcal{C}.$
We write an expression [equation (\ref{VBarDef})] that
involves products of dyadic Green's functions (DGFs),
one connecting $\vb r$ to a point $\vb x$ on $\mathcal{C}$
and a second connecting $\vb x$ to $\vb r^\prime.$
Then we evaluate the surface integral of this expression as
$\vb x$ ranges over all of $\mathcal{C}$. 
The answer we obtain depends on the relative positioning
of $\vb r$ and $\vb r^\prime$ with respect to $\mathcal{C}$:
If the two points lie both inside or both outside $\mathcal{C}$, 
the surface integral vanishes, while if the two points lie on 
opposite sides of $\mathcal{C}$ then the surface integral yields 
the derivative of a DGF connecting $\vb r$ to $\vb r^\prime$.
}
\label{IntegralIdentityCartoon}
\end{figure*}

In this Appendix we state and prove a new integral identity 
that underlies the stress-tensor derivation of the FSC
formulae presented in Section \ref{StressTensorFSCSection}.
To our knowledge, this integral identity is new and is stated, 
proved, and used for the first time in this paper, although
an identity bearing at least a superficial resemblance
appears in equation A.6 of Ref.~\cite{ScheelBuhmann2008}.

In Section \ref{StressTensorFSCSection}, we introduced a 
three-index integral kernel
$\mathcal{I}$ defined by
\begin{equation}
 \mathcal{I}_{ikl}(\vb r, \vb r^\prime)
  =\kappa^2 
   \oint_{\mathcal{C}} V_A(\vb r, \vb r^\prime, \vb x) 
   \, n_A(\vb x) \, d\vb x
\label{IdefAppendix}
\end{equation}
where the integration is over a closed surface in space (a topological
two-sphere) $\mathcal{C}$ and the integrand contains products of factors 
of dyadic Green's functions:
%====================================================================%
\numeq{VADef}
{ V_A(\vb r, \vb r^\prime, \vb x)
   =
   \GO_{ik}\GT_{lA} 
  -\frac{\delta_{i A}}{2} \GO_{Bk}\GT_{lB}
  -\CO_{ik}\CT_{lA} 
  +\frac{\delta_{iA}}{2} \CO_{Bk}\CT_{lB}.
}
%====================================================================%
Here the $\vb G$ and $\vb C$ dyadics are those defined by 
equations (\ref{DyadicGFG}) and (\ref{DyadicGFC}); we use
capital Roman letters ($A,B,\cdots$) to denote
contracted indices, and we are using a shorthand notation in 
which $\kappa$ arguments are suppressed and spatial arguments 
are replaced by superscripts,
$$\GO_{ij} \equiv G_{ij}(\kappa, \vb x-\vb r), 
   \qquad
  \GT_{ij} \equiv G_{ij}(\kappa, \vb r^\prime-\vb x).
$$
\begin{widetext}
We also defined a symmetrized version of $\mathcal{I}:$
%====================================================================%
\begin{align}
 \overline{\mathcal{I}}_{ikl}(\vb r, \vb r^\prime)
&=\mathcal{I}_{ikl}(\vb r, \vb r^\prime)
 +\mathcal{I}_{ilk}(\vb r^\prime, \vb r) \nn
&\equiv \kappa^2 \oint_{\mathcal C} \overline{V}_A(\vb x) \, n_A(\vb x) \, d\vb x
\label{IBarDef}
\end{align}
%====================================================================%
with
%====================================================================%
\numeq{VBarDef}
{ \overline{V}_A(\vb x)
 =\GO_{ik} \GT_{lA} + \GO_{kA} \GT_{il}
 - \delta_{iA}\GO_{Bk}\GT_{lB}
 -\CO_{ik} \CT_{lA} - \CO_{kA} \CT_{il} 
 + \delta_{iA}\CO_{Bk}\CT_{lB}.
}
%====================================================================%
The goal of this Appendix is to demonstrate that, by appealing
to the defining properties of the $\vb G$ and $\vb C$ 
dyadics, \textit{the surface integral in (\ref{IBarDef}) can be 
evaluated in closed form}, with the result 
(depicted schematically in Figure \ref{IntegralIdentityCartoon})
\numeq{IBarResult}
{\overline{\mathcal{I}}_{ikl}(\vb r, \vb r^\prime)
 =\begin{cases}
  0, 
  \qquad &\text{if both $\vb r, \vb r^\prime$ lie inside $\mathcal C$} 
  \\[8pt]
  %--------------------------------------------------------------------%
  \displaystyle{
  \pard{}{\vb r_i} G_{kl}(\vb r-\vb r^\prime) 
   }
  \qquad &\text{if $\vb r$ lies inside and $\vb r^\prime$ lies 
                outside $\mathcal C$}
  \\[12pt]
  %--------------------------------------------------------------------%
  \displaystyle{
  -
  \pard{}{\vb r_i} G_{kl}(\vb r-\vb r^\prime)
   }
  \qquad &\text{if $\vb r$ lies outside and $\vb r^\prime$ lies 
                inside $\mathcal C$}
  \\[8pt]
  %--------------------------------------------------------------------%
  0, 
  \qquad &\text{if both $\vb r, \vb r^\prime$ lie outside $\mathcal C$}.
  \end{cases}
}

[As far as we can tell, the symmetrization in equation \ref{IBarDef}
is necessary to achieve the compact form of equation (\ref{IBarResult});
our attempts to evaluate the non-symmetrized $\mathcal{I}$ in 
concise form were unsuccessful.]

For the purposes of this Appendix it will be convenient to
work in length units such that $\kappa=1.$ With this
convention, the $\vb G$ and $\vb C$ dyadics are related
to the scalar Green's function for the Helmholtz equation
according to 
\numeq{GCfromG02}
{ G_{ij}=\Big[\delta_{ij} - \PP{i}{j}\Big] G_0, \qquad
  C_{ij}=-\varepsilon_{ijk} \partial_k G_0,
}
and $G_0$ satisfies
\numeq{G0Def}
{ \Big[\partial_A \partial_A - 1 \Big] G_0(\vb r) = \delta(\vb r).
}
%
%====================================================================%

The evaluation of the surface integral (\ref{IBarDef}) now proceeds
in several stages.
%%%%%%%%%%%%%%%%%%%%%%%%%%%%%%%%%%%%%%%%%%%%%%%%%%
%%%%%%%%%%%%%%%%%%%%%%%%%%%%%%%%%%%%%%%%%%%%%%%%%%
%%%%%%%%%%%%%%%%%%%%%%%%%%%%%%%%%%%%%%%%%%%%%%%%%%
\subsection{Apply Divergence Theorem}
%%%%%%%%%%%%%%%%%%%%%%%%%%%%%%%%%%%%%%%%%%%%%%%%%%
The first step is to recast the surface integral in (\ref{IBarDef})
as a volume integral over the volume bounded by $\mathcal{C}$,
\begin{align*}
\oint_{\mathcal{C}} \overline{V}_A n_A \, dA 
&=\int_{\mathcal{V}} \partial_A \overline{V}_A \, dV 
  \qquad\Big( \mathcal{C} = \partial\mathcal{V} \Big)
\\ 
&=\int_{\mathcal{V}} W \, dV 
\end{align*}
where we put $W\equiv \partial_A \overline{V}_A.$

Each of the six terms in $\overline{V}_A$ 
contains two factors and hence contributes two terms
to $W$ (by the chain rule for differentiation). Terms
of the form $\partial_A C_{iA}$ vanish;
to each of the remaining (nonvanishing) terms we assign
a label, as tabulated in Table \ref{IBarProofTable1}.
Note that derivatives of $\GT_{ij}$ enter with a minus sign,
because we are differentiating with respect to $\vb x$, which 
enters the argument of $\GT_{ij}$ with a minus sign. 
%====================================================================%
\begin{table}[h]
\begin{center}
\caption{Terms in $W\equiv \partial_A \Vbar_A$.}
\renewcommand{\arraystretch}{2.5}
$$
 \begin{array}{|c|c|}\hline
   \text{\textbf{Label}}
&  \text{\textbf{Term}}
\\\hline
%--------------------------------------------------------------------%
 W^{\textbf{1}}
&  \Big[\partial_A \GO_{ik} \Big] 
   \Big[\GT_{lA}\Big] 
\\\hline
%--------------------------------------------------------------------%
 W^{\textbf{2}}
& -\Big[\GO_{ik} \Big] 
   \Big[\partial_A \GT_{lA} \Big] 
\\\hline
%--------------------------------------------------------------------%
 W^{\textbf{3}}
&  \Big[\partial_A \GO_{kA} \Big] 
   \Big[\GT_{il} \Big] 
\\\hline
%--------------------------------------------------------------------%
 W^{\textbf{4}}
& -\Big[\GO_{kA} \Big] 
   \Big[\partial_A \GT_{il} \Big] 
\\\hline
%--------------------------------------------------------------------%
 W^{\textbf{5}}
& -\Big[\partial_i \GO_{kA} \Big] 
   \Big[\GT_{lA} \Big] 
\\\hline
\end{array}
\qquad
 \begin{array}{|c|c|}\hline
   \text{\textbf{Label}}
&  \text{\textbf{Term}}
\\\hline
%--------------------------------------------------------------------%
 W^{\textbf{6}}
&  \Big[\GO_{kA} \Big] 
   \Big[\partial_i \GT_{lA} \Big]  
\\\hline
%--------------------------------------------------------------------%
 W^{\textbf{7}}
& -\Big[\partial_A \CO_{ik} \Big] 
   \Big[\CO_{lA} \Big] 
\\\hline
%--------------------------------------------------------------------%
 W^{\textbf{8}}
&  \Big[\CO_{kA} \Big] 
   \Big[\partial_A \CT_{il} \Big] 
\\\hline
%--------------------------------------------------------------------%
 W^{\textbf{9}}
&  \Big[\partial_i \CO_{Ak} \Big] 
   \Big[\CT_{lA} \Big] 
\\\hline
%--------------------------------------------------------------------%
 W^{\textbf{10}}
& -\Big[\CO_{Ak} \Big] 
   \Big[\partial_i \CT_{lA} \Big] 
\\\hline
%--------------------------------------------------------------------%
\end{array}$$
\label{IBarProofTable1}
\end{center}
\end{table}
%====================================================================%

%%%%%%%%%%%%%%%%%%%%%%%%%%%%%%%%%%%%%%%%%%%%%%%%%%
%%%%%%%%%%%%%%%%%%%%%%%%%%%%%%%%%%%%%%%%%%%%%%%%%%
%%%%%%%%%%%%%%%%%%%%%%%%%%%%%%%%%%%%%%%%%%%%%%%%%%
\subsection{Treatment of $W^{\textbf{2}}$ and $W^{\textbf{3}}$}
We first consider the terms labeled $W^{\textbf{2}}$
and $W^{\textbf{3}}$ in Table \ref{IBarProofTable1}. Starting 
with the first of these, we have
%====================================================================%
\begin{align}
 W^{\textbf{2}}
&=-\Big[\GO_{ik} \Big] 
   \Big[\partial_A \GT_{lA} \Big] 
\nonumber
%--------------------------------------------------------------------%
\intertext{Expand the second factor using (\ref{GCfromG02}):}
%--------------------------------------------------------------------%
&=-\Big[\GO_{ik} \Big] 
   \Big[\partial_l \big(1 - \PP{A}{A}\big) \GZT \Big]
\nonumber
%--------------------------------------------------------------------%
\intertext{Apply (\ref{G0Def}):}
&=+\Big[\GO_{ik} \Big] 
   \Big[\partial_l \delta(\textbf{2})\Big] \nonumber
\intertext{Integrate by parts:}
&\sim 
 +\Big[\partial_l \GO_{ik} \Big]
  \Big[\delta(\textbf{2})\Big]
\label{W2}
\end{align}
%====================================================================%
where $\sim$ means ``equivalent as long as we are 
underneath the volume-integration sign,'' and 
$\delta(\vb 2)$ is shorthand for $\delta(\vb r^\prime-\vb x).$
(Note that the minus
sign coming from the integration by parts is cancelled by the minus
sign coming from the fact that $\vb x$ enters $\vb 2$ with a minus
sign, as noted above.)

By analogous operations, we find
\numeq{W3}
{ W^{\textbf{3}} \sim 
 -\Big[\delta(\textbf{1})\Big]
  \Big[\partial_k \GT_{il} \Big]
}
where
$\delta(\vb 1)$ is shorthand for $\delta(\vb x-\vb r).$
We now set aside results (\ref{W2}) and (\ref{W3}) for 
future use.

%%%%%%%%%%%%%%%%%%%%%%%%%%%%%%%%%%%%%%%%%%%%%%%%%%
%%%%%%%%%%%%%%%%%%%%%%%%%%%%%%%%%%%%%%%%%%%%%%%%%%
%%%%%%%%%%%%%%%%%%%%%%%%%%%%%%%%%%%%%%%%%%%%%%%%%%
\subsection{Treatment of Remaining Terms}

\subsubsection{Rewrite in terms of $G_0$}

Turning next to the remaining 8 terms in Table 1, we begin
by using (\ref{GCfromG02}) to rewrite everything in terms of 
the scalar Green's function:
%====================================================================%
\begin{align}
&\hspace{-0.2in} 
W^{\textbf{1}} + W^{\textbf{4}} + W^{\textbf{5}} + W^{\textbf{6}}
\nn
&=\Big[\partial_A \GO_{ik} - \partial_i \GO_{kA} \Big]
  \Big[\GT_{lA}\Big] 
 +\Big[\GO_{kA} \Big] 
  \Big[\partial_i \GT_{lA} - \partial_A \GT_{il}\Big]  
\nonumber\\[5pt]
&=\Big[\delta_{ik} \partial_A \GZO - \delta_{kA} \partial_i \GZO \Big]
  \Big[\delta_{lA} \GZT - \PP{l}{A} \GZT \Big]
\nn
&\qquad
 +\Big[\delta_{kA} \GZO - \PP{k}{A} \GZO \Big]
  \Big[\delta_{lA} \partial_i \GZT - \delta_{il} \partial_A \GZT \Big]
\label{W1456}
\\[8pt]
%--------------------------------------------------------------------%
&\hspace{-0.2in} 
 W^{\textbf{7}} + W^{\textbf{8}} + W^{\textbf{9}} + W^{\textbf{10}}
\nn
&=
-\varepsilon_{ikB} \varepsilon_{lAC} 
  \Big[ \PP{A}{B} \GZO \Big]
  \Big[ \partial_{C} \GZT \Big]
 +\varepsilon_{kAB} \varepsilon_{ilC} 
  \Big[ \partial_{B} \GZO \Big]
  \Big[ \PP{A}{C} \GZT \Big]
\nn
&\,\,\,\,
 +\varepsilon_{AkB} \varepsilon_{lAC}
  \Big[ \PP{i}{B} \GZO \Big]
  \Big[ \partial_{C} \GZT \Big]
 -\varepsilon_{AkB} \varepsilon_{lAC}
  \Big[ \partial_B \GZO \Big]
  \Big[ \PP{i}{C} \GZT \Big].\nonumber
\end{align}
%====================================================================%
We can trade Levi-Civita symbols for Kronecker deltas using the 
standard identity
%====================================================================%
\begin{align*}
&\hspace{-0.1in}\varepsilon_{ABC} \varepsilon_{DEF} 
\\
&=\delta_{AD}\Big[\delta_{BE}\delta_{CF} - \delta_{BF}\delta_{CE}\Big]
 +\delta_{AE}\Big[\delta_{BF}\delta_{CD} - \delta_{BD}\delta_{CF}\Big]
 +\delta_{AF}\Big[\delta_{BD}\delta_{CE} - \delta_{BE}\delta_{CD}\Big].
\end{align*}
%====================================================================%
We then find  
%====================================================================%
\begin{align}
&\hspace{-0.2in}
 W^{\textbf{7}} + W^{\textbf{8}} + W^{\textbf{9}} + W^{\textbf{10}}
\nn
&=
  \Big[\PP{A}{A}\GZO\Big] 
  \Big[   \delta_{il} \partial_k \GZT 
        - \delta_{kl} \partial_i \GZT 
  \Big]
  -
  \Big[   \delta_{ik} \partial_l \GZO
        - \delta_{kl} \partial_i \GZO 
  \Big]
  \Big[\PP{A}{A}\GZT\Big] 
\nn
&\quad 
 -\delta_{il} \Big[ \PP{k}{A}    \GZO \Big] 
              \Big[ \partial_{A} \GZT \Big] 
 +\delta_{ik} \Big[ \partial_{A} \GZO \Big] 
              \Big[ \PP{l}{A}    \GZT \Big] 
\nn
&\quad
 -\Big[\partial_i \GZO\Big]
  \Big[\PP{k}{l}  \GZT\Big]
 +\Big[\PP{k}{l}  \GZO\Big]
  \Big[\partial_i \GZT\Big].
%--------------------------------------------------------------------%
\label{W78910}
\end{align}
%====================================================================%
\subsubsection{Label Individual Terms}

To proceed, we now assign a new label to each separate term in 
(\ref{W1456}) and (\ref{W78910}):
$$
   W^{\textbf{1}} + W^{\textbf{4}} + W^{\textbf{5}} + W^{\textbf{6}}
 + W^{\textbf{7}} + W^{\textbf{8}} + W^{\textbf{9}} + W^{\textbf{10}}
 =\sum_{\textbf{n}=\textbf{1}}^\textbf{16} X^{\textbf{n}}
$$
%====================================================================%
\begin{table}[h]
\begin{center}
\renewcommand{\arraystretch}{2.5}
\caption{Terms in $X$.}
$$
\begin{array}{|c|c|c|}\hline
   \text{\textbf{Label}}
&  \text{\textbf{Term}} 
\\\hline
%--------------------------------------------------------------------%
 X^{\textbf{1}}
&-\delta_{kl} 
  \Big[\partial_i \GZO \Big] 
  \Big[\GZT \Big]
\\\hline
%--------------------------------------------------------------------%
 X^{\textbf{2}}
& \Big[\partial_i \GZO \Big] 
  \Big[\PP{k}{l}\GZT \Big]
\\\hline
%--------------------------------------------------------------------%
 X^{\textbf{3}}
& \delta_{ik} 
  \Big[\partial_l \GZO \Big] 
  \Big[\GZT \Big]
\\\hline
%--------------------------------------------------------------------%
 X^{\textbf{4}}
&-\delta_{ik} 
  \Big[\partial_A \GZO \Big] 
  \Big[\PP{A}{l}\GZT \Big]
\\\hline
%--------------------------------------------------------------------%
 X^{\textbf{5}}
&-\delta_{il} 
  \Big[\GZO \Big] 
  \Big[\partial_k \GZT\Big]
\\\hline
%--------------------------------------------------------------------%
 X^{\textbf{6}}
& \delta_{kl} 
  \Big[\GZO \Big] 
  \Big[\partial_i \GZT\Big]
\\\hline
%--------------------------------------------------------------------%
 X^{\textbf{7}}
& \delta_{il} 
  \Big[\PP{A}{k} \GZO \Big] 
  \Big[\partial_A \GZT\Big]
\\\hline
%--------------------------------------------------------------------%
 X^{\textbf{8}}
&-\Big[\PP{k}{l} \GZO \Big] 
  \Big[\partial_i \GZT\Big]
\\\hline
%--------------------------------------------------------------------%
\end{array}
\qquad
\begin{array}{|c|c|c|}\hline
   \text{\textbf{Label}}
&  \text{\textbf{Term}} 
\\\hline
%--------------------------------------------------------------------%
 X^{\textbf{9}}
&-\delta_{kl} 
  \Big[\PP{A}{A} \GZO \Big] 
  \Big[\partial_i \GZT \Big]
\\\hline
%--------------------------------------------------------------------%
 X^{\textbf{10}}
&+\partial_{il}
  \Big[\PP{A}{A}  \GZO \Big] 
  \Big[\partial_k \GZT \Big]
\\\hline
%--------------------------------------------------------------------%
 X^{\textbf{11}}
&+\delta_{kl} 
  \Big[\partial_i \GZO \Big] 
  \Big[ \PP{A}{A} \GZT \Big]
\\\hline
%--------------------------------------------------------------------%
 X^{\textbf{12}}
&-\delta_{ik} 
  \Big[\partial_l \GZO \Big] 
  \Big[\PP{A}{A}\GZT \Big]
\\\hline
%--------------------------------------------------------------------%
 X^{\textbf{13}}
&-\delta_{il} 
  \Big[\PP{A}{k} \GZO \Big] 
  \Big[\partial_A \GZT\Big]
\\\hline
%--------------------------------------------------------------------%
 X^{\textbf{14}}
&+\delta_{ik} 
  \Big[\partial_A \GZO \Big] 
  \Big[\PP{l}{A} \GZT\Big]
\\\hline
%--------------------------------------------------------------------%
 X^{\textbf{15}}
&-\Big[\partial_i \GZO \Big] 
  \Big[\PP{k}{l} \GZT\Big]
\\\hline
%--------------------------------------------------------------------%
 X^{\textbf{16}}
& \Big[\PP{k}{l} \GZO \Big] 
  \Big[\partial_i \GZT\Big]
\\\hline
%--------------------------------------------------------------------%
\end{array}$$
\end{center}
\end{table}
%====================================================================%
%

%%%%%%%%%%%%%%%%%%%%%%%%%%%%%%%%%%%%%%%%%%%%%%%%%%
%%%%%%%%%%%%%%%%%%%%%%%%%%%%%%%%%%%%%%%%%%%%%%%%%%
%%%%%%%%%%%%%%%%%%%%%%%%%%%%%%%%%%%%%%%%%%%%%%%%%%
\subsubsection{Recombine Terms}

We first note the obvious cancellations:
$$\begin{array}{lclclcl}
 X^{\textbf{2}} + X^{\textbf{15}} 
 &= 
 &0
 &\qquad
 &X^{\textbf{4}} + X^{\textbf{14}} 
 &= 
 &0
\\
 X^{\textbf{7}} + X^{\textbf{13}} 
 &=
 &0
 &\qquad 
 &X^{\textbf{8}} + X^{\textbf{16}} 
 &=
 &0
\end{array}$$
Summing and appropriately recombining the remaining terms, we find
%====================================================================%
\begin{align*}
\hspace{-0.2in} 
  X^{\textbf{1}} + X^{\textbf{3}}  + X^{\textbf{11}}  + X^{\textbf{12}}
%--------------------------------------------------------------------%
&=\Big[ \delta_{kl} \partial_i \GZO 
       -\delta_{ik} \partial_l \GZO 
  \Big]
  \Big[ \PP{A}{A} \GZT - \GZT \Big]
%--------------------------------------------------------------------%
\intertext{and}
%--------------------------------------------------------------------%
  X^{\textbf{5}} + X^{\textbf{6}} + X^{\textbf{9}} + X^{\textbf{10}} 
&=\Big[ \PP{A}{A} \GZO - \GZO \Big]
  \Big[ \delta_{il} \partial_k \GZO 
       -\delta_{kl} \partial_i \GZO 
  \Big].
\end{align*}
%====================================================================%
%
From (\ref{GCfromG02}) it follows that
$$ \Big[ \delta_{kl} \partial_i G_0
        -\delta_{ik} \partial_l G_0
   \Big]
  =
   \Big[ \partial_{i} G_{kl} - \partial_l G_{ik} \Big].
$$
Using this and (\ref{G0Def}), 
we can rewrite the previous two equations in the form
%====================================================================%
\begin{align}
\hspace{-0.2in} 
  X^{\textbf{1}} + X^{\textbf{3}}  + X^{\textbf{11}}  + X^{\textbf{12}}
%--------------------------------------------------------------------%
&=\Big[ \partial_i \GO_{kl}
       -\partial_l \GO_{ik} 
  \Big]
  \Big[ \delta(\vb 2) \Big]
\label{X131112}\\
%--------------------------------------------------------------------%
\intertext{and}
%--------------------------------------------------------------------%
  X^{\textbf{5}} + X^{\textbf{6}} + X^{\textbf{9}} + X^{\textbf{10}} 
&=\Big[ \delta(\vb 1) \Big]
  \Big[ \partial_k \GT_{il} 
       -\partial_i \GT_{kl} \Big].
\label{X56910}
\end{align}

%%%%%%%%%%%%%%%%%%%%%%%%%%%%%%%%%%%%%%%%%%%%%%%%%%
%%%%%%%%%%%%%%%%%%%%%%%%%%%%%%%%%%%%%%%%%%%%%%%%%%
%%%%%%%%%%%%%%%%%%%%%%%%%%%%%%%%%%%%%%%%%%%%%%%%%%
\subsection{Final Steps}

Finally, we combine equations (\ref{W2}), (\ref{W3}), 
(\ref{X131112}) and (\ref{X56910}) to obtain
%====================================================================%
\begin{align*}
 \overline{\mathcal{I}}_{ikl}
&=\int_{\mathcal V} \Big\{
  W^{\textbf{2}} + W^{\textbf{3}}
 +X^{\textbf{1}} + X^{\textbf{3}}  + X^{\textbf{11}}  + X^{\textbf{12}}
 +X^{\textbf{5}} + X^{\textbf{6}}  + X^{\textbf{9}}  + X^{\textbf{10}}
 \Big\}\, d\vb x
\\
%--------------------------------------------------------------------%
&=\int_{\mathcal V}
  \Big\{ 
    \Big[ \partial_i \GO_{kl} \Big] 
    \delta(\vb 2)  
   -
    \delta(\vb 1)
    \Big[ \partial_i \GT_{kl} \Big]
  \Big\} d\vb x
\end{align*}
%====================================================================%
Writing out the function arguments, this reads
\numeq{IBarFinal}
{
 \overline{\mathcal{I}}_{ikl}(\vb r, \vb r^\prime)
=\int_{\mathcal V}
  \bigg\{ 
    \Big[ \partial_i G_{kl}(\vb x-\vb r) \Big]
    \delta(\vb r^\prime - \vb x)
   -
    \delta(\vb x-\vb r)
    \Big[ \partial_i G_{kl}(\vb r^\prime-\vb x) \Big]
  \bigg\}d\vb x.
}
We now proceed on a case-by-case basis depending on the positions
of $\vb r, \vb r^\prime$. 
\begin{itemize}
 \item First, if $\vb r$ and $\vb r^\prime$ both lie outside 
       the bounding surface $\mathcal C$, then neither $\delta$ 
       function contributes and we have $\overline{\mathcal{I}}=0.$
 \item If $\vb r$ lies inside $\mathcal C$ while 
       $\vb r^\prime$ lies outside $\mathcal C$, 
       then only the second $\delta$ function contributes, and we find
       \begin{align*}
         \mathcal{I}_{ikl}(\vb r, \vb r^\prime)
         &= -\partial_{i} G_{kl}(\vb r^\prime-\vb r)
         \\
         &= +\partial_{i} G_{kl}(\vb r-\vb r^\prime)
       \end{align*}
 \item If $\vb r$ lies outside $\mathcal C$ while 
       $\vb r^\prime$ lies inside $\mathcal C$, 
       then only the first $\delta$ function contributes, and we find
       \begin{align*}
         \mathcal{I}_{ikl}(\vb r, \vb r^\prime)
         &= \partial_{i} G_{kl}(\vb r^\prime-\vb r)
         \\
         &= -\partial_{i} G_{kl}(\vb r-\vb r^\prime).
       \end{align*}
 \item Finally, if both $\vb r$ and $\vb r^\prime$ lie inside 
       $\mathcal C$ then both $\delta$ functions contribute, their 
       contributions cancel, and we find 
       $\overline{\mathcal{I}}=0$.
\end{itemize}

The result (\ref{IBarResult}) is thus established, and our proof
is complete.

\subsection{The $\overline{\mathcal{J}}$ Kernel}

The development of Section \ref{DielectricStressTensorDerivationSection}
also makes reference to a version
of the $\mathcal{I}$ kernel defined in analogy to 
equations (\ref{IdefAppendix}) and (\ref{VADef}), but with the
``$GG-CC$'' structure of (\ref{VADef}) replaced by a 
``$CG+GC$'' structure:
\numeq{JdefAppendix}
{
 \mathcal{J}_{ikl}(\vb r, \vb r^\prime)
  =\kappa^2 
   \oint_{\mathcal{C}} Y_A(\vb r, \vb r^\prime, \vb x)
   \, n_A(\vb x) \, d\vb x
}
\numeq{XADef}
{ Y_A(\vb r, \vb r^\prime, \vb x)
   =
   \CO_{ik}\GT_{lA} 
  -\frac{\delta_{i A}}{2} \CO_{Bk}\GT_{lB}
  +\GO_{ik}\CT_{lA} 
  -\frac{\delta_{iA}}{2} \GO_{Bk}\CT_{lB}.
}
We also define a symmetrized version defined in analogy to 
(\ref{IBarDef}):
$$ \overline{\mathcal{J}}_{ikl}(\vb r, \vb r^\prime)
  =  \mathcal{J}_{ikl}(\vb r, \vb r^\prime)
   + \mathcal{J}_{ilk}(\vb r^\prime, \vb r).
$$
In Appendix \ref{DGFAppendix}
%(Equation \ref{ReciprocalCurlIdentities}) 
we noted that the curl 
operation takes $\vb G$ into $\vb C$ and $\vb C$ into $-\vb G$.
[Technically, in the latter case there is an additional 
$\delta$ function, which we neglect for reasons
discussed in conjunction with equations (\ref{MErMMr})
above.]
With this observation we see that (\ref{XADef}) is obtained
from (\ref{VADef}) simply by taking the curl with respect to
the $k$ index, and thus that the $\overline{\mathcal{J}}$ 
kernel is the result of the same operation applied to the
$\overline{\mathcal{I}}$ kernel:
\begin{align*}
\overline{\mathcal{J}}_{ikl}(\vb r, \vb r^\prime)
&=\frac{1}{\kappa} \varepsilon_{kAB} \partial_{A} \overline{\mathcal{I}}_{iBl}
(\vb r, \vb r^\prime)
\\[10pt]
&=\begin{cases}
  0, 
  \qquad &\text{if both $\vb r, \vb r^\prime$ lie inside $\mathcal C$} 
  \\[8pt]
  %--------------------------------------------------------------------%
  \displaystyle{
  \pard{}{\vb r_i} C_{kl}(\vb r-\vb r^\prime) 
   }
  \qquad &\text{if $\vb r$ lies inside and $\vb r^\prime$ lies 
                outside $\mathcal C$}
  \\[12pt]
  %--------------------------------------------------------------------%
  \displaystyle{
  -\pard{}{\vb r_i} C_{kl}(\vb r-\vb r^\prime)
   }
  \qquad &\text{if $\vb r$ lies outside and $\vb r^\prime$ lies 
                inside $\mathcal C$}
  \\[8pt]
  %--------------------------------------------------------------------%
  0, 
  \qquad &\text{if both $\vb r, \vb r^\prime$ lie outside $\mathcal C$}.
  \end{cases}
\end{align*}
\end{widetext}

%**********************************************************************
%**********************************************************************
%**********************************************************************

%\bibliography{PRA0212}
%merlin.mbs 2010-03-15 4.21a (PWD, AO, DPC)

\end{document}